\documentclass[a4paper,12pt]{article}
\usepackage[utf8]{inputenc}

\usepackage{comment}

\usepackage[margin=2.5cm]{geometry}
\usepackage{graphicx}
\usepackage{caption}

\usepackage{subfig}

\usepackage{hyperref}

\usepackage{amsmath}

\usepackage{color}

\usepackage{authblk}

\begin{document}

\title{Quasinormal modes of slowly rotating Kerr-Newman black holes using the double series method %
}

\author[1]{Jose Luis Bl\'azquez-Salcedo\thanks{\href{mailto:jlblaz01@ucm.es}{jlblaz01@ucm.es}}}
\author[2]{Fech Scen Khoo\thanks{\href{mailto:fech.scen.khoo@uni-oldenburg.de}{fech.scen.khoo@uni-oldenburg.de}}} 
\affil[1]{Departamento de F\'isica Te\'orica and IPARCOS, Facultad de Ciencias F\'isicas, Universidad Complutense de Madrid, Spain}
\affil[2]{Institute of  Physics, University of Oldenburg, D-26111 Oldenburg, Germany}

\date{\today}

\maketitle

\begin{abstract}
We calculate the spectrum of quasinormal modes of slowly rotating Kerr-Newman black holes. Using a perturbative double expansion method, second order in rotation and first order in non-radial perturbations, we obtain the system of equations that describe polar-led and axial-led perturbations. We analyse gravitational, electromagnetic and scalar {fundamental} modes, focusing on the $\mathrm{l}=2$ perturbations. We reproduce previous results and check that isospectrality between axial and polar-led perturbations is approximately satisfied with good accuracy. Our results show that the slow rotation approximation can be used to estimate with reasonable precision the spectrum of configurations up to 50-60$\%$ of the extremal angular momentum.
\end{abstract}

\section{Introduction}

{Since} the first detection {GW150914} of black hole coalescence by the LIGO/Virgo Collaboration \cite{LV1}, 
such merger {systems in the} strong gravity regime provide unique opportunities to test general relativity (GR) as well as alternative gravity theories (see e.g. \cite{Faraoni:2010pgm,Berti:2015itd,CANTATA:2021ktz}). 
The reasons for considering gravity theories beyond GR %
{are multiple, ranging from the problem of the quantization of gravity via the issue of the presence of singularities all the way to the presence of the dark components in cosmology.}
{Alternative gravity theories often involve an additional scalar degree of freedom.
Depending on its coupling to gravity and matter, its presence can give rise to interesting physical phenomena such as spontaneous scalarization of neutron stars \cite{Damour:1993hw} or black holes with scalar hair \cite{Kanti:1995vq,Doneva:2017bvd,Silva:2017uqg,Antoniou:2017acq}}.

{Quasinormal modes (QNMs) emitted in the ringdown of merger events will allow to detect deviations from GR with increasing precision in future observations
{from such as the Einstein Telescope 
\cite{Punturo:2010zz}}.
Therefore the prediction of QNMs of compact objects in alternative gravity theories is of significant relevance. 
The QNMs of the Schwarzschild and Kerr black holes are well studied (see e.g. \cite{Kokkotas:1999bd,Nollert:1999ji,Rezzolla:2003ua,Berti:2009kk,Konoplya:2011qq, Hatsuda:2020egs}).
In alternative gravity theories the calculation of QNMs has so far been restricted to the static case (see e.g. \cite{Blazquez-Salcedo:2018pxo}), except for the very recent extension to slow rotation for the case of Einstein-dilaton-Gauss-Bonnet gravity}
{in the small coupling approximation}
\cite{Pierini:2021jxd,Pierini:2022eim},
{dynamical Chern-Simons gravity \cite{Wagle:2021tam}},
Einstein-bumblebee gravity \cite{Liu:2022dcn},
and higher-derivative gravity
\cite{Cano:2020cao, Cano:2021myl}.

Indeed, the inclusion of rotation represents a challenging task for the determination of QNMs. This is seen already when electromagnetic charge is taken into account, extending the Kerr black holes to the Kerr-Newman (KN) black holes.
KN black holes represent the most general (asymptotically flat) electro-vac black hole solutions of GR. 

The KN metric was discovered in \cite{Newman:1965my}. 
A more recent review of the KN metric can be found in \cite{Adamo:2014baa}.
According to the no-hair theorems {\cite{Chrusciel:2012jk},  KN black holes are} described by only three parameters, namely the mass, angular momentum, and charge.

Due to the technical obstructions presented by the {set of} coupled electromagnetic and gravitational perturbation {equations} \cite{Chandrasekhar:1985kt}, the {resulting} difficulties in the computations of the QNM spectrum of KN {black holes have} only been circumvented in recent years %
\cite{Pani:2013ija,Pani:2013wsa}, albeit in a slow-rotation approximation.
{On the other hand,} QNMs of weakly charged KN black holes for arbitrary spins have been computed in \cite{Mark:2014aja}.
A major purpose for {this} endeavour {has been the quest to establish} linear mode stability of the KN spacetime \cite{Mashhoon:1985cya}.
Strong evidence for {linear mode} stability was {finally} provided by Dias et al., first in \cite{Dias:2015wqa},
who considered the full rotation of the KN black holes without any approximations.
Starting from the Newman-Penrose formalism \cite{Newman:1961qr}, the authors solved the coupled PDE system (also identified in \cite{Chandrasekhar:1985kt}) using numerical methods \cite{Dias:2015nua}.
This work was then followed by a series of analyses concerning the KN QNMs \cite{Carullo:2021oxn, Dias:2021yju, Dias:2022oqm}. 

In the present work, we consider ordinary Einstein-Maxwell theory minimally coupled with a massless Klein-Gordon scalar field. 
{This represents a first step towards deriving the formalism and testing the resulting perturbation equations to calculate the QNMs of rotating black holes in various alternative gravity theories. In this sense the presently studied theory might be considered as a particular} limiting case of %
{Einstein-Maxwell-scalar theory, which received a lot of interest recently as a toy model that exhibits spontaneous scalarization of black holes, albeit of the charge induced case (see e.g. \cite{Herdeiro:2018wub}).}

{In particular,} we {here} generalize the work in \cite{Pani:2013ija,Pani:2013wsa}
{obtained} in the slow rotation limit, to second order in rotation.
At first order in rotation, there {arises} no mixing in parity (i.e., axial polar mixing) in the perturbations.
However, at second order in rotation, where the QNMs receive a corresponding higher order correction as well, such a complication {due to parity mixing} occurs.
Solving the equations then becomes {highly} non-trivial when axial and polar perturbations are {no longer} separable.
Furthermore, we quantitatively compare our results to the work of Dias et al. to find out how far our second order slow rotation approximation {matches} their full rotating scheme.  
On top of that, we inspect the isospectrality of the QNM spectrum using our results of axial and polar $\mathrm{l}=2$ perturbations. 
This serves as a verification of our numerical method, as we find good agreement between these two types of perturbations within our slow rotation approximation.

In the next Section \ref{setup}, we present the general setup of the problem, where
we state the theory and its field equations, followed by the slow rotation ansatz in second order.
We also provide the solutions of the background functions at each order.
In Section \ref{non-rad-perturb}, we detail out the ansatz for non-radial perturbations, the necessary projections to apply for the polar- and axial-led perturbations for a generic angular number $\mathrm{l}$, before writing down the final sets of perturbation functions specifically for $\mathrm{l}=2$ which we need to solve. 
The wave behavior at the horizon and close to infinity is also explicitly given.
In Section \ref{results} we show our results for the gravitational, electromagnetic, and scalar quasinormal modes, and the respective isospectrality.
The Appendix \ref{append_coeff} {shows} the corresponding coefficients in our formulae for QNMs in slow rotation up to second order.

\section{General setting}
\label{setup}

\subsection{Theory and field equations}

In order to fix notation, we will present here the field equations. First, consider the standard action 
with an additional scalar field $\Phi$ minimally coupled to gravity,
\begin{eqnarray}
			S=&&\frac{1}{2\kappa}\int d^4x \sqrt{-g} 
		\Big[R - F_{\mu\nu}F^{\mu\nu} - \frac{1}{2} \partial_\mu \Phi \, \partial^\mu \Phi 
		 \Big] \, ,
   \label{eq:quadratic} 
\end{eqnarray}
with the electromagnetic field strength tensor $F_{\mu\nu}=\partial_\mu A_\nu - \partial_\nu A_\mu$.	
The Einstein equations are
\begin{eqnarray}
\mathcal{G}_{\mu\nu} = G_{\mu\nu} - T_{\mu\nu} = 0 \, ,
\end{eqnarray}
where $G_{\mu\nu}$ is the standard Einstein tensor and $T_{\mu\nu}$ is the total stress energy tensor,
composed of the scalar and electromagnetic stress energy tensors,
\begin{eqnarray}
T_{\mu\nu} &=& T_{\mu\nu}^{(\Phi)} + T_{\mu\nu}^{(EM)} \, , \\
T_{\mu\nu}^{(\Phi)} &=& \frac{1}{2}\partial_{\mu}\Phi\,\partial_{\nu}\Phi - \frac{1}{4}g_{\mu\nu}  (\partial \Phi)^2  \, , \\
T_{\mu\nu}^{(EM)} &=& 
 2 \left( F_{\mu\sigma} F_{\nu}^{\,\sigma} - \frac{1}{4} g_{\mu\nu} F^2 \right)  
 \, .
\end{eqnarray}
Additionally, we have the field equations {of} the electromagnetic field and the scalar field,
\begin{eqnarray}
 \mathcal{F}^{\nu} =\nabla_{\mu} (\sqrt{-g} F^{\mu\nu} ) &=& 0 \, , \\
 \mathcal{S} =\frac{1}{\sqrt{-g}} \partial_{\mu} (\sqrt{-g} g^{\mu\nu} \partial_{\nu} \Phi)  &=& 0 \, .
\end{eqnarray}

\subsection{Kerr-Newman solution in Boyer-Lindquist coordinates}

In Boyer-Lindquist coordinates $(\hat{r},\hat{\theta})$, by defining
the following quantities,
\begin{eqnarray}
\rho^2 = \hat{r}^2 + a^2\cos(\hat{\theta}) \, , \, \,
\Delta = \hat{r}^2 - 2 M \hat{r} + a^2 + Q^2 \, ,
\end{eqnarray}
the KN metric is expressed as
\begin{eqnarray}
ds^2 = -\frac{\Delta}{\rho^2}\left( dt - a\sin^2{(\hat{\theta})} d\phi \right)^2  +\frac{\sin^2{(\hat{\theta}})}{\rho^2}\left( a dt - (\hat{r}^2+a^2) d\phi \right)^2
+\rho^2 \left( \frac{d\hat{r}^2}{\Delta} + d\theta \right)^2 \, ,
\end{eqnarray}
and the electromagnetic potential is
\begin{eqnarray}
A = \frac{Q\hat{r}}{\rho^2}(dt -a\sin^2{(\hat{\theta})}d\phi)\, .
\end{eqnarray}
The angular momentum is given by $J=aM$. The black hole horizons are found where $\Delta=0$, meaning
\begin{equation}
    \hat{r}_{\pm} = M \pm \sqrt{M^2-a^2-Q^2} \, .
\end{equation}
Here we will only be interested in the external horizon, $r_+$. The area of the black hole horizon is
\begin{eqnarray}
A_H = 4\pi (r_+^2 + a^2) \, ,
\end{eqnarray}
 and its angular velocity is
\begin{eqnarray}
\Omega_H = \frac{a}{r^2_+ + a^2}\, .
\end{eqnarray}
Of great importance is the extremal KN limit, found when $r_+=r_-$, meaning that mass, charge and angular velocity satisfy the following relation
\begin{eqnarray}
{M}^2-a^2-Q^2=0 \, .
\end{eqnarray}

\subsection{Slow rotation ansatz in Schwarzschild-like coordinates}

The metric for a stationary and axially symmetric geometry, up to second order in rotation (see \cite{Maselli:2015tta}) is
\begin{eqnarray}
ds^2 &=& -e^{2\nu}\left[1+\epsilon_r^2 2\left(h_0(r)+h_2(r)P_2(\theta)\right)\right]dt^2 + e^{2\lambda}\left[1+\frac{1}{r}\epsilon_r^2 2\left(m_0(r)+m_2(r)P_2(\theta)\right)\right]dr^2 \nonumber \\ 
&+& r^2\left[1+\epsilon_r^2 2\left(k_0(r)+k_2(r)P_2(\theta)\right)\right]
\left\{ d\theta^2+\sin^2{(\theta)}\left[d\phi-\epsilon_rw(r)dt\right]^2 \right\} \ .
\label{metric_1}
\end{eqnarray}
Here we introduce the perturbation parameter $\epsilon_r$. This is a slow-rotation perturbation parameter ($\epsilon_r \ll 1$). The metric is written up to second order terms in rotation. Hence, note that the angular dependence is explicitly given by the Legendre polynomial,
\begin{eqnarray}
P_2(\theta) = \frac{1}{2}\left(3\cos^2{(\theta)}-1\right) \ .
\end{eqnarray}

We choose the gauge $k_0(r)=0$ (Schwarzschild-like coordinates). 
We use the following expressions for the static functions,   
\begin{eqnarray}
   && e^{2\nu} \equiv f(r) \, , \\ 
   && e^{-2\lambda} \equiv 1-\frac{2m(r)}{r}  \ .
\end{eqnarray}
We introduce the function $v_2$ redefining $k_2=h_2-v_2$.

Since the metric is stationary and axially symmetric, the gauge field $A$ must have these symmetries in order to be consistent. Up to second order in rotation, this means
\begin{eqnarray}
A = \left( a_0(r) + \epsilon_r^2 (c_0(r) + c_2(r)P_2(\theta)) \right) dt + \epsilon_r b(r) \sin^2(\theta) d\phi  \ .
\label{EMfield_1}
\end{eqnarray}

In our case the scalar field is trivial, but generically we have the expression
\begin{eqnarray}
\Phi = \varphi(r) + \epsilon_r^2 \left(\varphi_{20}(r)+\varphi_{22}(r)P_2(\theta)\right) \ .
\label{scalar_1}
\end{eqnarray}

The killing vector is given by $\partial_t$. 
The horizon is located at the surface $r=r_H$ for which $f(r_H)=0$. Because of the Schwarzschild coordinates, the horizon area is simply
\begin{eqnarray}
A_H = 4\pi r_H^2\, ,
\end{eqnarray}
and
the horizon velocity $\Omega_H$ is
\begin{eqnarray}
\Omega_H = w(r_H)\, .
\end{eqnarray}

\subsection{Kerr-Newmann in the Schwarzschild-like coordinates}

We here provide the corresponding solutions for the functions in the metric and gauge field expanded up to second order in rotation.

\subsubsection{Static: Reissner-Nordstr\"om solution}

The well-known Reissner-Nordstr\"om solution is described by three radial functions:
\begin{eqnarray}
a_0 = \frac{Q}{r} \ ,  \\
f = 1-{\frac {r_H}{r}}+{\frac {{{Q}}^{2}}{{r}^{2}}}-{\frac {{{Q}}^{2}}{rr_H}}
\ ,  \\
m = M_0 - {\frac {{{Q}}^{2}}{2\,r}} = {\frac {r_H}{2}}+{\frac {{{Q}}^{2}}{2\,r_H}}-{\frac {{{Q}}^{2}}{2\,r}} \ ,
\end{eqnarray}
where $M_0 = {\frac {r_H}{2}}+{\frac {{{Q}}^{2}}{2\,r_H}}$ is the static mass and $Q$ is the electric charge.

\subsubsection{First order Kerr-Newman}

The first order perturbation in angular momentum is given by the inertial dragging function $w$, and the magnetic field $b$, which are radial functions given by
\begin{eqnarray}
w &=&  \frac{2 J}{r^3} \left( 1 - 
\frac { {{Q}}^{2}r_H}{ \left( {{Q}}^{2}+{{r_H}}^{2} \right) {r}} \right) 
\, ,
\\
b &=& -\,{\frac {2 J{Q}\,r_H}{ \left( {{Q}}^{2}+r_H
^{2} \right) r}}
\, ,
\end{eqnarray}
where $J$ is the angular momentum of the configuration. This implies that the horizon angular velocity is 
\begin{eqnarray}
\Omega_H=\frac{2J}{r_H^3}\left( 1 - 
\frac {Q^2}{ Q^2+r_H^{2}} \right) .
\end{eqnarray}

\subsubsection{Second order Kerr-Newman}

The second order is given by the following radial functions:
{\small
\begin{eqnarray}
m_0 &=& 
\delta M - \frac{J^2(Q^2+3r_H^2)(3Q^2+r_H^2)}{3(Q^2+r_H^2)^2r^3} + \frac{7J^2Q^2r_H}{3(Q^2+r_H^2)r^4}-\frac{4J^2Q^4r_H^2}{3(Q^2+r_H^2)^2r^5}
\, , \\
h_0 &=& 
\frac{J^2}{(Q^2+r_H^2)(Q^2-r_Hr)} \left( 1 + \frac{r_H}{r} + \frac{r_H^2}{r^2} - \frac{Q^2r_H(3Q^2+5r_H^2)}{3(Q^2+r_H^2)r^3} + \frac{2r_H^2Q^4}{3(Q^2+r_H^2)r^4} \right)
\, , \\
h_2 &=&
\frac{2J^2r_H}{(Q^2+r_H^2)r^3}+\frac{J^2(3Q^4-2Q^2r_H^2+3r_H^4)}{3(Q^2+r_H^2)^2r^4}-\frac{10J^2Q^2r_H}{3(Q^2+r_H^2)r^5}+\frac{8J^2Q^4r_H^2}{3(Q^2+r_H^2)^2r^6}
\, , \\
\nu_2 &=& 
-\frac{J^2(Q^2+3r_H^2)(3Q^2+r_H^2)}{3(Q^2+r_H^2)^2r^4}+\frac{4J^2Q^2r_H}{3(Q^2+r_H^2)r^5}
\, , \\
m_2 &=&
-\frac{2J^2r_H}{(Q^2+r_H^2)r^2}
+ \frac{J^2(3Q^2+7r_H^2)(7Q^2+3r_H^2)}{3(Q^2+r_H^2)^2r^3}
- \frac{J^2(15Q^4+98Q^2r_H^2+15r_H^4)}{3(Q^2+r_H^2)r_Hr^4} \nonumber \\ &&
+ \frac{J^2Q^2(61Q^4+170Q^2r_H^2+61r_H^4)}{3(Q^2+r_H^2)^2r^5}
- \frac{26J^2Q^4r_H}{(Q^2+r_H^2)r^6}
+ \frac{32J^2Q^6r_H^2}{3(Q^2+r_H^2)^2r^7}
\, , \\
c_0 &=&
\frac{2J^2Qr_H}{3(Q^2+r_H^2)r^4} - \frac{2J^2Q^3r_H^2}{3(Q^2+r_H^2)^2r^5}
\, ,\\
c_2 &=&
-\frac{4J^2Qr_H^2}{(Q^2+r_H^2)^2r^3}
-\frac{2J^2Qr_H}{3(Q^2+r_H^2)r^4}
+\frac{2J^2Q(3Q^4+8Q^2r_H^2+3r_H^2)}{3(Q^2+r_H^2)^2r^5}\nonumber \\ &&
-\frac{14J^2Q^3r_H}{3(Q^2+r_H^2)r^6}
+\frac{8J^2Q^5r_H^2}{3(Q^2+r_H^2)^2r^7} \, .
\end{eqnarray}
}
The second order correction to the mass is
\begin{eqnarray}
\delta M = {\frac {{J}^{2}}{r_H\, \left( {{Q}}^{2}+r_H^{2}
 \right) }}\, .
\end{eqnarray}
This means that, up to second order in rotation, the total mass is thus
\begin{eqnarray}
{M} = M_0 + \epsilon_r^2\delta M \, .
\end{eqnarray}
Let us note that in the previous solutions we have chosen the electromagnetic gauge so that the electric charge is not modified at second order in rotation by choosing the integration constant so that $c_0$ has a vanishing $r^{-1}$ term.

The ergosurface is given by the implicit equation
\begin{eqnarray}
g_{tt}(r,\theta)=-f(r)\left[1+\epsilon_r^2 2\left(h_0(r)+h_2(r)P_2(\theta)\right)\right]+\epsilon_r^2 r^2w(r)^2 \sin^2(\theta) = 0 \, ,
\end{eqnarray}
which has the following solution up to second order in rotation
\begin{eqnarray}
r_e = r_H \left(1 +  \epsilon_r^2\frac{J^2}{r_H^4}\frac{4}{\left(1-\left(\frac{Q}{r_H}\right)^4\right)\left(1+\left(\frac{Q}{r_H}\right)^2\right)}\sin^2(\theta)\right)\, .
\end{eqnarray}

\subsubsection{Transformation to Boyer-Lindquist coordinates}

For completeness, let us write here the necessary functions that define the transformation from the Schwarzschild coordinates we are using to the Boyer-Lindquist coordinates in which we wrote the KN solution at the beginning. These are three functions, that determine the transformation of the radial and angular coordinates:
\begin{eqnarray}
\hat{r} = r + \epsilon_r^2 \left(\zeta_0 + \zeta_2 P_2\right) \, , \\
\hat{\theta} = \theta + \epsilon_r^2 \zeta_\theta \sin(\theta)\cos(\theta) \, , 
\end{eqnarray}
where
\begin{eqnarray}
\zeta_0 = \frac{a^2}{3}\left( \frac{Q^2}{2r^3} - \frac{M_0}{r^2} - \frac{1}{r} \right) \, ,
\\
\zeta_2 = \frac{a^2}{3}\left( \frac{1}{r} + \frac{M_0}{r^2} - \frac{6M_0^2+Q^2}{r^3} + \frac{7M_0Q^2}{r^4} - \frac{2Q^4}{r^5} \right)
\, ,
\\
\zeta_\theta = a^2\left( \frac{Q^2}{2r^4} - \frac{M_0}{r^3} - \frac{M_0}{2r^2}  \right)
\, .
\end{eqnarray}
Note that up to second order we have $a = \frac{J}{M_0}$.

\section{Non-radial perturbations}
\label{non-rad-perturb}

\subsection{Ansatz}
\label{sec_ansatz}

Here we perturb the previous stationary and axially symmetric configurations (slow rotation approximation, second order in $\epsilon_r$).
The background metric $g^{(sr)}_{\mu\nu}$ is given by equation (\ref{metric_1}),
and
the background EM field $A^{(sr)}_{\mu}$ is given by equation (\ref{EMfield_1}),
while the background scalar field $\varphi^{(sr)}$ is trivial\footnote{The superscript $(sr)$ denotes slow rotation.}. Following \cite{Kojima:1992ie,Pani:2012bp,Pani:2013wsa,Pierini:2021jxd}, we will consider first order non-radial, time dependent perturbations, which we will decompose in tensor spherical harmonics. We will also assume a harmonic time dependence. Hence, introducing the control parameter $\epsilon_q$ for these perturbations, the three dynamical fields $g_{\mu\nu},A_{\mu},\Phi$ in the theory with their axial and polar perturbation parts are,
\\
for the metric:
\begin{eqnarray}
g_{\mu\nu} &=& g^{(sr)}_{\mu\nu} + \epsilon_q \delta h_{\mu\nu}(t,r,\theta,\phi) \, , \quad  \text{where} \\
\delta h_{\mu\nu} &=& \delta h^{(A)}_{\mu\nu} + \delta h^{(P)}_{\mu\nu}  \nonumber \\ 
&=& e^{-i\omega t} \sum_{\mathrm{l},\mathrm{m}} \left( h^{(A)}_{\mu\nu}[\mathrm{l},\mathrm{m}](r) \cdot \mathcal{Y}^{(A)}_{\mu\nu}[\mathrm{l},\mathrm{m}](\theta,\phi) + h^{(P)}_{\mu\nu}[\mathrm{l},\mathrm{m}](r) \cdot \mathcal{Y}^{(P)}_{\mu\nu}[\mathrm{l},\mathrm{m}](\theta,\phi) \right)
\, ,
\end{eqnarray}
for the gauge field:
\begin{eqnarray}
A_{\mu} &=& A^{(sr)}_{\mu} + \epsilon_q \delta A_{\mu}(t,r,\theta,\phi)
\, , 
\quad  \text{where}
\\
\delta A_{\mu} &=& \delta A^{(A)}_{\mu} + \delta A^{(P)}_{\mu}  \nonumber \\ 
&=& e^{-i\omega t} \sum_{\mathrm{l},\mathrm{m}} \left( q^{(A)}_{\mu}[\mathrm{l},\mathrm{m}](r) \cdot \hat{{Y}}^{(A)}_{\mu}[\mathrm{l},\mathrm{m}](\theta,\phi) + q^{(P)}_{\mu}[\mathrm{l},\mathrm{m}](r) \cdot \hat{{Y}}^{(P)}_{\mu}[\mathrm{l},\mathrm{m}](\theta,\phi) \right)
\, , 
\end{eqnarray}
for the scalar field:
\begin{eqnarray}
\Phi &=& \varphi^{(sr)} + \epsilon_q \delta\varphi(t,r,\theta,\phi) = \varphi^{(sr)} + \epsilon_q \delta\varphi^{(P)} \, , 
\quad  \text{where}
\\
\delta\varphi^{(P)} &=& e^{-i\omega t} \sum_{\mathrm{l},\mathrm{m}} \Phi_{1[\mathrm{l},\mathrm{m}]}(r)Y_{[\mathrm{l},\mathrm{m}]}(\theta,\phi) \, .
\end{eqnarray}
The superscript $(A)$ corresponds to axial and $(P)$ to polar perturbations. The functions $\mathcal{Y}^{(A,P)}_{\mu\nu}[\mathrm{l},\mathrm{m}](\theta,\phi)$ are the respective axial and polar components of the tensor spherical harmonics, while the functions $\hat{Y}^{(A,P)}_{\mu}[\mathrm{l},\mathrm{m}](\theta,\phi)$ are the respective axial and polar components of the vector spherical harmonics. These can be written as a function of the scalar spherical harmonics $Y_{[\mathrm{l},\mathrm{m}]}(\theta,\phi)$. They depend on two quantum numbers, $\mathrm{l}$ and $\mathrm{m}$. As usual we have $\mathrm{l}=\left\{0,1,2, ...\right\}$ and for each $\mathrm{l}$, $\mathrm{m}=\left\{-\mathrm{l},-\mathrm{l}+1,...,\mathrm{l}-1,\mathrm{l}\right\}$. Since the background solution is axially symmetric, the axis of rotation being $z$, perturbations with different $\mathrm{m}$ number will be independent of each other. Hence we can drop the summation over $\mathrm{m}$, 
and
we will consider perturbations with fixed values of $\mathrm{m}$, meaning that the possible values of $\mathrm{l}$ are $\mathrm{l}=\left\{|\mathrm{m}|,|\mathrm{m}|+1,|\mathrm{m}|+2,...\right\}$.

The non-zero components of $\delta h^{(A)}_{\mu\nu}$ and $\delta A^{(A)}_{\mu}$ 
for axial perturbations
are
\begin{eqnarray}
&&\delta h^{(A)}_{t\theta} = - e^{-i\omega t} \sum_{\mathrm{l}} h_{0[\mathrm{l},\mathrm{m}]}(r)\frac{\partial_\phi Y_{[\mathrm{l},\mathrm{m}]}(\theta,\phi)}{\sin{\theta}} \, , \\
&&\delta h^{(A)}_{t\phi} = e^{-i\omega t} \sum_{\mathrm{l}} h_{0[\mathrm{l},\mathrm{m}]}(r) \sin{\theta} \partial_\theta Y_{[\mathrm{l},\mathrm{m}]}(\theta,\phi) \, ,  \\
&&\delta h^{(A)}_{r\theta} = - e^{-i\omega t} \sum_{\mathrm{l}} h_{1[\mathrm{l},\mathrm{m}]}(r)\frac{\partial_\phi Y_{[\mathrm{l},\mathrm{m}]}(\theta,\phi)}{\sin{\theta}} \, ,  \\
&&\delta h^{(A)}_{r\phi} = e^{-i\omega t} \sum_{\mathrm{l}} h_{1[\mathrm{l},\mathrm{m}]}(r) \sin{\theta} \partial_\theta Y_{[\mathrm{l},\mathrm{m}]}(\theta,\phi) \, , \\
&&\delta A^{(A)}_{\theta} = - e^{-i\omega t} \sum_{\mathrm{l}} W_{2[\mathrm{l},\mathrm{m}]}(r)\frac{\partial_\phi Y_{[\mathrm{l},\mathrm{m}]}(\theta,\phi)}{\sin{\theta}}    \, , \\
&&\delta A^{(A)}_{\phi} = e^{-i\omega t} \sum_{\mathrm{l}} W_{2[\mathrm{l},\mathrm{m}]}(r) \sin{\theta} \partial_\theta Y_{[\mathrm{l},\mathrm{m}]}(\theta,\phi) \, .  
\end{eqnarray}
Therefore, the axial perturbations are described by the radial functions $$\left\{h_{0[\mathrm{l},\mathrm{m}]},h_{1[\mathrm{l},\mathrm{m}]},W_{2[\mathrm{l},\mathrm{m}]}\right\}.$$ 
In the coming section,
we introduce a notation $a_{[\mathrm{l},\mathrm{m}]}$ as a vector whose components are these axial perturbation functions and the corresponding first order radial derivatives.

The non-zero components of $\delta h^{(P)}_{\mu\nu}$ and $\delta A^{(P)}_{\mu}$ 
for polar perturbations
are
\begin{eqnarray}
&&\delta h^{(P)}_{tt} = - f(r) e^{-i\omega t} \sum_{\mathrm{l}} N_{[\mathrm{l},\mathrm{m}]}(r)Y_{[\mathrm{l},\mathrm{m}]}(\theta,\phi) \,  , \\
&&\delta h^{(P)}_{tr} = - e^{-i\omega t} \sum_{\mathrm{l}} H_{1[\mathrm{l},\mathrm{m}]}(r)Y_{[\mathrm{l},\mathrm{m}]}(\theta,\phi) \,  , \\
&&\delta h^{(P)}_{rr} = - \frac{1}{1-2m/r} e^{-i\omega t} \sum_{\mathrm{l}} L_{[\mathrm{l},\mathrm{m}]}(r)Y_{[\mathrm{l},\mathrm{m}]}(\theta,\phi) \,  , \\
&&\delta h^{(P)}_{\theta\theta} = r^2 e^{-i\omega t} \sum_{\mathrm{l}} T_{[\mathrm{l},\mathrm{m}]}(r)Y_{[\mathrm{l},\mathrm{m}]}(\theta,\phi) \,  , \\
&&\delta h^{(P)}_{\phi\phi} = r^2 e^{-i\omega t} \sum_{\mathrm{l}} T_{[\mathrm{l},\mathrm{m}]}(r)\sin^2{\theta} Y_{[\mathrm{l},\mathrm{m}]}(\theta,\phi) \, , \\
&&\delta A^{(P)}_{t} = e^{-i\omega t} \sum_{\mathrm{l}} A_{0[\mathrm{l},\mathrm{m}]}(r) Y_{[\mathrm{l},\mathrm{m}]}(\theta,\phi) \, , \\
&&\delta A^{(P)}_{r} = e^{-i\omega t} \sum_{\mathrm{l}} W_{[\mathrm{l},\mathrm{m}]}(r) Y_{[\mathrm{l},\mathrm{m}]}(\theta,\phi) \, , \\
&&\delta A^{(P)}_{\theta} = e^{-i\omega t} \sum_{\mathrm{l}} V_{[\mathrm{l},\mathrm{m}]}(r) \partial_\theta Y_{[\mathrm{l},\mathrm{m}]}(\theta,\phi) \, , \\
&&\delta A^{(P)}_{\phi} = e^{-i\omega t} \sum_{\mathrm{l}} V_{[\mathrm{l},\mathrm{m}]}(r) \partial_\phi Y_{[\mathrm{l},\mathrm{m}]}(\theta,\phi) \, .
\end{eqnarray}
Hence the polar perturbations of the KN metric are described by the radial functions $$\left\{N_{[\mathrm{l},\mathrm{m}]},H_{1[\mathrm{l},\mathrm{m}]},L_{[\mathrm{l},\mathrm{m}]},T_{[\mathrm{l},\mathrm{m}]},A_{0[\mathrm{l},\mathrm{m}]},W_{[\mathrm{l},\mathrm{m}]},V_{[\mathrm{l},\mathrm{m}]}\right\}.$$  
Similarly, we introduce later a notation
$p_{[\mathrm{l},\mathrm{m}]}$ as a vector composed of these polar perturbation functions and the corresponding first order radial derivatives.

For a spherically symmetric background we know that perturbations with different $\mathrm{l}$ number are independent of each other. In addition, perturbations with different parity under reflections (axial-polar) are independent of each other. So in the spherical case, we could drop the summation on $\mathrm{l}$, and  consider axial and polar perturbations separately. 
However, for a rotating configuration, this is no longer the case -- perturbations with different $\mathrm{l}$ number and parity will mix with each other.

\subsection{Field equations}

We substitute the ansatz
discussed in the previous section 
into the field equations. We use Maple for these computations. To first order in the $\epsilon_q$ parameter for non-radial perturbations, and to second order in the $\epsilon_r$ perturbation parameter for slow rotation approximation, 
the field equations for the metric, electromagnetic, and scalar field would be:
\begin{eqnarray}
\mathcal{G}_{\mu\nu} = \mathcal{G}_{\mu\nu}^{(sr)} + \epsilon_q\delta\mathcal{G}_{\mu\nu}e^{-i\omega t}=0 \, , \\
\mathcal{F}_{\mu} = \mathcal{F}^{(sr)}_{\mu} + \epsilon_q\delta\mathcal{F}_{\mu}e^{-i\omega t}=0 \, , \\
\mathcal{S} = \mathcal{S}^{(sr)} + \epsilon_q\delta\mathcal{S}e^{-i\omega t}=0 \, .
\end{eqnarray}
Since the background is the slowly rotating Kerr-Newman solution, then $\mathcal{G}_{\mu\nu}^{(sr)}=0$, $\mathcal{F}^{(sr)}_{\mu}=0$, and $\mathcal{S}^{(sr)}=0$.

On the other hand, when we include terms in the order of $\epsilon_r^2$, the partial differential equations that result from $\delta\mathcal{G}_{\mu\nu}=0$, $\delta\mathcal{F}_{\mu}=0$ and $\delta\mathcal{S}=0$ can no longer be separated into a product of $\text{radial}$ and $\text{angular}$ components. However, following \cite{Kojima:1992ie} we can proceed in two steps: 1) Decomposing the field equations in spherical harmonics, and 2) Truncating the equations in the slow rotation approximation.

First, we assume that the field equations can be decomposed in spherical harmonics. We know that certain components of the field equations transform like a scalar, a vector or a tensor under rotations \cite{Ferrari:2020nzo}. So, in practice, to extract the radial field equations for the perturbations, we project the components of the field equations to the respective scalar, vector or tensor spherical harmonics. That is:

\

$\bullet$ \quad \textbf{Scalar components:}

\

This is obvious {in} the case {of} the scalar equation, 
\begin{eqnarray}
&&\mathcal{S}{[\mathrm{l},\mathrm{m}]}(r)= \iint d\Omega \, Y^*_{[\mathrm{l},\mathrm{m}]}(\theta,\phi)\delta\mathcal{S} \, .
\end{eqnarray}
Here $Y^*_{[\mathrm{l},\mathrm{m}]}(\theta,\phi)$ is the complex conjugate of $Y_{[\mathrm{l},\mathrm{m}]}(\theta,\phi)$ and we are integrating over the 2-sphere, where $d\Omega=\sin(\theta)d\theta d\phi$ with $\theta\in[0,\pi]$ and $\phi\in[0,2\pi]$.

Regarding the Einstein equations (which {concern} a tensor), the components that transform like a scalar are the following polar components:
\begin{eqnarray}
&&\mathcal{P}_{tt}[\mathrm{l},\mathrm{m}](r) = \iint d\Omega \, Y^*_{[\mathrm{l},\mathrm{m}]}(\theta,\phi)\delta\mathcal{G}_{tt} \, , \\
&&\mathcal{P}_{tr}[\mathrm{l},\mathrm{m}](r) = \iint d\Omega \, Y^*_{[\mathrm{l},\mathrm{m}]}(\theta,\phi)\delta\mathcal{G}_{tr} \, , \\
&&\mathcal{P}_{rr}[\mathrm{l},\mathrm{m}](r) = \iint d\Omega \, Y^*_{[\mathrm{l},\mathrm{m}]}(\theta,\phi)\delta\mathcal{G}_{rr} \, , \\
&&\mathcal{P}_{+}[\mathrm{l},\mathrm{m}](r) = \iint d\Omega \, Y^*_{[\mathrm{l},\mathrm{m}]}(\theta,\phi)
\left(
\delta\mathcal{G}_{\theta\theta}+\frac{1}{\sin^2(\theta)}\delta\mathcal{G}_{\phi\phi}
\right) \, .
\end{eqnarray}

Regarding the Maxwell equations (which {concern} a vector), the components that transform like a scalar are the following polar components:
\begin{eqnarray}
&&\mathcal{P}_{t}[\mathrm{l},\mathrm{m}](r) = \iint d\Omega\, Y^*_{[\mathrm{l},\mathrm{m}]}(\theta,\phi)\delta\mathcal{F}_{t} \, , \\
&&\mathcal{P}_{r}[\mathrm{l},\mathrm{m}](r) = \iint d\Omega\, Y^*_{[\mathrm{l},\mathrm{m}]}(\theta,\phi)\delta\mathcal{F}_{r}  \, .
\end{eqnarray}

\

$\bullet$ \quad \textbf{Vector components:}

\

Define the axial and polar vector spherical harmonics,
\begin{eqnarray}
&& \Vec{Y}_{a,[\mathrm{l},\mathrm{m}]}=\left(-\frac{1}{\sin(\theta)}\partial_\phi Y_{[\mathrm{l},\mathrm{m}]},
\, 
\sin(\theta)\partial_\theta Y_{[\mathrm{l},\mathrm{m}]}\right) \, , \\
&& \Vec{Y}_{p,[\mathrm{l},\mathrm{m}]}=\left(\partial_\theta Y_{[\mathrm{l},\mathrm{m}]},
\,
\partial_\phi Y_{[\mathrm{l},\mathrm{m}]}\right) \, ,
\end{eqnarray}
and the auxiliary vectors
\begin{eqnarray}
\vec{V}_{t} = \left( \delta\mathcal{G}_{t\theta} , \delta\mathcal{G}_{t\phi} \right) \, , \\
\vec{V}_{r} = \left( \delta\mathcal{G}_{r\theta} , \delta\mathcal{G}_{r\phi} \right) \, , \\
\vec{F} = \left( \delta\mathcal{F}_{\theta} , \delta\mathcal{F}_{\phi} \right) \, .
\end{eqnarray}
Projecting the auxiliary vectors onto these spherical harmonics\footnote{The 2-sphere metric is $\begin{pmatrix}
 1 & 0  \\ 
 0 & \sin^2(\theta)   
\end{pmatrix}$.}, we get
the axial components,
\begin{eqnarray}
&&\mathcal{A}_{vt}[\mathrm{l},\mathrm{m}](r) = \iint d\Omega 
\left(
-\frac{1}{\sin(\theta)}\partial_\phi Y^*_{[\mathrm{l},\mathrm{m}]}\delta\mathcal{G}_{t\theta}+\frac{1}{\sin(\theta)}\partial_\theta Y_{[\mathrm{l},\mathrm{m}]}\delta\mathcal{G}_{t\phi}  \right) \, , \\
&&\mathcal{A}_{vr}[\mathrm{l},\mathrm{m}](r) = \iint d\Omega 
\left(
-\frac{1}{\sin(\theta)}\partial_\phi Y^*_{[\mathrm{l},\mathrm{m}]}\delta\mathcal{G}_{r\theta}+\frac{1}{\sin(\theta)}\partial_\theta Y_{[\mathrm{l},\mathrm{m}]}\delta\mathcal{G}_{r\phi}  \right)
 \, ,\\
&&\mathcal{A}_{F}[\mathrm{l},\mathrm{m}](r) = \iint d\Omega 
\left(
-\frac{1}{\sin(\theta)}\partial_\phi Y^*_{[\mathrm{l},\mathrm{m}]}\delta\mathcal{F}_{\theta}+\frac{1}{\sin(\theta)}\partial_\theta Y_{[\mathrm{l},\mathrm{m}]}\delta\mathcal{F}_{\phi}  \right)
 \, ,
\end{eqnarray}
and the polar components,
\begin{eqnarray}
&&\mathcal{P}_{vt}[\mathrm{l},\mathrm{m}](r) = \iint d\Omega 
\left(
\partial_\theta Y^*_{[\mathrm{l},\mathrm{m}]}\delta\mathcal{G}_{t\theta}+\frac{1}{\sin^2(\theta)}\partial_\phi Y^*_{[\mathrm{l},\mathrm{m}]}\delta\mathcal{G}_{t\phi}\right) \, , \\
&&\mathcal{P}_{vr}[\mathrm{l},\mathrm{m}](r) = \iint d\Omega 
\left(
\partial_\theta Y^*_{[\mathrm{l},\mathrm{m}]}\delta\mathcal{G}_{r\theta}+\frac{1}{\sin^2(\theta)}\partial_\phi Y^*_{[\mathrm{l},\mathrm{m}]}\delta\mathcal{G}_{r\phi} \right) \, , \\
&&\mathcal{P}_{F}[\mathrm{l},\mathrm{m}](r) = \iint d\Omega 
\left(
\partial_\theta Y^*_{[\mathrm{l},\mathrm{m}]}\delta\mathcal{F}_{\theta}+\frac{1}{\sin^2(\theta)}\partial_\phi Y^*_{[\mathrm{l},\mathrm{m}]}\delta\mathcal{F}_{\phi} \right) \, .
\end{eqnarray}

\

$\bullet$ \quad \textbf{Tensor components:}

\

We define the tensor spherical harmonics on the 2-sphere,
\begin{eqnarray}
&& \hat{Y}_{a,[\mathrm{l},\mathrm{m}]}=\frac{1}{2}
\begin{pmatrix}
 W_{[\mathrm{l},\mathrm{m}]} & X_{[\mathrm{l},\mathrm{m}]}  \\ 
 X_{[\mathrm{l},\mathrm{m}]} & -\sin^2(\theta)W_{[\mathrm{l},\mathrm{m}]}
\end{pmatrix}
\, , \\
&& \hat{Y}_{p,[\mathrm{l},\mathrm{m}]}=\frac{1}{2}
\begin{pmatrix}
-\sin^{-1}(\theta) X_{[\mathrm{l},\mathrm{m}]} & \sin(\theta)W_{[\mathrm{l},\mathrm{m}]}  \\ 
 \sin(\theta)W_{[\mathrm{l},\mathrm{m}]} & \sin(\theta)X_{[\mathrm{l},\mathrm{m}]}
\end{pmatrix}
 \, ,
\end{eqnarray}
with
\begin{eqnarray}
X_{[\mathrm{l},\mathrm{m}]} &=& 2\partial^2_{\theta\phi}Y_{[\mathrm{l},\mathrm{m}]}-2\cot{(\theta)}\partial_\phi Y_{[\mathrm{l},\mathrm{m}]} \, , \\
W_{[\mathrm{l},\mathrm{m}]} &=& \partial^2_{\theta}Y_{[\mathrm{l},\mathrm{m}]}-\cot{(\theta)}\partial_\theta Y_{[\mathrm{l},\mathrm{m}]} - \frac{1}{\sin^2(\theta)}\partial^2_{\phi}Y_{[\mathrm{l},\mathrm{m}]} \, ,
\end{eqnarray}
and an auxiliary tensor
\begin{eqnarray}
\hat{T} = 
\begin{pmatrix}
 \delta\mathcal{G}_{\theta\theta} & \delta\mathcal{G}_{\theta\phi}  \\ 
 \delta\mathcal{G}_{\phi\theta} & \delta\mathcal{G}_{\phi\phi} 
\end{pmatrix}\, .
\end{eqnarray}
It is useful to define
\begin{eqnarray}
\delta\mathcal{G}_{-} = \delta\mathcal{G}_{\theta\theta} - \frac{1}{\sin^2{(\theta)}} \delta\mathcal{G}_{\phi\phi} \, .
\end{eqnarray}
Then the projection of $\hat{T}$ onto the tensor spherical harmonics results in
\begin{eqnarray}
&&\mathcal{A}_{-}[\mathrm{l},\mathrm{m}](r) = \iint d\Omega 
\left(
\frac{2}{\sin^2(\theta)}W^*_{[\mathrm{l},\mathrm{m}]}\delta\mathcal{G}_{\theta\phi}-\frac{1}{\sin(\theta)} X^*_{[\mathrm{l},\mathrm{m}]}\delta\mathcal{G}_{-}\right)  \, , \\
&&\mathcal{P}_{-}[\mathrm{l},\mathrm{m}](r) = \iint d\Omega 
\left(
W^*_{[\mathrm{l},\mathrm{m}]}\delta\mathcal{G}_{-}+\frac{2}{\sin^2(\theta)} X^*_{[\mathrm{l},\mathrm{m}]}\delta\mathcal{G}_{\theta\phi}\right) \, . 
\end{eqnarray}

After the projections, we are left with a system of radial equations,
which are polar or axial.
Hereafter, we shall drop the explicit dependence on $r$ from the equations.

For polar perturbations, 
the Einstein equations imply
\begin{eqnarray}
&& \mathcal{P}_{tt}[\mathrm{l},\mathrm{m}]=0 \, , \, 
\mathcal{P}_{tr}[\mathrm{l},\mathrm{m}]=0 \, , \, 
\mathcal{P}_{rr}[\mathrm{l},\mathrm{m}]=0 \, , \nonumber \\
&& \mathcal{P}_{vt}[\mathrm{l},\mathrm{m}]=0 \, , \,
\mathcal{P}_{vr}[\mathrm{l},\mathrm{m}]=0 \, , \,
\mathcal{P}_{+}[\mathrm{l},\mathrm{m}]=0 \, , \,
\mathcal{P}_{-}[\mathrm{l},\mathrm{m}]=0 \, ,
\end{eqnarray}
the Maxwell field equation implies
\begin{eqnarray}
&& \mathcal{P}_{t}[\mathrm{l},\mathrm{m}]=0 \, , \, 
\mathcal{P}_{r}[\mathrm{l},\mathrm{m}]=0 \, ,
\mathcal{P}_{F}[\mathrm{l},\mathrm{m}]=0 \, ,
\end{eqnarray}
and the scalar field equation implies
\begin{eqnarray}
\mathcal{S}[\mathrm{l},\mathrm{m}]=0 \, .
\end{eqnarray}

For axial perturbations, 
the Einstein field equations imply
\begin{eqnarray}
\mathcal{A}_{vt}[\mathrm{l},\mathrm{m}]=0 \, , \,
\mathcal{A}_{vr}[\mathrm{l},\mathrm{m}]=0 \, , \, 
\mathcal{A}_{-}[\mathrm{l},\mathrm{m}]=0 \, ,
\end{eqnarray}
and the Maxwell equation implies
\begin{eqnarray}
\mathcal{A}_{F}[\mathrm{l},\mathrm{m}]=0 \, .
\end{eqnarray}

In total there are 4 axial equations and 11 polar equations from our system of equations.
For simplicity, we will refer to the 4 axial equations as $\mathcal{A}[\mathrm{l},\mathrm{m}]=0$, and $\mathcal{P}[\mathrm{l},\mathrm{m}]=0$ for the rest of the polar equations.
\\

The second step is the truncation of this system of equations. Note that for fixed values of $[\mathrm{l},\mathrm{m}]$, these equations form a system of homogeneous linear equations that can be written more explicitly as
\begin{eqnarray}
\mathcal{P}[\mathrm{l},\mathrm{m}] &=& 
\hat{\alpha}_{p} \circ p_{[\mathrm{l},\mathrm{m}]} 
+ \epsilon_r 
\left(
  \mathrm{m} \cdot \hat{\beta}^0_{p} \circ p_{[\mathrm{l},\mathrm{m}]} 
+ \hat{\beta}^+_{p} \circ a_{[\mathrm{l+1},\mathrm{m}]}
+ \hat{\beta}^-_{p} \circ a_{[\mathrm{l-1},\mathrm{m}]}
\right)
\nonumber \\
&+& \epsilon_r^2 
\left(
   \hat{\gamma}^+_{p} \circ p_{[\mathrm{l+2},\mathrm{m}]}
+  \hat{\gamma}^0_{p} \circ p_{[\mathrm{l},\mathrm{m}]}
+  \hat{\gamma}^-_{p} \circ p_{[\mathrm{l-2},\mathrm{m}]}
\right) \, , 
\label{pol_gen}
\\
\mathcal{A}[\mathrm{l},\mathrm{m}] &=& 
\hat{\alpha}_{a} \circ a_{[\mathrm{l},\mathrm{m}]} 
+ \epsilon_r 
\left(
  \mathrm{m} \cdot \hat{\beta}^0_{a} \circ a_{[\mathrm{l},\mathrm{m}]} 
+  \hat{\beta}^+_{a} \circ p_{[\mathrm{l+1},\mathrm{m}]}
+ \hat{\beta}^-_{a} \circ p_{[\mathrm{l-1},\mathrm{m}]}
\right)
\nonumber \\
&+& \epsilon_r^2 
\left(
   \hat{\gamma}^+_{a} \circ a_{[\mathrm{l+2},\mathrm{m}]}
+  \hat{\gamma}^0_{a} \circ a_{[\mathrm{l},\mathrm{m}]}
+  \hat{\gamma}^-_{a} \circ a_{[\mathrm{l-2},\mathrm{m}]}
\right) \, .
\label{ax_gen}
\end{eqnarray}
We have defined the differential operators $\hat{\alpha}_{p,a},\hat{\beta}^{\pm, 0}_{p,a},\hat{\gamma}^{\pm ,0}_{p,a}$
\footnote{More specifically, the operators $\hat{\alpha}_{p,a}$ are linear combinations of radial derivatives, and the coefficients depend on the zero order functions $\left\{f(r),m(r),...\right\}$ (apart from the eigenfrequency $\omega$ and the $[\mathrm{l},\mathrm{m}]$ angular numbers). The operators $\hat{\beta}^{\pm,0}_{p,a}$ are similar, 
in addition they are also proportional to the first order functions $\left\{w(r),...\right\}$ and their derivatives. {It is similar} for the operators $\hat{\gamma}^{\pm,0}_{p,a}$, they depend linearly on the second order functions $\left\{m_0(r),h_0(r),v_2(r),h_2(r),...\right\}$ and {additionally on} their derivatives.}.
These are linear operators in $r$, with coefficients that depend on $(r,\mathrm{l},\mathrm{m},\omega)$. From these systems we can see that the functions $p_{[\mathrm{l},\mathrm{m}]}$ are in principle coupled with $p_{[\mathrm{l}\pm2,\mathrm{m}]}$ and $a_{[\mathrm{l}\pm1,\mathrm{m}]}$, while the $a_{[\mathrm{l},\mathrm{m}]}$ functions are coupled with $a_{[\mathrm{l}\pm2,\mathrm{m}]}$ and $p_{[\mathrm{l}\pm1,\mathrm{m}]}$.

For a static background ($\epsilon_r=0$), the axial and polar perturbations with different $\mathrm{l}$ number decouple from each other, meaning that we just have 
\begin{eqnarray}
\mathcal{P}[\mathrm{l},\mathrm{m}]=\hat{\alpha}_{p} \circ p_{[\mathrm{l},\mathrm{m}]} =0 \, , \mathcal{A}[\mathrm{l},\mathrm{m}]=\hat{\alpha}_{a} \circ a_{[\mathrm{l},\mathrm{m}]} =0 \, . 
\end{eqnarray}

For a rotating background, the situation changes drastically. 
For the system to be consistent we have to consider a tower of equations, for example $\mathcal{P}[\mathrm{l},\mathrm{m}]=0$, $\mathcal{A}[\mathrm{l}\pm1,\mathrm{m}]=0$, $\mathcal{P}[\mathrm{l}\pm2,\mathrm{m}]=0$, etc.

To consistently truncate this tower of equations, first note that we are looking for quasinormal modes with eigenfrequencies $\omega$ that should smoothly connect to the quasinormal modes of the static background as we decrease smoothly the angular momentum $J$:
\begin{eqnarray}
\omega_{[\mathrm{l},\mathrm{m}]} = \omega^{(0)}_{[\mathrm{l},\mathrm{m}]}
+ \epsilon_r \delta\omega^{(1)}_{[\mathrm{l},\mathrm{m}]}
+ \epsilon_r^2 \delta\omega^{(2)}_{[\mathrm{l},\mathrm{m}]} \, .
\end{eqnarray}
The same is true for the perturbation functions: as we decrease smoothly the angular momentum $J$, the closer the perturbation functions should be to the ones of a static background. This means that
\begin{eqnarray}
p_{[\mathrm{l},\mathrm{m}]} = p^{(0)}_{[\mathrm{l},\mathrm{m}]}
+ \epsilon_r \delta p^{(1)}_{[\mathrm{l},\mathrm{m}]}
+ \epsilon_r^2 \delta p^{(2)}_{[\mathrm{l},\mathrm{m}]} \, , \\
a_{[\mathrm{l},\mathrm{m}]} = a^{(0)}_{[\mathrm{l},\mathrm{m}]}
+ \epsilon_r \delta a^{(1)}_{[\mathrm{l},\mathrm{m}]}
+ \epsilon_r^2 \delta a^{(2)}_{[\mathrm{l},\mathrm{m}]} \, .
\end{eqnarray}

Hence, for small enough values of $\epsilon_r$, we could expect a family of polar-led perturbations and a family of axial-led perturbations\footnote{We are leaving out modes that do not have an analytical static limit.}.
In this way, for fixed values of $\mathrm{l},\mathrm{m}$, it is possible to decouple the equations into two sets, the polar-led and the axial-led perturbations.

\subsubsection{Polar-led perturbations}

We are interested in obtaining the modes connected to the polar $\omega^{(0)}_{[\mathrm{l},\mathrm{m}]}$ modes of the static limit. The dominating perturbation functions in this case are $p_{[\mathrm{l},\mathrm{m}]}$, and they satisfy in the static limit the equations $\mathcal{P}[\mathrm{l},\mathrm{m}]=0$ (see Eq.(\ref{pol_gen})), so at second order in $\epsilon_r$ we should solve
\begin{eqnarray}
\hat{\alpha}_{p} \circ p_{[\mathrm{l},\mathrm{m}]} 
&+& \epsilon_r 
\left(
  \mathrm{m} \cdot \hat{\beta}^0_{p} \circ p_{[\mathrm{l},\mathrm{m}]} 
+ \hat{\beta}^+_{p} \circ a_{[\mathrm{l+1},\mathrm{m}]}
+ \hat{\beta}^-_{p} \circ a_{[\mathrm{l-1},\mathrm{m}]}
\right)
\nonumber \\
&+& \epsilon_r^2 
\left(
   \hat{\gamma}^+_{p} \circ p_{[\mathrm{l+2},\mathrm{m}]}
+  \hat{\gamma}^0_{p} \circ p_{[\mathrm{l},\mathrm{m}]}
+  \hat{\gamma}^-_{p} \circ p_{[\mathrm{l-2},\mathrm{m}]}
\right) =0
\, .
\label{eq_alphap}
\end{eqnarray}
The perturbation functions $a_{[\mathrm{l}\pm1,\mathrm{m}]}$ enter at first order in $\epsilon_r$. These perturbation functions, on the other hand, must satisfy $\mathcal{A}[\mathrm{l}\pm1,\mathrm{m}]=0$. Since they enter as a first order term in Eq.(\ref{eq_alphap}),
we can drop the $\epsilon_r^2$ terms in $\mathcal{A}[\mathrm{l}\pm1,\mathrm{m}]=0$, as they would produce higher order corrections $\epsilon_r^3$.
\begin{eqnarray}
 \hat{\alpha}_{a} \circ a_{[\mathrm{l}+1,\mathrm{m}]} 
+ \epsilon_r 
\left(
  \mathrm{m} \cdot \hat{\beta}^0_{a} \circ a_{[\mathrm{l}+1,\mathrm{m}]} 
+  \hat{\beta}^+_{a} \circ p_{[\mathrm{l+2},\mathrm{m}]}
+ \hat{\beta}^-_{a} \circ p_{[\mathrm{l},\mathrm{m}]}
\right) = 0 \, , \\
 \hat{\alpha}_{a} \circ a_{[\mathrm{l}-1,\mathrm{m}]} 
+ \epsilon_r 
\left(
  \mathrm{m} \cdot \hat{\beta}^0_{a} \circ a_{[\mathrm{l}-1,\mathrm{m}]} 
+  \hat{\beta}^+_{a} \circ p_{[\mathrm{l},\mathrm{m}]}
+ \hat{\beta}^-_{a} \circ p_{[\mathrm{l-2},\mathrm{m}]}
\right) = 0 \, .
\end{eqnarray}
Next, the perturbation functions $p_{[\mathrm{l}\pm2,\mathrm{m}]}$ enter Eq.(\ref{eq_alphap}) at second order in $\epsilon_r$. 
This means we should solve the $\mathcal{P}[\mathrm{l}\pm2,\mathrm{m}]=0$ equations. 
By the same argument as before, we can drop the corrections in the order of $\epsilon_r$ and $\epsilon_r^2$ in these equations, since they carry corrections of order $\epsilon_r^3$ or higher. 
Hence, the equations we need are only 
\begin{eqnarray}
\label{lin_homo_ode1}
\hat{\alpha}_{p} \circ p_{[\mathrm{l}+2,\mathrm{m}]} =0
\, , 
\\
\hat{\alpha}_{p} \circ p_{[\mathrm{l}-2,\mathrm{m}]} =0
\, .
\label{lin_homo_ode2}
\end{eqnarray}
Recall that {Eq. (\ref{lin_homo_ode1}) and (\ref{lin_homo_ode2}) are linear equations, 
therefore a trivial solution is
}
\begin{eqnarray}
p_{[\mathrm{l}\pm2,\mathrm{m}]}=0 \, .
\end{eqnarray}
This {solution} simplifies greatly the system of equations we have to consider, meaning, for polar-led perturbations, the equations we need to solve boil down to
\begin{eqnarray}
&& \hat{\alpha}_{p} \circ p_{[\mathrm{l},\mathrm{m}]} 
+ \epsilon_r 
\left(
  \mathrm{m} \cdot \hat{\beta}^0_{p} \circ p_{[\mathrm{l},\mathrm{m}]} 
+ \hat{\beta}^+_{p} \circ a_{[\mathrm{l+1},\mathrm{m}]}
+ \hat{\beta}^-_{p} \circ a_{[\mathrm{l-1},\mathrm{m}]}
\right) + \epsilon_r^2 
\hat{\gamma}^0_{p} \circ p_{[\mathrm{l},\mathrm{m}]}
=0
\, , \, \, \, \, \, \, \, \,  \, \,  \\
&& \hat{\alpha}_{a} \circ a_{[\mathrm{l}+1,\mathrm{m}]} 
+ \epsilon_r 
\left(
  \mathrm{m} \cdot \hat{\beta}^0_{a} \circ a_{[\mathrm{l}+1,\mathrm{m}]} 
+ \hat{\beta}^-_{a} \circ p_{[\mathrm{l},\mathrm{m}]}
\right) = 0 \, , \\
&& \hat{\alpha}_{a} \circ a_{[\mathrm{l}-1,\mathrm{m}]} 
+ \epsilon_r 
\left(
  \mathrm{m} \cdot \hat{\beta}^0_{a} \circ a_{[\mathrm{l}-1,\mathrm{m}]} 
+  \hat{\beta}^+_{a} \circ p_{[\mathrm{l},\mathrm{m}]}
\right) = 0 \, .
\end{eqnarray}

\subsubsection{Axial-led perturbations}

An analogous argument works in the case of axial-led perturbations. It is possible to show that in this case the system of equations can be reduced to
\begin{eqnarray}
&& \hat{\alpha}_{a} \circ a_{[\mathrm{l},\mathrm{m}]} 
+ \epsilon_r 
\left(
  \mathrm{m} \cdot \hat{\beta}^0_{a} \circ a_{[\mathrm{l},\mathrm{m}]} 
+ \hat{\beta}^+_{a} \circ p_{[\mathrm{l+1},\mathrm{m}]}
+ \hat{\beta}^-_{a} \circ p_{[\mathrm{l-1},\mathrm{m}]}
\right) + \epsilon_r^2 
\hat{\gamma}^0_{a} \circ a_{[\mathrm{l},\mathrm{m}]}
=0
\, ,  \, \, \, \, \, \, \, \, \, \,    \\
&& \hat{\alpha}_{p} \circ p_{[\mathrm{l}+1,\mathrm{m}]} 
+ \epsilon_r 
\left(
  \mathrm{m} \cdot \hat{\beta}^0_{p} \circ p_{[\mathrm{l}+1,\mathrm{m}]} 
+ \hat{\beta}^-_{p} \circ a_{[\mathrm{l},\mathrm{m}]}
\right) = 0 \, , \\
&& \hat{\alpha}_{p} \circ p_{[\mathrm{l}-1,\mathrm{m}]} 
+ \epsilon_r 
\left(
  \mathrm{m} \cdot \hat{\beta}^0_{p} \circ p_{[\mathrm{l}-1,\mathrm{m}]} 
+  \hat{\beta}^+_{p} \circ a_{[\mathrm{l},\mathrm{m}]}
\right) = 0 \, .
\end{eqnarray}



\subsection{Perturbation equations for $\mathrm{l}=2$ modes}

The system of equations described in the previous section can be further simplified\footnote{Note that this is in principle an over-determined system, as we have more ODEs than the dependent variables.}. In the following, we will focus on the $\mathrm{l}=2$ perturbations with $|\mathrm{m}|=0,1,2$. 

First we fix the remaining gauge freedom in the perturbations by choosing $h_{0[\mathrm{l},\mathrm{m}]}=0$ and $V_{[2,\mathrm{m}]}=0$ for polar-led perturbations, and $N_{[1,\mathrm{m}]}=0, V_{[1,\mathrm{m}]}=0, V_{[3,\mathrm{m}]}=0$ for the axial-led perturbations. Then in both cases the minimal system of equations has the following generic structure
\begin{eqnarray}
\vec{z}_i \, ' = \mathbf{M}_i \vec{z}_i \, ,
\end{eqnarray}
where $i=p,a$ for polar and axial-led perturbations respectively.
For polar-led perturbations we have
\begin{eqnarray}
   \vec{z}_p = && \Big[  T_{[2,\mathrm{m}]} \, ,
           H_{1[2,\mathrm{m}]} \, ,
           h_{0[3,\mathrm{m}]} \, ,
           h_{1[3,\mathrm{m}]} \, ,
           A_{0[2,\mathrm{m}]} \, ,
           A_{0[2,\mathrm{m}]}' \, , \nonumber \\
    &&       W_{2[1,\mathrm{m}]} \, ,
           W_{2[1,\mathrm{m}]}' \, ,
           W_{2[3,\mathrm{m}]} \, ,
           W_{2[3,\mathrm{m}]}' \, ,
           \Phi_{1[2,\mathrm{m}]} \, ,
           \Phi_{1[2,\mathrm{m}]}' \Big]^T \, .
\end{eqnarray}
$\mathbf{M}_p$ is a $12 \times 12$ matrix whose elements are functions of the background metric (up to second order in rotation), the angular numbers $\mathrm{l}$ and $\mathrm{m}$, and the eigenfrequency $\omega$. 
For axial-led perturbations ($i=a$) we have
\begin{eqnarray}
   \vec{z}_a =  && \Big[ 
   h_{0[2,\mathrm{m}]} \, ,
   h_{1[2,\mathrm{m}]} \, ,
   L_{[1,\mathrm{m}]} \, ,
   T_{[3,\mathrm{m}]} \, ,
    H_{1[3,\mathrm{m}]} \, ,
           W_{2[2,\mathrm{m}]} \, ,
           W_{2[2,\mathrm{m}]}' \, ,\nonumber \\
      &&        A_{0[1,\mathrm{m}]} \, ,
           A_{0[1,\mathrm{m}]}' \, ,
           A_{0[3,\mathrm{m}]} \, ,
           A_{0[3,\mathrm{m}]}' \, , 
        \Phi_{1[1,\mathrm{m}]} \, ,
           \Phi_{1[1,\mathrm{m}]}' \, ,
           \Phi_{1[3,\mathrm{m}]} \, ,
           \Phi_{1[3,\mathrm{m}]}' \Big]^T \, .
\end{eqnarray}
Here $\mathbf{M}_a$ is a $15 \times 15$ matrix, whose elements are similar to the polar case. Note that for $|\mathrm{m}|=2$, the polar functions $L_{[1,\mathrm{m}]}$, $A_{0[1,\mathrm{m}]}$ and $\Phi_{1[1,\mathrm{m}]}$ vanish, and the system simplifies.

\subsubsection{Asymptotic behaviour of the wave solution}

The quasinormal modes of a black hole configuration are determined by solutions of the previous system of equations that behave as an outgoing wave far enough from the black hole, and as an ingoing wave as we approach the horizon. To parametrize this behavior it is convenient to introduce the tortoise coordinate.

The tortoise coordinate for the slowly rotating configuration can be written as
\begin{eqnarray}
 \frac{dR^*}{dr} = \sqrt{\frac{g_{rr}}{-g_{tt}+\epsilon_r^2 w^2 r^2 \sin^2(\theta)}} \, ,  
\end{eqnarray}
which as we approach infinity reads
\begin{eqnarray}
\frac{dR^*}{dr} = 1 + \frac{2M_0}{r} + \epsilon_r^2\frac{2\delta M}{r}  = 1 + \frac{2M}{r}\, .
\end{eqnarray}
The outgoing wave solution of the polar-led perturbations reads
\begin{eqnarray}
           T_{[2,\mathrm{m}]} &=& e^{i\omega R^* } \left( T^{p+}_0 + \frac{T^{p+}_1}{r} + ... \right) \, ,\\
           H_{1[2,\mathrm{m}]} &=& i\omega r e^{i\omega R^* } \left( H^{p+}_{10} + \frac{H^{p+}_{11}}{r} + ... \right)\, ,\\
           h_{0[3,\mathrm{m}]} &=&  r e^{i\omega R^* } \left( u^{p+}_{00} + \frac{u^{p+}_{01}}{r} + ... \right)\, ,\\
           h_{1[3,\mathrm{m}]} &=&  r e^{i\omega R^* } \left( u^{p+}_{10} + \frac{u^{p+}_{11}}{r} + ... \right)\, ,\\
           A_{0[2,\mathrm{m}]} &=&  e^{i\omega R^* } \left( A^{p+}_{00} + \frac{A^{p+}_{01}}{r} + ... \right)\, ,\\
           W_{2[1,\mathrm{m}]} &=&  e^{i\omega R^* } \left( b^{p+}_{20} + \frac{b^{p+}_{21}}{r} + ... \right)\, ,\\
           W_{2[3,\mathrm{m}]} &=&  e^{i\omega R^* } \left( u^{p+}_{20} + \frac{u^{p+}_{21}}{r} + ... \right)\, ,\\
           \Phi_{1[2,\mathrm{m}]} &=& \frac{1}{r} e^{i\omega R^* } \left( P^{p+}_{0} + \frac{P^{p+}_{1}}{r} + ... \right)\, .
\end{eqnarray}
If one asks for the above expansions to satisfy the perturbation equations, this translates into a tower of algebraic relations for the coefficients of the expansions. 
It can be worked out that, when $|\mathrm{m}|<2$, only six of the constant coefficients of the expansions are free. These are: the polar and axial gravitational amplitudes $T^{p+}_0$ and $u^{p+}_{00}$; the polar electromagnetic amplitude $A^{p+}_{00}$, and two axial electromagnetic amplitudes $b^{p+}_{20}$ and $u^{p+}_{20}$; the scalar amplitude $P^{p+}_{0}$. All other constant coefficients in the series expansions are fixed by the perturbation equations in terms of these six amplitudes (and certainly they depend in general on the parameters of the background, angular numbers and eigenfrequency), for example, $H^{p+}_{10}=T^{p+}_0$, $u^{p+}_{10}=-u^{p+}_{00}$, etc. When $|\mathrm{m}|=2$, there are only five free constant coefficients, since $b^{p+}_{20}$ vanishes.

For axial-led perturbations, the expansion looks similar
\begin{eqnarray}
           h_{0[2,\mathrm{m}]} &=&  r e^{i\omega R^* } \left( h^{a+}_{00} + \frac{h^{a+}_{01}}{r} + ... \right) \, ,\\
           h_{1[2,\mathrm{m}]} &=&  r e^{i\omega R^* } \left( h^{a+}_{10} + \frac{h^{a+}_{11}}{r} + ... \right)\, ,\\
           T_{[3,\mathrm{m}]} &=& e^{i\omega R^* } \left( u^{a+}_{00} + \frac{u^{a+}_{01}}{r} + ... \right)\, ,\\
           H_{1[3,\mathrm{m}]} &=& i\omega r e^{i\omega R^* } \left( u^{a+}_{10} + \frac{u^{a+}_{11}}{r} + ... \right)\, ,\\
           L_{[1,\mathrm{m}]} &=& \frac{1}{r^3} e^{i\omega R^* } \left( b^{a+}_{00} + \frac{b^{a+}_{01}}{r} + ... \right)\, ,\\
           W_{2[2,\mathrm{m}]} &=& e^{i\omega R^* } \left( W^{a+}_{20} + \frac{W^{a+}_{21}}{r} + ... \right)\, ,\\
           A_{0[1,\mathrm{m}]} &=& e^{i\omega R^* } \left( b^{a+}_{20} + \frac{b^{a+}_{21}}{r} + ... \right)\, ,\\
           A_{0[3,\mathrm{m}]} &=& e^{i\omega R^* } \left( u^{a+}_{20} + \frac{u^{a+}_{21}}{r} + ... \right)\, ,\\
           \Phi_{1[1,\mathrm{m}]} &=& \frac{1}{r} e^{i\omega R^* } \left( b^{a+}_{30} + \frac{b^{a+}_{31}}{r} + ... \right)\, ,\\
           \Phi_{1[3,\mathrm{m}]} &=& \frac{1}{r} e^{i\omega R^* } \left( u^{a+}_{30} + \frac{u^{a+}_{31}}{r} + ... \right)\, .
\end{eqnarray}
Again we insert this expansion in the equations, 
for $|\mathrm{m}|<2$, only seven of the constant coefficients from the above expansion are free: the axial and polar gravitational amplitudes $h^{a+}_{00}$ and $u^{a+}_{00}$; the axial electromagnetic amplitude $W^{a+}_{20}$, and polar electromagnetic amplitudes $b^{a+}_{20}$ and $u^{a+}_{20}$; and the scalar amplitudes $b^{a+}_{30}$ and $u^{a+}_{30}$. The other coefficients are fixed by the perturbation equations consistently in terms of these amplitudes, and depend on the background, angular numbers and eigenfrequencies, for example, $h^{a+}_{10}=-h^{a+}_{00}$, $u^{a+}_{10}=u^{a+}_{00}$, etc. For $|\mathrm{m}|=2$ there are only five free constants, since $b^{a+}_{20}$ and  $b^{a+}_{30}$ vanish.

Next we study the ingoing wave behaviour at the horizon. The tortoise coordinate close to the horizon reads
\begin{eqnarray}
\frac{dR^*}{dr} = \frac{r_H^3}{(r_H^2-Q^2)(r-r_H)} + \epsilon_r^2 \frac{2J^2r_H^3(Q^2+3r_H^2)}{(r_H^4-Q^4)^2(r-r_H)}  \, .
\end{eqnarray}
The ingoing wave solution of the polar-led perturbations has the form
\begin{eqnarray}
           T_{[2,\mathrm{m}]} &=& e^{-i(\omega-\mathrm{m}\Omega_H) R^* } \left( T^{p-}_0 + T^{p-}_1(r-r_H) + ... \right)\, ,\\
           H_{1[2,\mathrm{m}]} &=& i\omega \frac{r_H}{r-r_H} e^{-i(\omega-\mathrm{m}\Omega_H) R^* }  \left( H^{p-}_{10} + H^{p-}_{11}(r-r_H) + ... \right)\, ,\\
           h_{0[3,\mathrm{m}]} &=& e^{-i(\omega-\mathrm{m}\Omega_H) R^* }  \left( u^{p-}_{00} + u^{p-}_{01}(r-r_H) + ... \right)\, ,\\
           h_{1[3,\mathrm{m}]} &=& \frac{r_H}{r-r_H} e^{-i(\omega-\mathrm{m}\Omega_H) R^* }  \left( u^{p-}_{10} + u^{p-}_{11}(r-r_H) + ... \right)\, ,\\
           A_{0[2,\mathrm{m}]} &=&  e^{-i(\omega-\mathrm{m}\Omega_H) R^* }  \left( A^{p-}_{00} + A^{p-}_{01}(r-r_H) + ... \right)\, ,\\
           W_{2[1,\mathrm{m}]} &=&  e^{-i(\omega-\mathrm{m}\Omega_H) R^* }  \left( b^{p-}_{20} + b^{p-}_{21}(r-r_H) + ... \right)\, ,\\
           W_{2[3,\mathrm{m}]} &=&  e^{-i(\omega-\mathrm{m}\Omega_H) R^* }  \left( u^{p-}_{20} + u^{p-}_{21}(r-r_H) + ... \right)\, ,\\
           \Phi_{1[2,\mathrm{m}]} &=& \frac{1}{r} e^{-i(\omega-\mathrm{m}\Omega_H) R^* }  \left( P^{p-}_{0} + P^{p-}_{1}(r-r_H) + ... \right)\, .
\end{eqnarray}
Note that the angular velocity of the horizon $\Omega_H$ enters the exponential behavior because of the inertial dragging that the wave suffers close to the horizon. 
Following a similar procedure at infinity, we substitute the horizon expansions into the equations and read the tower of algebraic relations for the expansion coefficients that result. 
Only the following coefficients {result} to be free: $T^{p-}_0$, $u^{p-}_{00}$, $A^{p-}_{00}$, $u^{p-}_{20}$ and $P^{p-}_{0}$ for $|\mathrm{m}|=2$, plus $b^{p-}_{20}$ when $|\mathrm{m}|<2$.

For axial-led perturbations, the expansion at the horizon is
\begin{eqnarray}
           h_{0[2,\mathrm{m}]} &=& e^{-i(\omega-\mathrm{m}\Omega_H) R^* }  \left( h^{a-}_{00} + h^{a-}_{01}(r-r_H) + ... \right)\, ,\\
           h_{1[2,\mathrm{m}]} &=& \frac{r_H}{r-r_H} e^{-i(\omega-\mathrm{m}\Omega_H) R^* }  \left( h^{a-}_{10} + h^{a-}_{11}(r-r_H) + ... \right)\, ,\\
           T_{[3,\mathrm{m}]} &=& e^{-i(\omega-\mathrm{m}\Omega_H) R^* } \left( T^{a-}_0 + T^{a-}_1(r-r_H) + ... \right)\, ,\\
           H_{1[3,\mathrm{m}]} &=& i\omega \frac{r_H}{r-r_H} e^{-i(\omega-\mathrm{m}\Omega_H) R^* }  \left( H^{a-}_{10} + H^{a-}_{11}(r-r_H) + ... \right)\, ,\\
           L_{[1,\mathrm{m}]} &=& e^{-i(\omega-\mathrm{m}\Omega_H) R^* } \left( L^{a-}_{0} + L^{a-}_{1}(r-r_H) + ... \right)\, ,\\
           W_{2[2,\mathrm{m}]} &=& e^{-i(\omega-\mathrm{m}\Omega_H) R^* }  \left( W^{a-}_{20} + W^{a-}_{21}(r-r_H) + ... \right)\, ,\\
           A_{0[1,\mathrm{m}]} &=& e^{-i(\omega-\mathrm{m}\Omega_H) R^* }  \left( b^{a-}_{20} + b^{a-}_{21}(r-r_H) + ... \right)\, ,\\
           A_{0[3,\mathrm{m}]} &=& e^{-i(\omega-\mathrm{m}\Omega_H) R^* }  \left( u^{a-}_{20} + u^{a-}_{21}(r-r_H) + ... \right)\, ,\\
           \Phi_{1[1,\mathrm{m}]} &=& \frac{1}{r} e^{-i(\omega-\mathrm{m}\Omega_H) R^* }  \left( b^{a-}_{30} + b^{a-}_{31}(r-r_H) + ... \right)\, ,\\
           \Phi_{1[3,\mathrm{m}]} &=& \frac{1}{r} e^{-i(\omega-\mathrm{m}\Omega_H) R^* }  \left( u^{a-}_{30} + u^{a-}_{31}(r-r_H) + ... \right)\, .
\end{eqnarray}
Here only the following amplitudes are free: $h^{a-}_{00}$, $u^{a-}_{00}$, $W^{a-}_{20}$, $u^{a-}_{20}$ and $u^{a-}_{30}$ for $|\mathrm{m}|=2$, 
plus $b^{a-}_{20}$ and $b^{a-}_{30}$ when $|\mathrm{m}|<2$.

\section{Results}
\label{results}

\subsection{Numerical method}

We fix the parameters of the background. In this case (of slowly rotating Kerr-Newman {black holes}), it means choosing the horizon radius, electric charge and angular momentum. For a type of perturbation and a particular value of eigenfrequency $\omega$, we evaluate the asymptotic expansions at some points $r_1 > r_H$ and $r_2<\infty$. With the boundary conditions, we integrate the full perturbation equations, producing as many independent solutions as the free constants in the expansions. We integrate the equations using a forth order Runge-Kutta method in a compactified coordinate $x=1-r_H/r$. In particular, the parameters we choose are 
\begin{eqnarray}
1-r_H/r_1 = 10^{-2} \, , \quad 1-r_H/r_2 = 0.9 \, .
\end{eqnarray}
These will reproduce as well the Reissner-Nordstr\"om modes with 5 significant figures, as a sanity check.

We study if a linear combination of these solutions matches at an intermediate point $r_1<r_j<r_2$. For this we construct a determinant (Wronskian) evaluating the functions at $r=r_j$\footnote{Typically we choose $r_j=2r_H$.}. The zeroes of the determinant correspond to the quasinormal mode frequencies. We implement a simple Newton method to calculate the zeroes of this determinant.

With this numerical method we can calculate the quasinormal modes for slowly rotating black holes. In practice, the numerical procedure is only consistent for small values of the angular momentum $J$. For very large values of the angular momentum, the series expansion no longer satisfies the equations. This means, if $J$ is too large, the integrated solutions of the equations gain some numerical noise. 
This method only allows us to obtain the quasinormal modes up to second order in rotation, hence what we should look for are expressions of the form
\begin{eqnarray}
\omega_{[\mathrm{l},\mathrm{m}]} = \omega^{(0)}_{[\mathrm{l},\mathrm{m}]}
+ \epsilon_r \delta\omega^{(1)}_{[\mathrm{l},\mathrm{m}]}
+ \epsilon_r^2 \delta\omega^{(2)}_{[\mathrm{l},\mathrm{m}]} \, ,
\label{omega_expansion}
\end{eqnarray}
for the eigenfrequencies. 

In practice, to extract the linear and quadratic corrections to the quasinormal mode frequency Eq. (\ref{omega_expansion}), what we do is to generate the modes for different values of the angular momentum, considering $|J/M^2| \ll 1$. 

To simplify the calculations and the presentation of the results, we take advantage of the following symmetries:

- The results are independent of the sign of $Q$, hence we focus on $Q>0$.

- The modes for $(J,-\mathrm{m})$ are the same as for $(-J,\mathrm{m})$. Hence we focus on $\mathrm{m}=0,1,2$ with positive and negative values of $J$.

- We focus on the modes $\omega$ with positive real part. As for the modes with negative real part (say $\omega_-$), let us point out that the following relation holds:
$\omega_-(\pm J,\mathrm{m}) =-\omega^*(\mp J,\mathrm{m})$.
Hence all the modes with negative real part can be obtained from the modes with positive real part, as we consider the co- and counter-rotating cases.

In the coming sections we will discuss the results we obtain with this method.
We present the results of quasinormal modes computed for $\mathrm{l}=2, \mathrm{m}=2,1,0$, covering both the co- and counter-rotating black holes in the slow-rotation limit.
We show three families of modes of gravitational, electromagnetic, and scalar nature.
The results are fitted on a quadratic expression
$f(\xi) = a + b\xi + c\xi^2$, where $\xi$ is the scaled angular momentum, $J/M^2$ (see Eq. (\ref{quad_fit}) in Appendix \ref{append_coeff}). 
The coefficients $a,b,c$ of our quadratic fit {for} each case can be found in the appendix as well.
Note that the scaled angular momentum takes its maximum value in the extremal limit, that is, 
\begin{eqnarray}
\xi_{ext}=\frac{J_{ext}}{M^2_{ext}} = \sqrt{\frac{1- \left(\frac{Q}{r_H}\right)^2}{1+ \left(\frac{Q}{r_H} \right)^2}} \, .
\end{eqnarray}

\subsection{Gravitational modes}

\begin{figure}[h!]
		\centering
		\subfloat[]{
    		\includegraphics[width=0.37\textwidth,angle=-90]{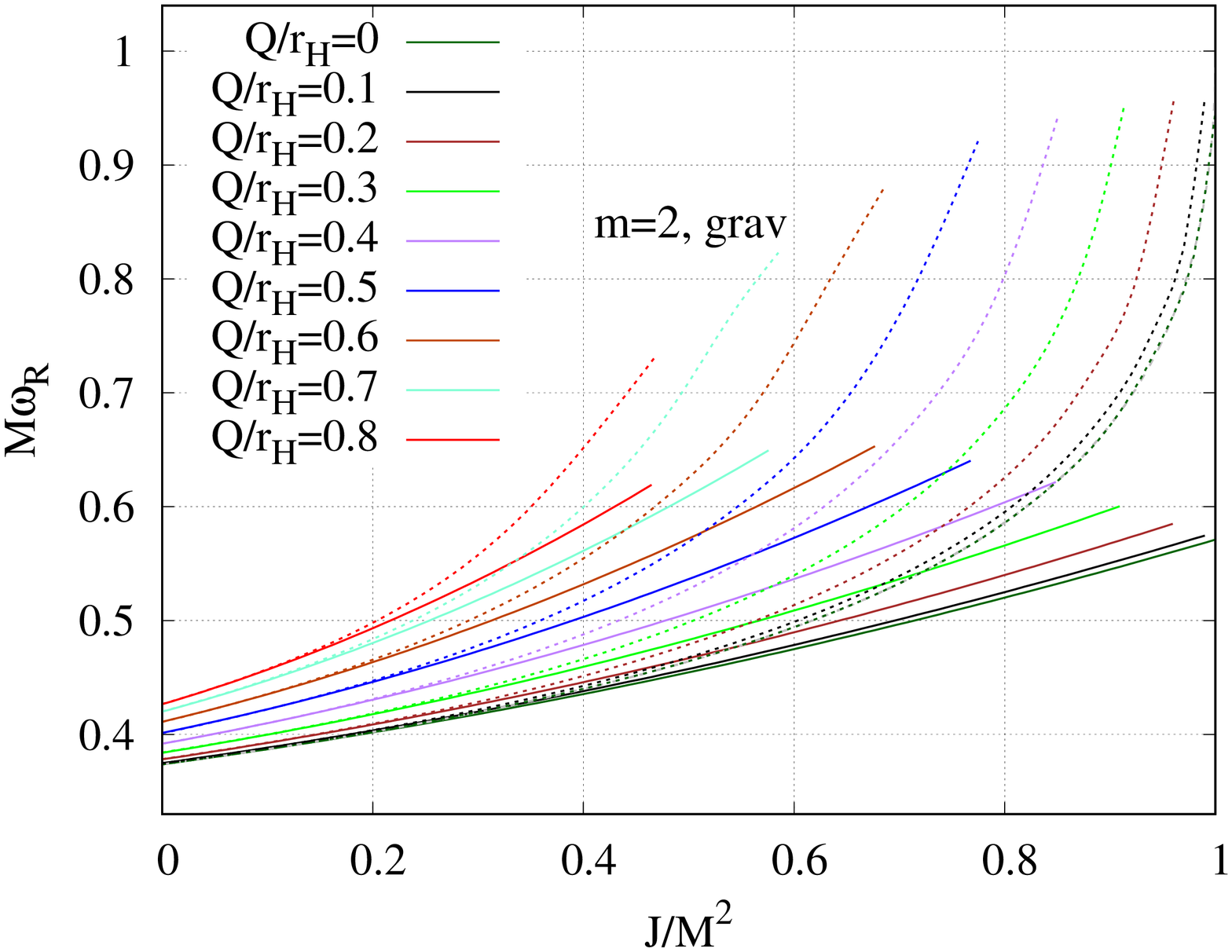}
		}
		\subfloat[]{
	    	\includegraphics[width=0.37\textwidth,angle=-90]{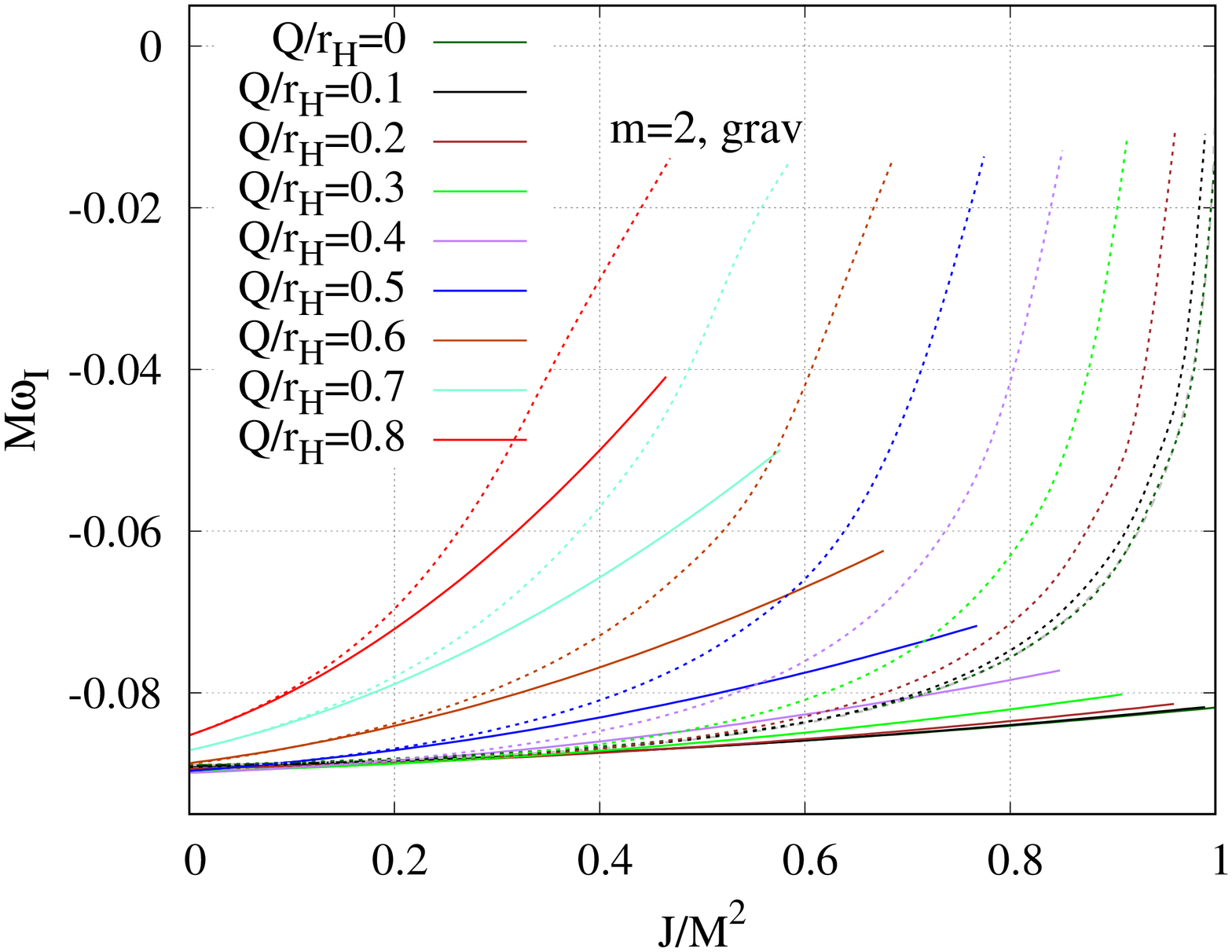}
		} \\
		\subfloat[]{
			\includegraphics[width=0.37\textwidth,angle=-90]{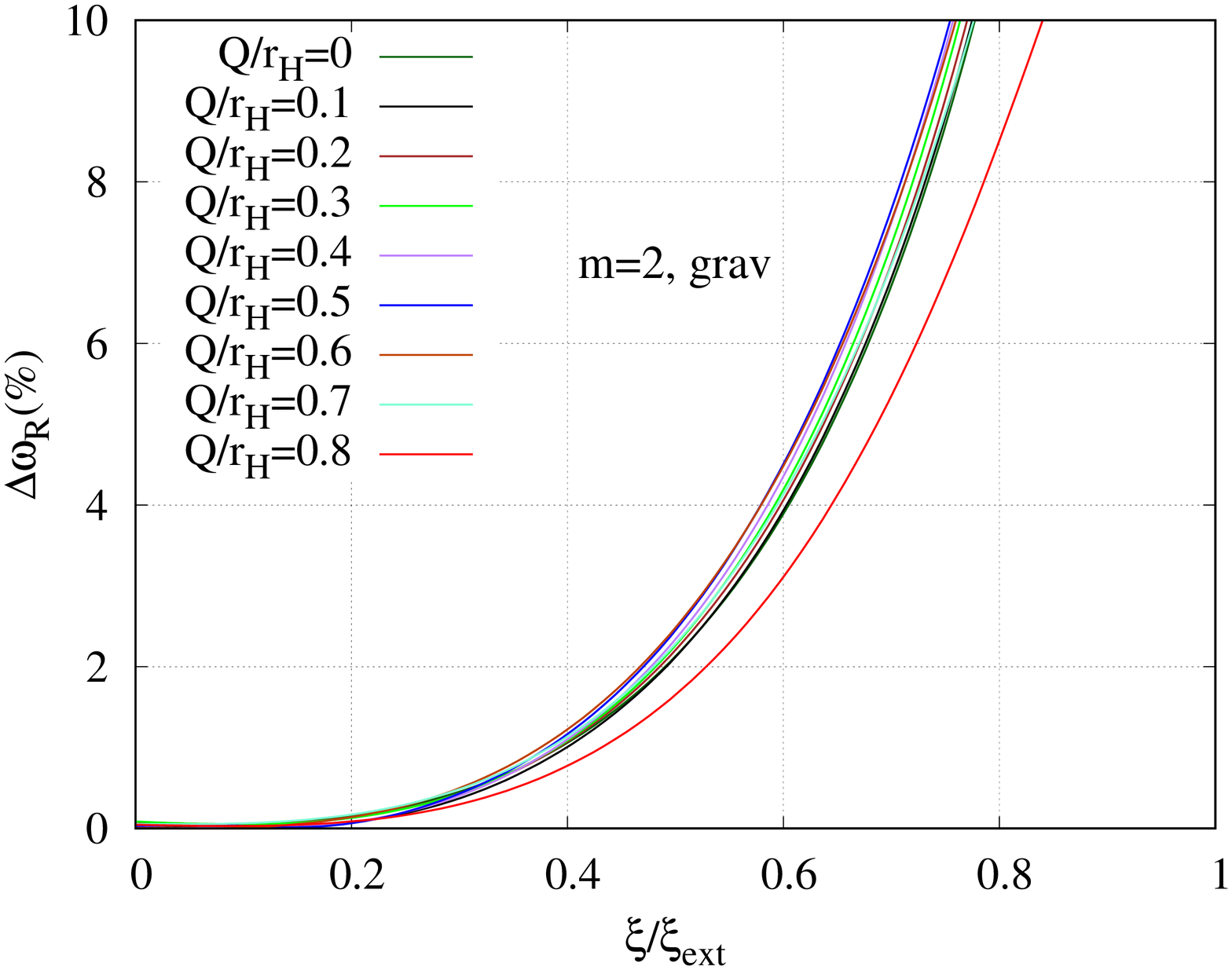}
		}
		\subfloat[]{
		\includegraphics[width=0.37\textwidth,angle=-90]{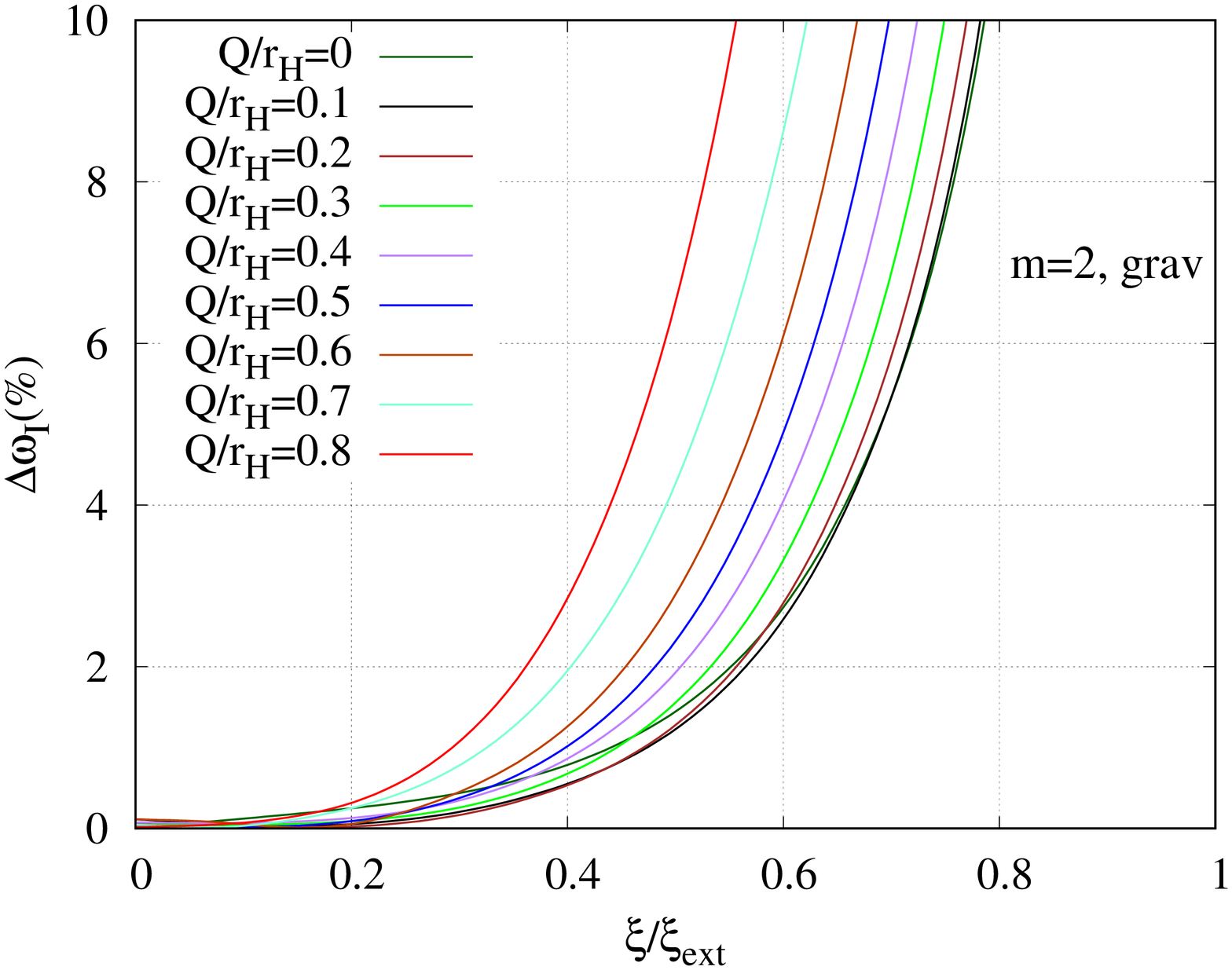}
		}
		\caption{Gravitational $\mathrm{l} = 2, \mathrm{m}=2$ quasinormal modes: 
  \\
  Real frequency $\omega_R$ (a) and imaginary frequency $\omega_I$ (b) scaled with mass $M$ versus angular momentum $J$ scaled with inverse mass-squared. Different values of charge $Q$ divided by the Schwarzschild radius $r_H$ are in different colors. The exact Kerr solution is in dotted grey, and in colors and dotted are the exact solutions {for} different charges.
  \\
  Difference in the gravitational $\mathrm{l} = 2, \mathrm{m}=2$ quasinormal modes between our quadratic fit and formula (\ref{mode_formula}): Real frequency $\omega_R$ (c) and imaginary frequency $\omega_I$ (d) versus the dimensionless angular momentum $\xi/\xi_{ext}$. Different values of charge $Q$ divided by the Schwarzschild radius are in different colors.}
		\label{Fig_l2m2grav_pol_formula}
	\end{figure}

\begin{figure}[]
		\centering
		\subfloat[]{
		\includegraphics[width=0.37\textwidth,angle=-90]{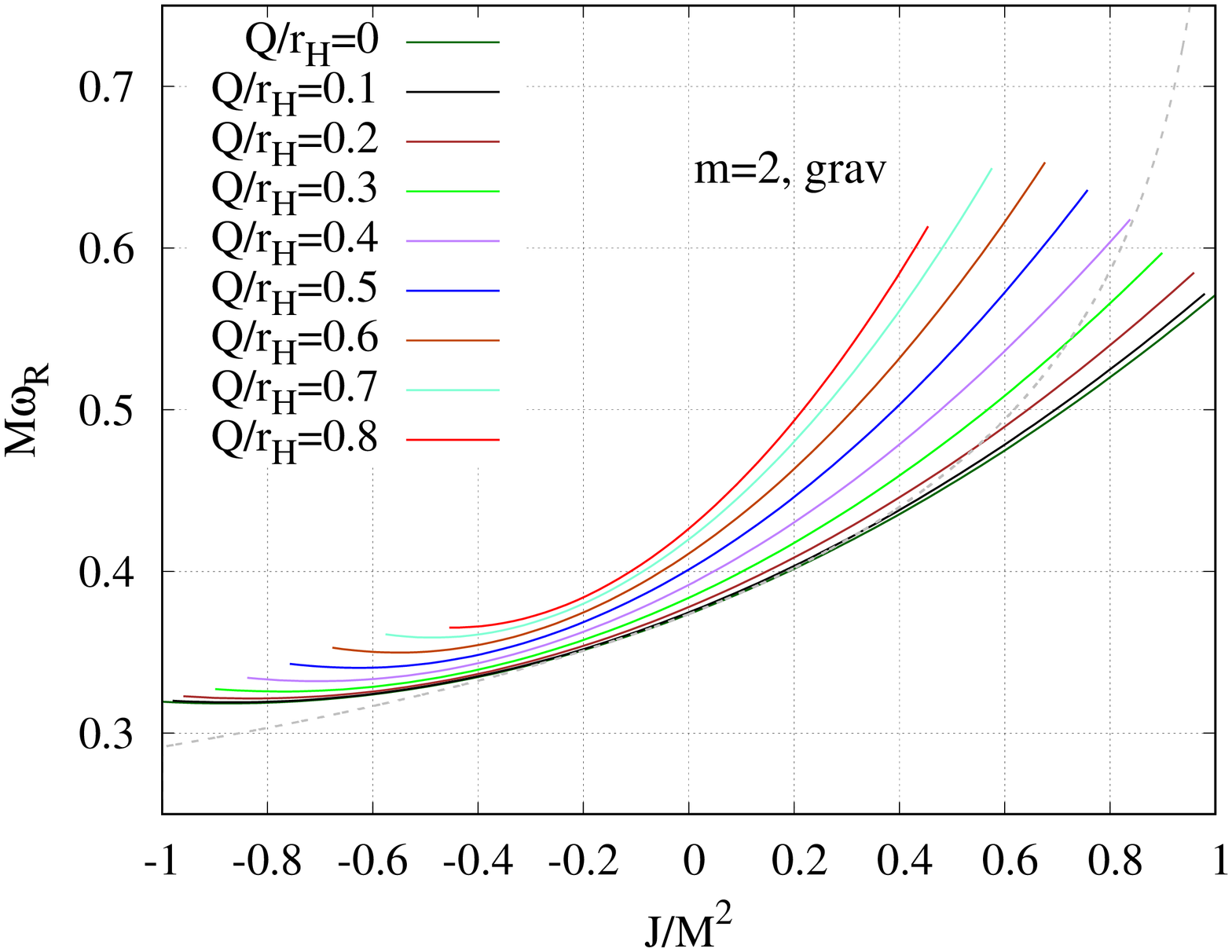}
		}
		\subfloat[]{
		\includegraphics[width=0.37\textwidth,angle=-90]{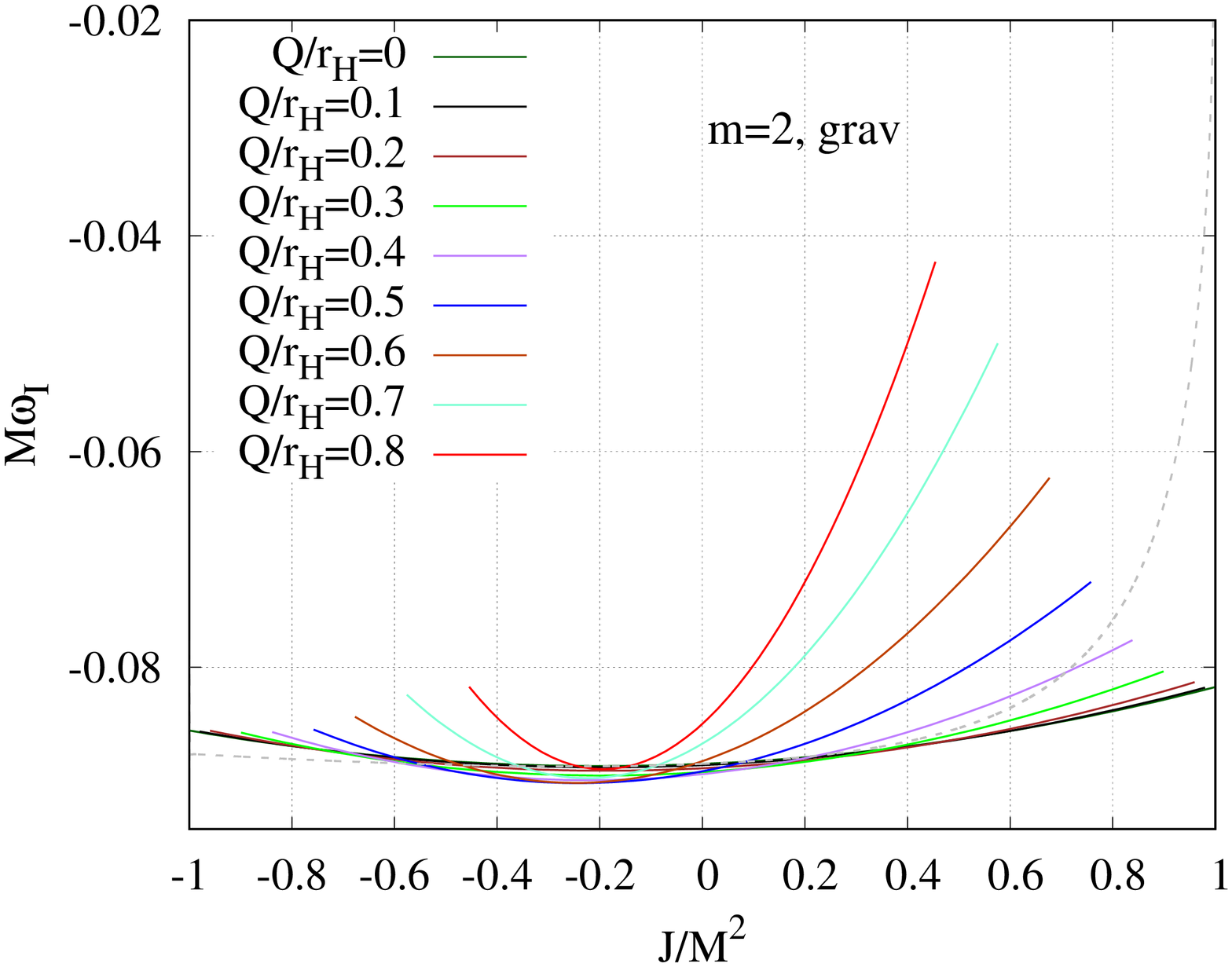}
		} \\
		\subfloat[]{
		\includegraphics[width=0.37\textwidth,angle=-90]{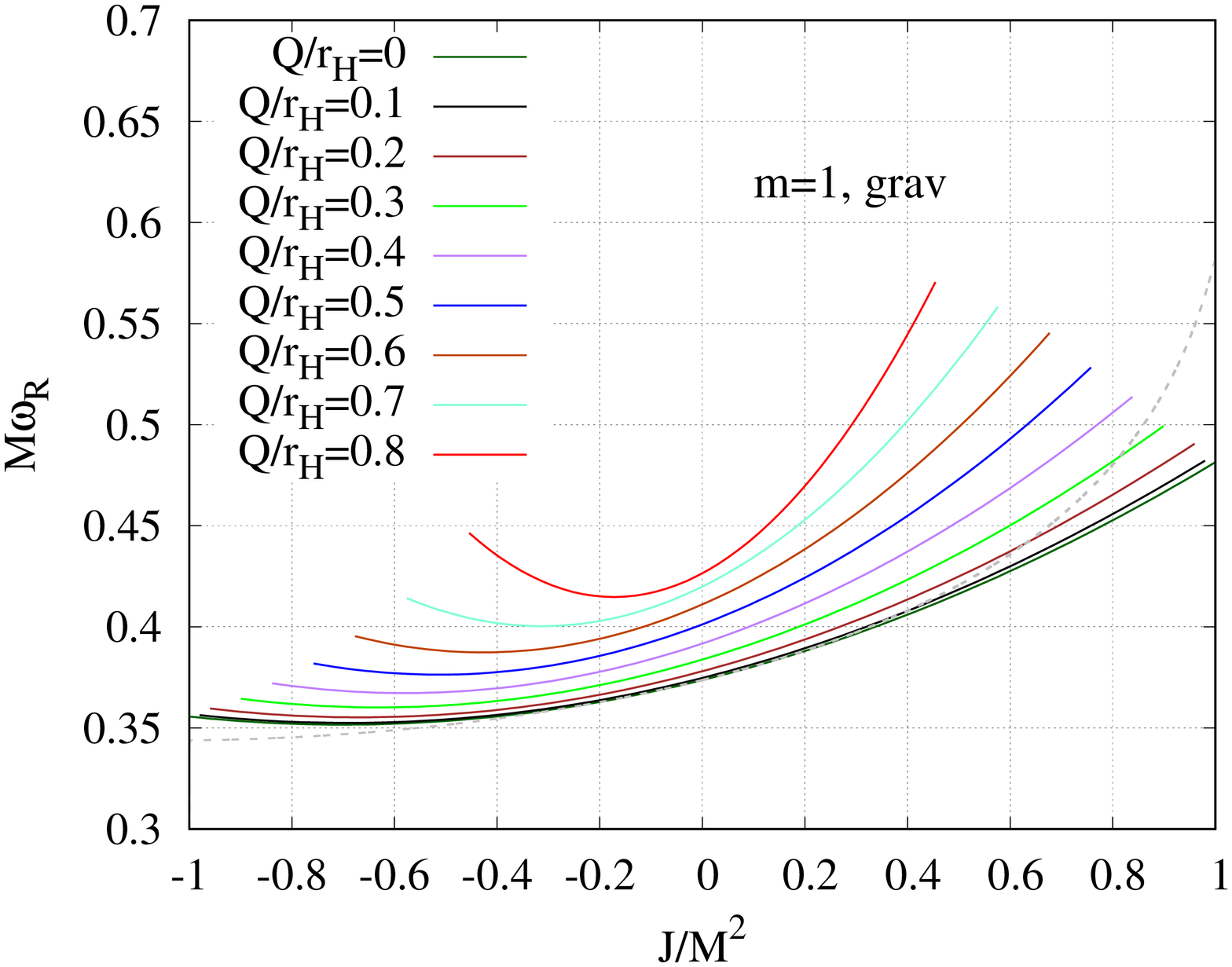}
		} 
		\subfloat[]{
		\includegraphics[width=0.37\textwidth,angle=-90]{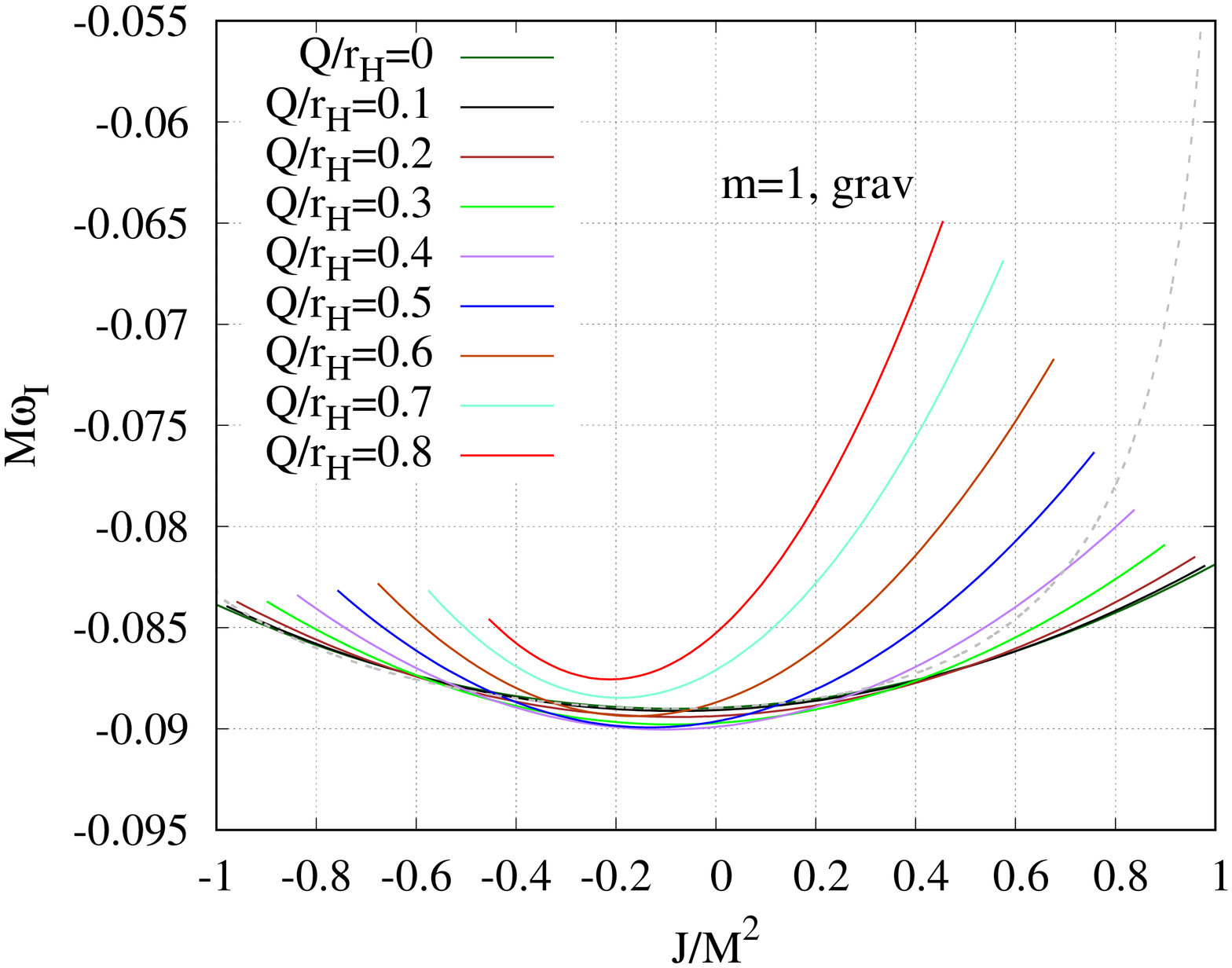}
		} \\
		\subfloat[]{
		\includegraphics[width=0.37\textwidth,angle=-90]{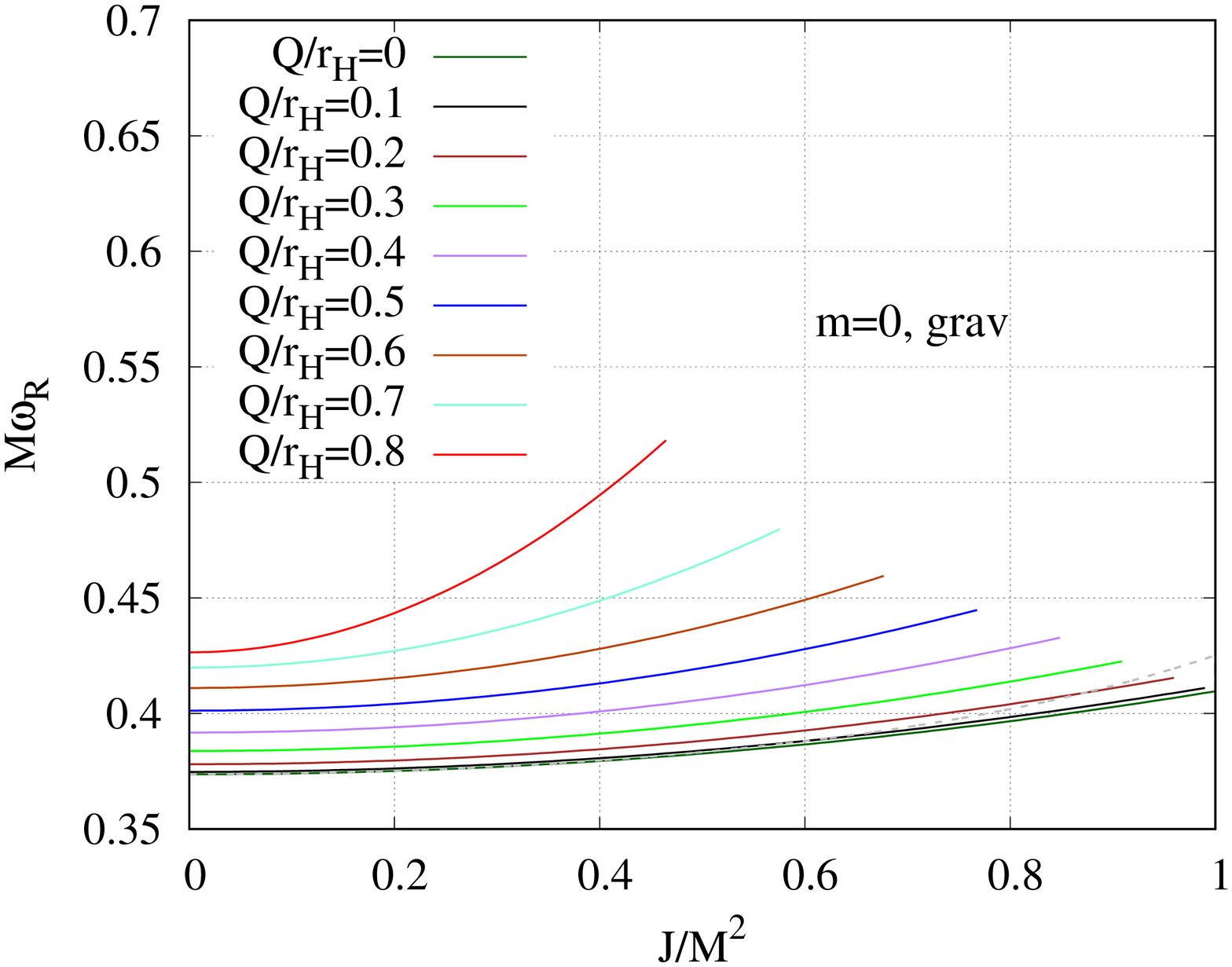}
		}
		\subfloat[]{
		\includegraphics[width=0.37\textwidth,angle=-90]{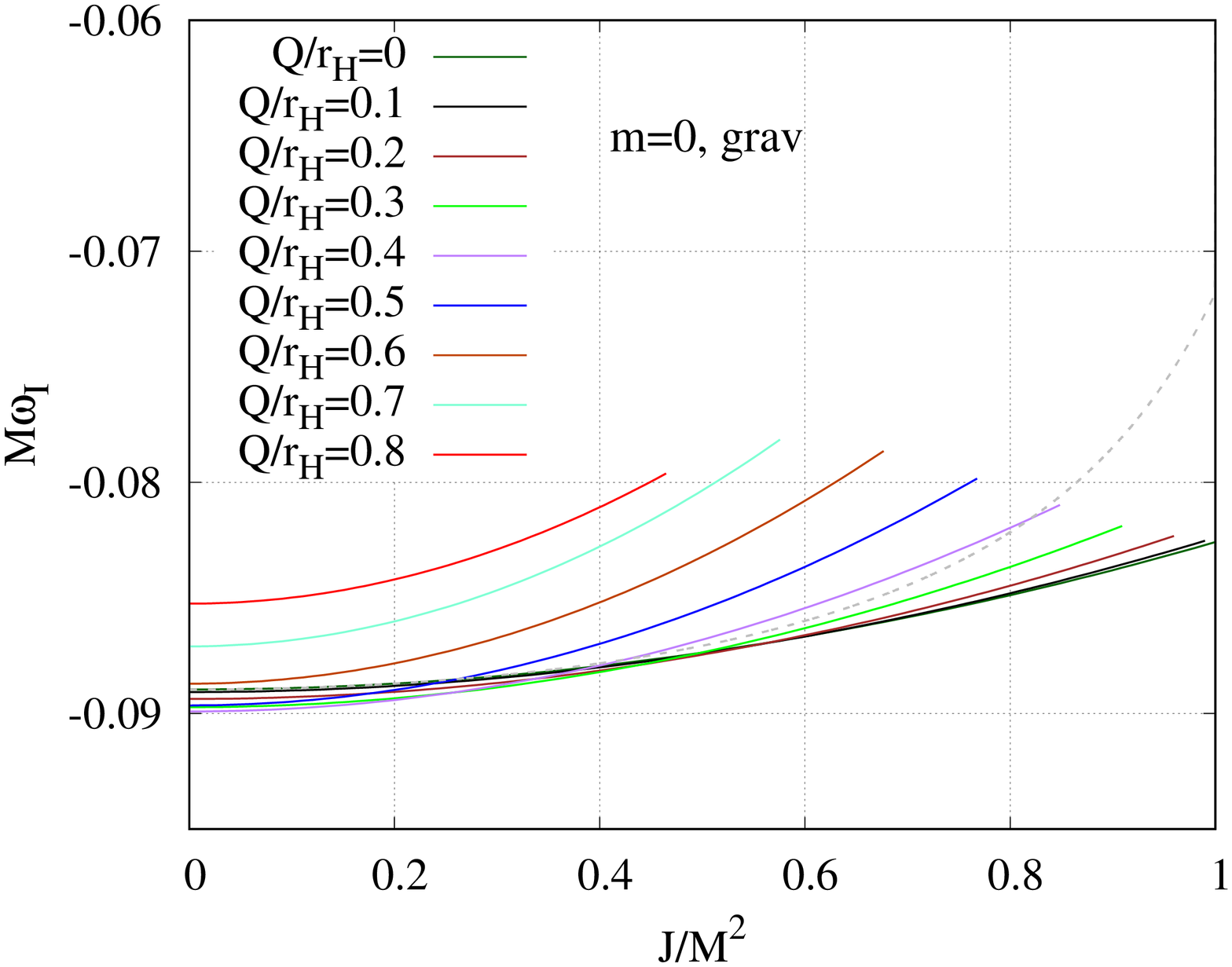}
		}
		\caption{Gravitational $\mathrm{l} = 2, \mathrm{m}=2,1,0$ quasinormal modes: Real frequency $\omega_R$ (a),(c),(e) and imaginary frequency $\omega_I$ (b),(d),(f) scaled with mass $M$ versus angular momentum $J$ scaled with inverse mass-squared. Different values of charge $Q$ divided by the Schwarzschild radius $r_H$ are in different colors, with the exact Kerr solution in dotted grey. }
		\label{Fig_l2grav_pol}
	\end{figure}

Let us start here with the $\mathrm{l}=\mathrm{m}=2$ gravitational-led modes that are expected to dominate the ringdown phase. These are shown in Figure 
 \ref{Fig_l2m2grav_pol_formula}
 and Figures
 \ref{Fig_l2grav_pol}(a),(b). 
 The corresponding coefficients for the quadratic expression are given in Tables \ref{m2_grav}, \ref{m1_grav} and \ref{m0_grav} in Appendix \ref{append_coeff}.

In Figures \ref{Fig_l2m2grav_pol_formula}(a) and (b) we show, respectively, the mass-scaled real and imaginary parts of the frequency as a function of the scaled angular momentum $J/M^2$. Different values of the scaled charge $Q/r_H$ are shown in different colors. Solid lines represent the values of the modes calculated using the second order slow rotation approximation. As a comparison, the exact Kerr-Newman modes are shown with dashed curves\footnote{
In particular, for the calculation of the exact modes we use the formula found in \cite{Carullo:2021oxn}, that is
\begin{equation}
    X = X_{0} \left( \frac{\sum_{k,j=0}^3 b_{k,j} \left(\frac{J}{M^2} \right)^k \left(\frac{Q}{M} \right)^j}{\sum_{k,j=0}^3 c_{k,j} \left(\frac{J}{M^2}\right)^k \left(\frac{Q}{M} \right)^j} \right) \, ,
    \label{mode_formula}
\end{equation}
where $X=(M\omega_R, M\omega_I)$, and $X_0$ are the corresponding Schwarzschild values. The 
coefficients $b_{k,j}, c_{k,j} $ can be found in \cite{Carullo:2021oxn}.
}.
Figures \ref{Fig_l2m2grav_pol_formula}(a) and (b) show an excellent quantitative agreement in the slow rotation limit, and overall a good prediction of the qualitative behaviour (both real and imaginary frequencies increase as the black hole spins up).

Since the maximum scaled angular momentum shrinks as the charge increases, it is convenient to normalize the scaled angular momentum in terms of the extremal value, $\xi/\xi_{ext}$. In Figures \ref{Fig_l2m2grav_pol_formula}(c) and (d) we show the deviation of the slow rotation approximation with respect to the exact quasinormal mode values, as a function of $\xi/\xi_{ext}$. Both figures show that the slow rotation approximation is in very good agreement with the exact values, within a $3\%$ deviation in the real part for $\xi/\xi_{ext}<0.5$, this deviation being rather insensitive to the electric charge. In fact it increases beyond $10\%$ only for very large values of $\xi$ (larger than $0.8\xi_{ext}$) where the slow rotation approximation is not expected to apply anyway. The imaginary part on the other hand is not as independent of the charge. A clear tendency is exhibited where
the higher the dimensionless charge the faster it is
for the deviation to exceed $10\%$. In any case, we find that the deviation is within $7\%$ or smaller for $\xi/\xi_{ext}<0.5$, again showing that the slow rotation approximation can cover a good range of the parameter space.

\begin{figure}[]
		\centering
		\subfloat[]{
		\includegraphics[width=0.37\textwidth,angle=-90]{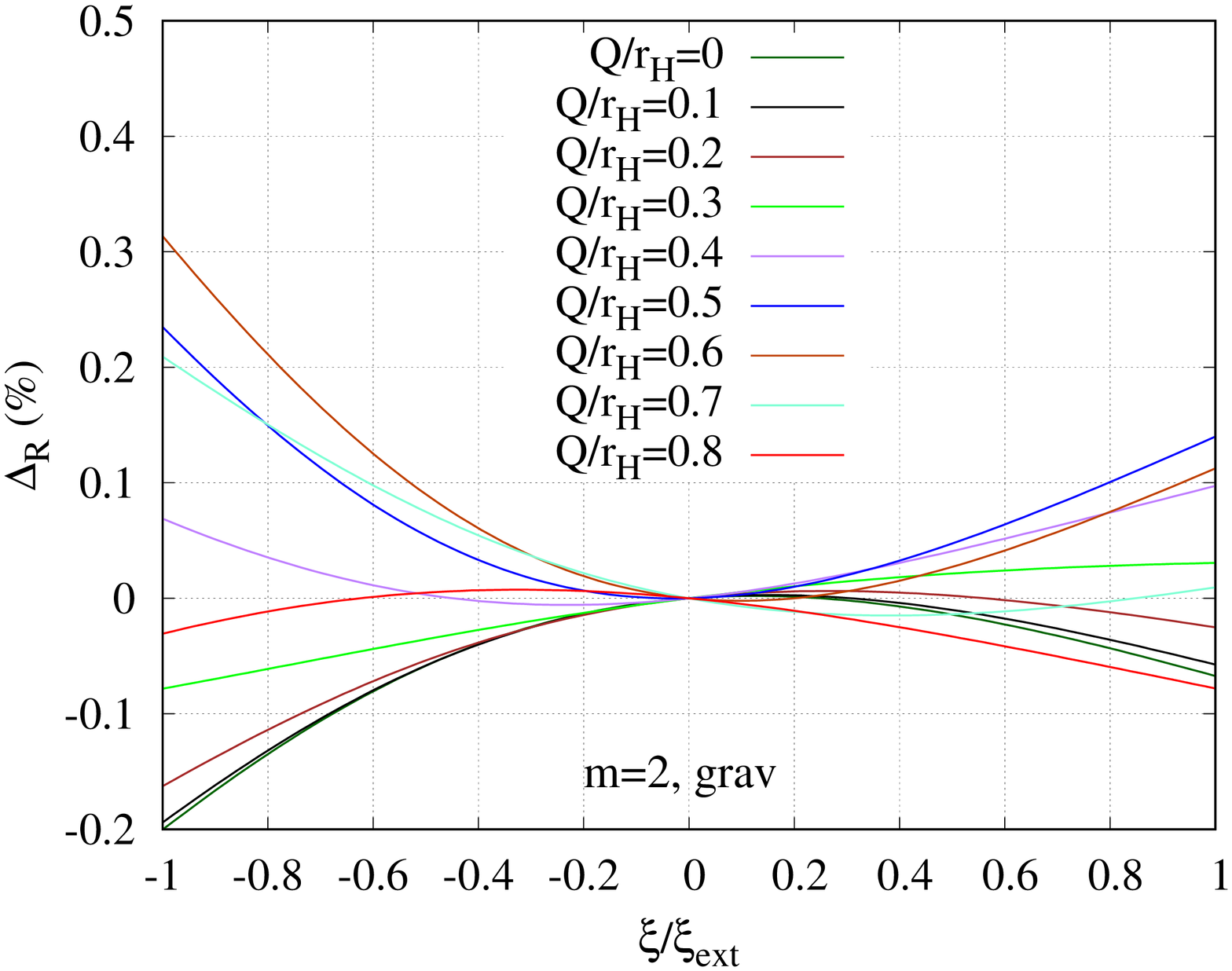}
		}
		\subfloat[]{
		\includegraphics[width=0.37\textwidth,angle=-90]{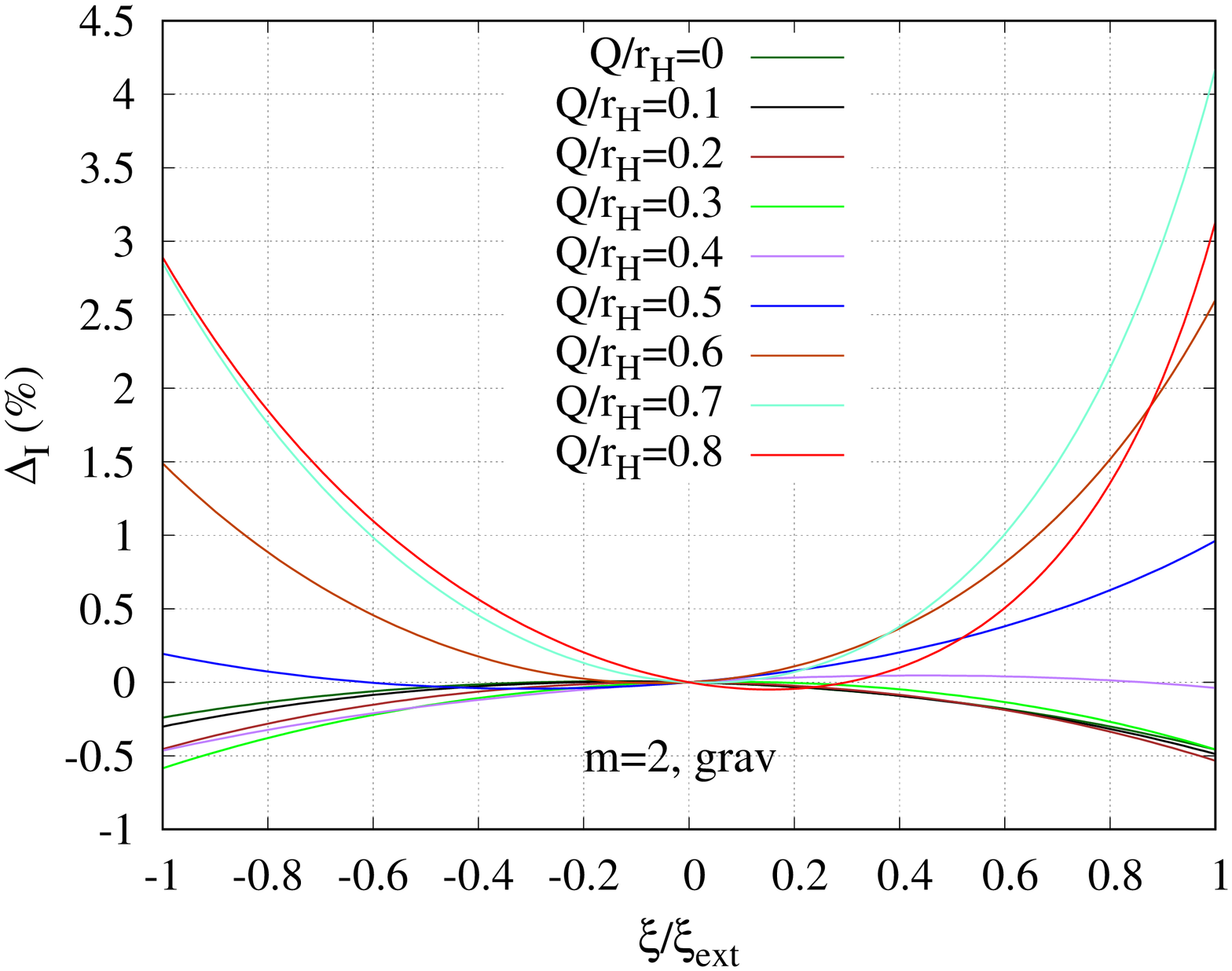}
		} \\
		\subfloat[]{
		\includegraphics[width=0.37\textwidth,angle=-90]{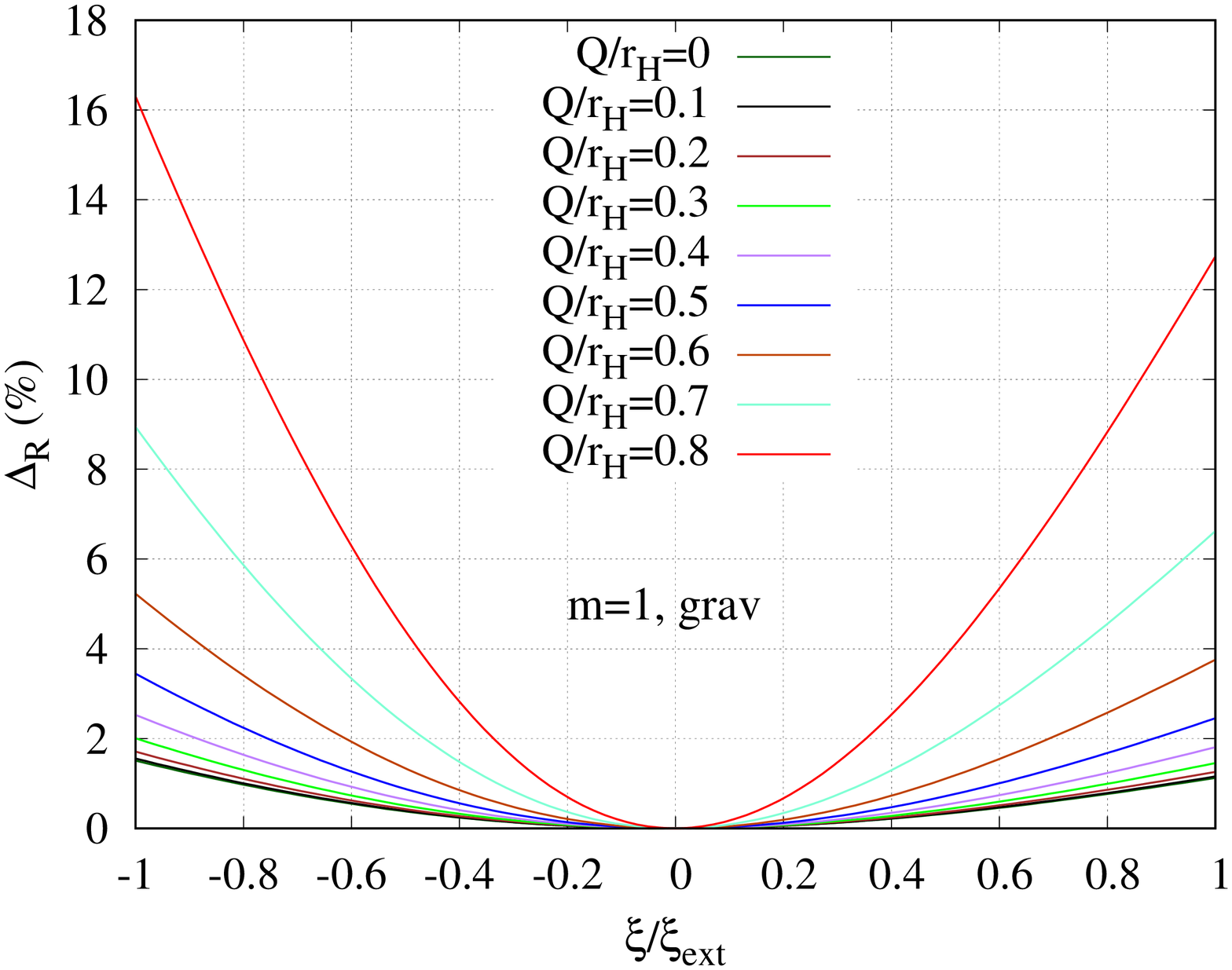}
		}
		\subfloat[]{
		\includegraphics[width=0.37\textwidth,angle=-90]{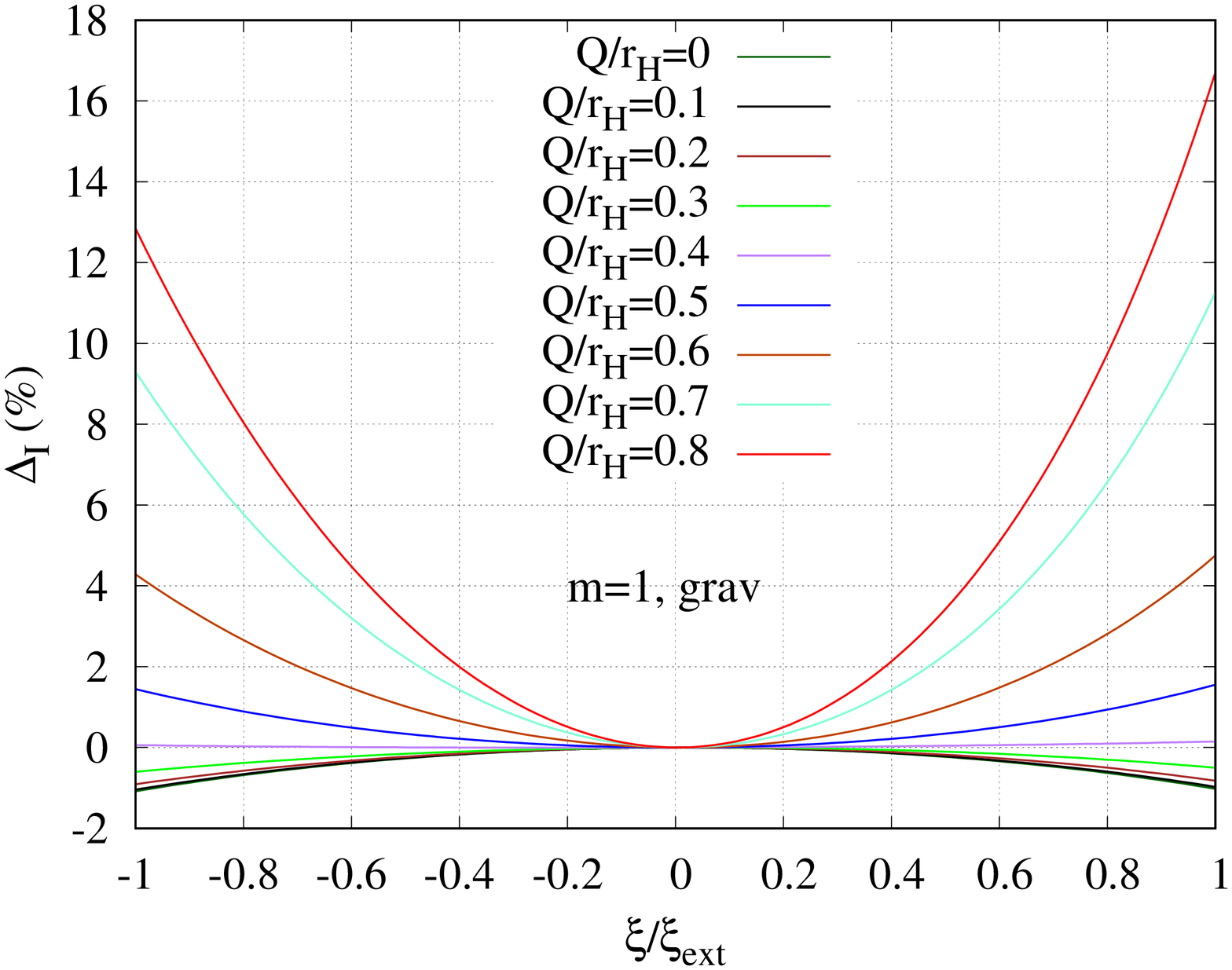}
		} \\
		\subfloat[]{
		\includegraphics[width=0.37\textwidth,angle=-90]{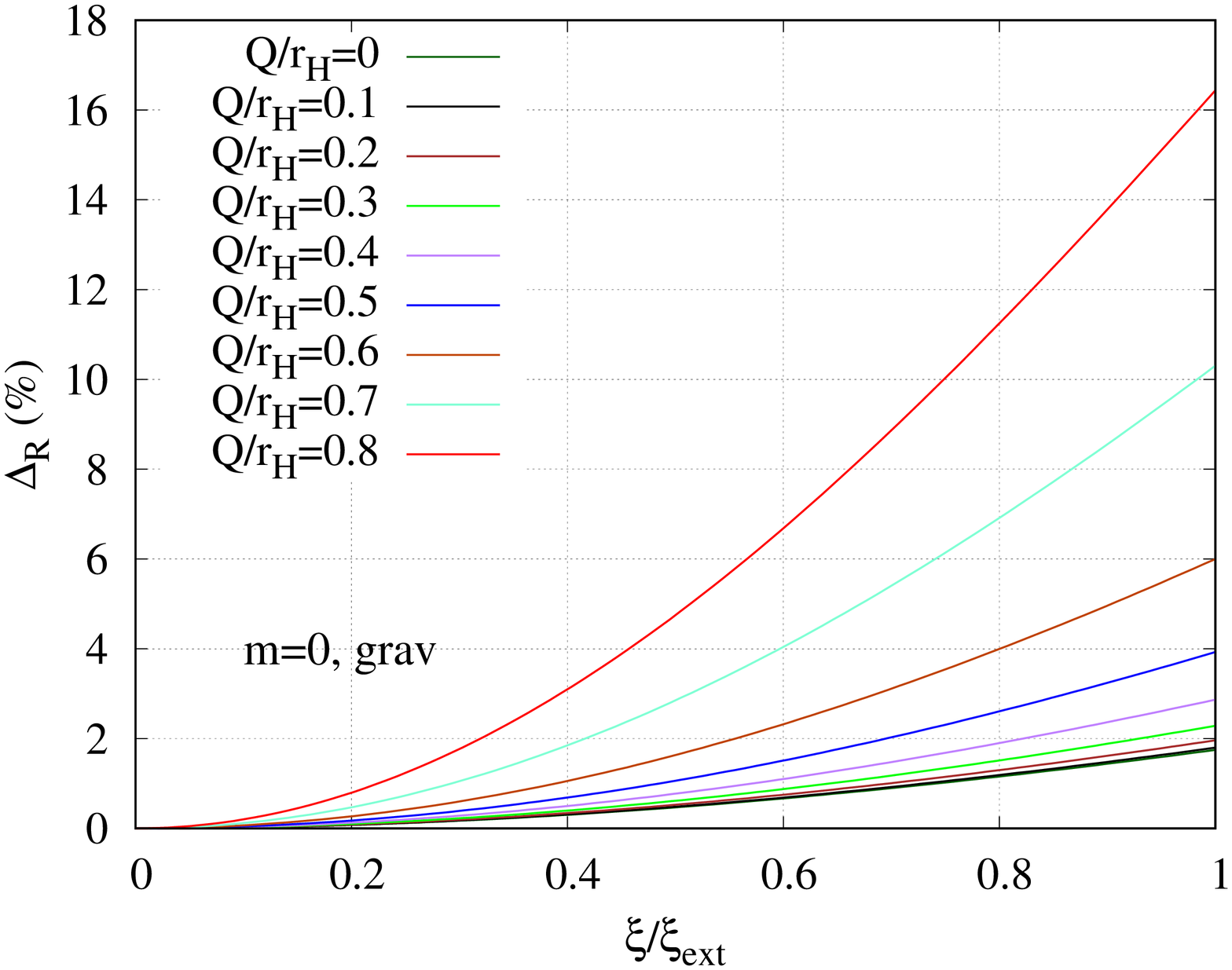}
		}
		\subfloat[]{
		\includegraphics[width=0.37\textwidth,angle=-90]{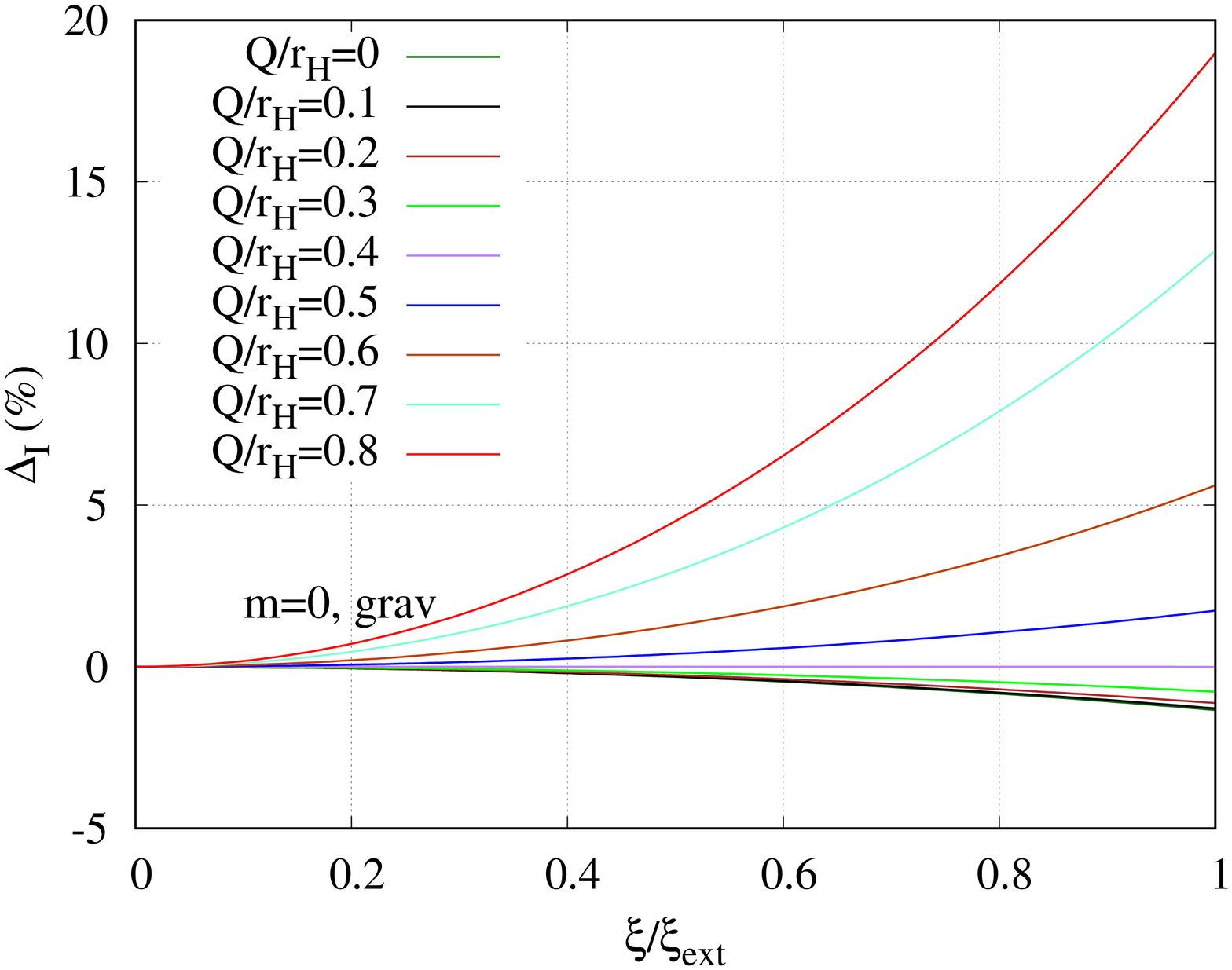}
		}
		\caption{Isospectrality for gravitational $\mathrm{l} = 2, \mathrm{m}=2,1,0$ quasinormal modes: Percentage of difference in the real frequency $\omega_R$ (a),(c),(e) and imaginary frequency $\omega_I$ (b),(d),(f) versus the dimensionless angular momentum $\xi$ scaled with its extremal value. Different values of charge $Q$ divided by the Schwarzschild radius $r_H$ are in different colors. }
		\label{Fig_l2grav_ISO}
	\end{figure}	

In Figures \ref{Fig_l2grav_pol} we show the quasinormal modes for $\mathrm{m}=2,1,0$ in the slow rotation approximation, including the negative values of the angular momentum (a compact way of showing also the modes for negative values of $\mathrm{m}$). 

In Figure \ref{Fig_l2grav_pol}(a), 
the real frequency of the $\mathrm{l}= 2, \mathrm{m}=2$ gravitational modes scaled with the black hole mass $M$ is shown with respect to the angular momentum $J$ divided by the square of the black hole mass.
The colors represent the different charges scaled with the Schwarzschild radius $r_H$, 
with Kerr given by the grey dotted line. 
As the positive $J/M^{2}$ grows,
the scaled real frequency tends to increase for the different charges.
While going down the negative $J/M^{2}$ from 0, the scaled real frequency decreases before rising slightly again.
The rise of the real frequency is {stronger for} larger scaled charge.
For the imaginary frequency similarly scaled by the mass 
a more drastic increase is observed with higher charge, {as seen} in Figure 
\ref{Fig_l2grav_pol}(b).
Note that for Kerr {black  holes}, in our slow-rotating approximation, the resulting modes coincide with the exact Kerr modes up to $J/M^2=\pm 0.5$.
	
The spectrum of the gravitational modes for $\mathrm{m}=1$
is given in Figure 
\ref{Fig_l2grav_pol}(c) and (d).
The scaled real frequencies are quite distinct {for} the different dimensionless charges.
The scaled imaginary frequencies of this family of modes are {approximately} bounded by $-0.09$, a slightly larger value as compared to the gravitational $\mathrm{m}=2$ imaginary frequencies.
{Again}, the slowly-rotating modes for $Q/r_H=0$ agree fairly well with the exact Kerr modes up to $J/M^2=\pm 0.5$.

The scaled real frequency for the gravitational modes of $\mathrm{m}=0$ (Figure \ref{Fig_l2grav_pol}(e)) appears to be less sensitive as compared to the $\mathrm{m}=1,2$ gravitational modes --
the real frequency rises {more slowly with increasing} $J/M^2$.
Figure \ref{Fig_l2grav_pol}(f) shows the scaled imaginary frequency part of the mode.
For dimensionless angular momentum $J/M^2 < 0.2$,
the values of the scaled imaginary frequency are very close to each other for the dimensionless charges smaller than 0.6. 
In our approximation for slow rotation, 
the real frequency modes are in good
agreement with the exact Kerr modes up to $J/M^2=0.6$, while for the imaginary frequencies, they agree up to $J/M^2=0.5$.

Furthermore, we examine the isospectrality of the polar and axial modes. {As seen} in Figure \ref{Fig_l2grav_ISO},
the isospectrality gradually breaks as the value of the dimensionless angular momentum increases.
The departure from isospectrality becomes more significant for the larger dimensionless charges.
The gravitational $\mathrm{m}=2$ modes, in particular the real frequency part of the modes exhibit the smallest deviation in isospectrality, {remaining} under 0.3\% as shown in Figure \ref{Fig_l2grav_ISO}(a). Isospectrality of the imaginary part is satisfied up to a small deviation of $3\%$ as shown in Figure \ref{Fig_l2grav_ISO}(b). For the other values of $\mathrm{m}$, isospectrality deteriorates as the charge increases, where the deviation is always below 20\% for the charges considered.

Regarding the coefficients of the quadratic relations (shown in Tables \ref{m2_grav}, \ref{m1_grav} and \ref{m0_grav} in Appendix \ref{append_coeff}), it is interesting to note that the linear terms are proportional to the angular number $\mathrm{m}$: for $\mathrm{m}=0$ the linear coefficients vanish, while the linear coefficients for $\mathrm{m}=2$ always double the linear coefficients for $\mathrm{m}=1$. In general, we observe a tendency of the coefficients to increase with the value of the charge.

\subsection{Electromagnetic modes}

\begin{figure}[]
		\centering
		\subfloat[]{
		\includegraphics[width=0.37\textwidth,angle=-90]{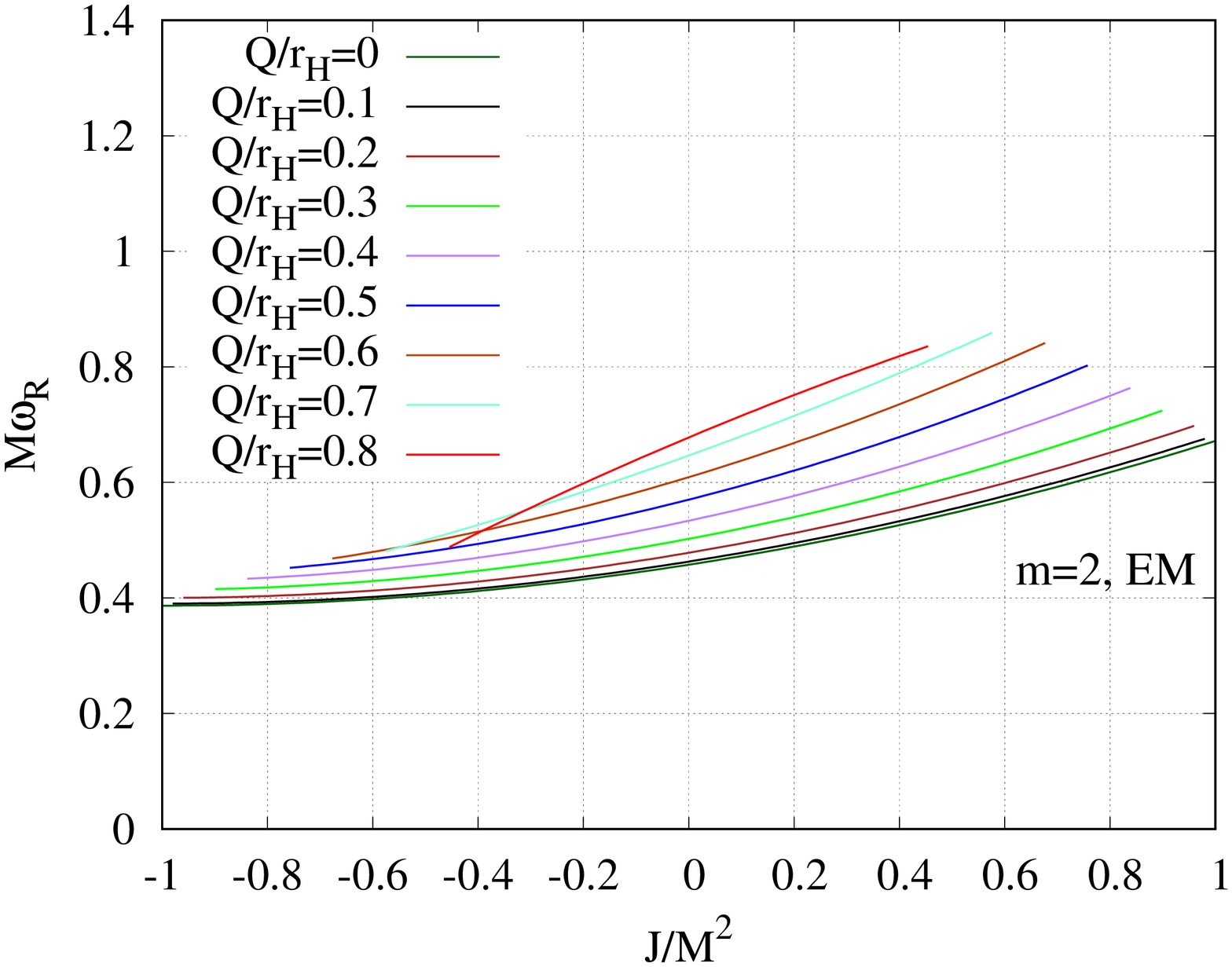}
		}
		\subfloat[]{
		\includegraphics[width=0.37\textwidth,angle=-90]{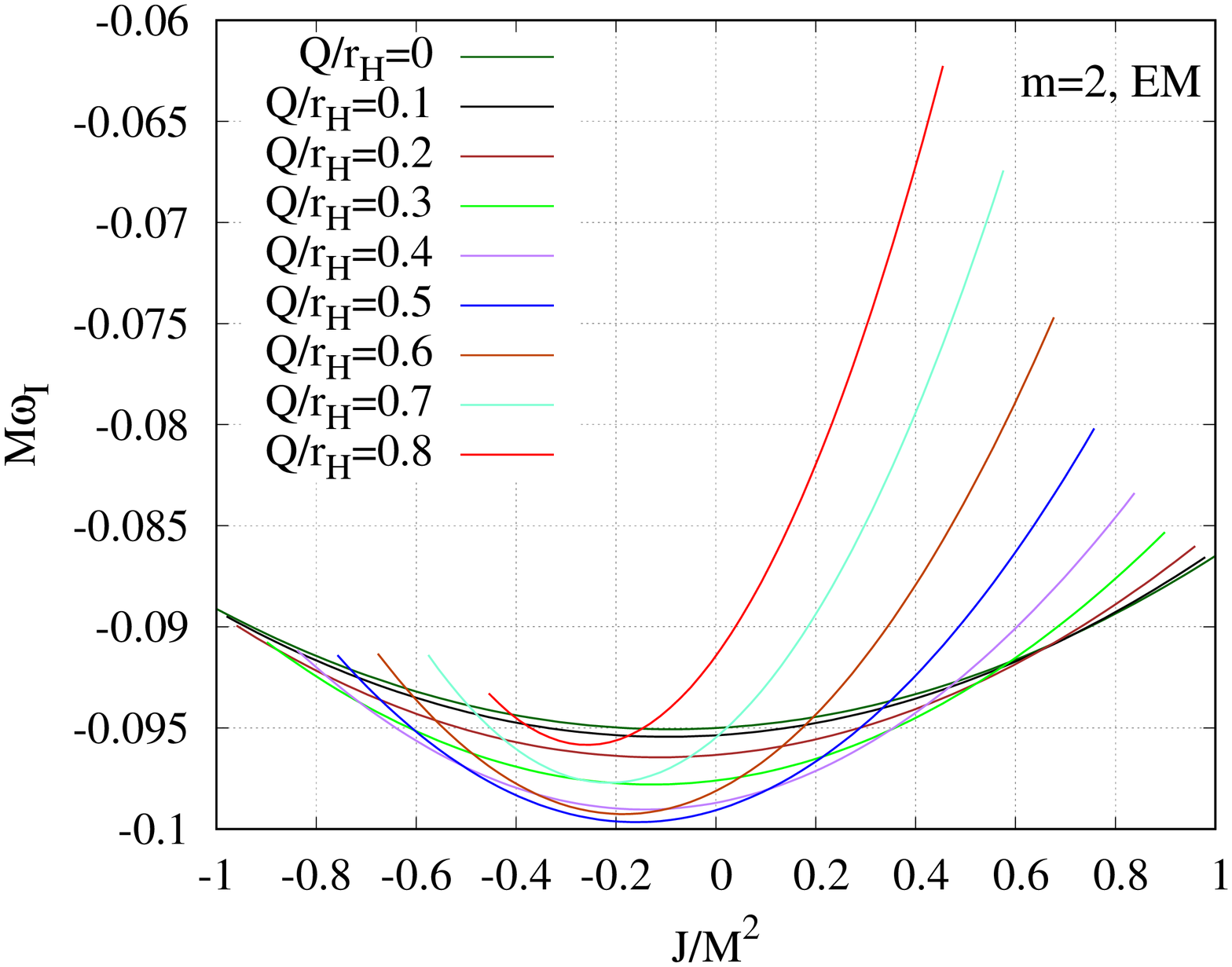}
		} \\
		\subfloat[]{
		\includegraphics[width=0.37\textwidth,angle=-90]{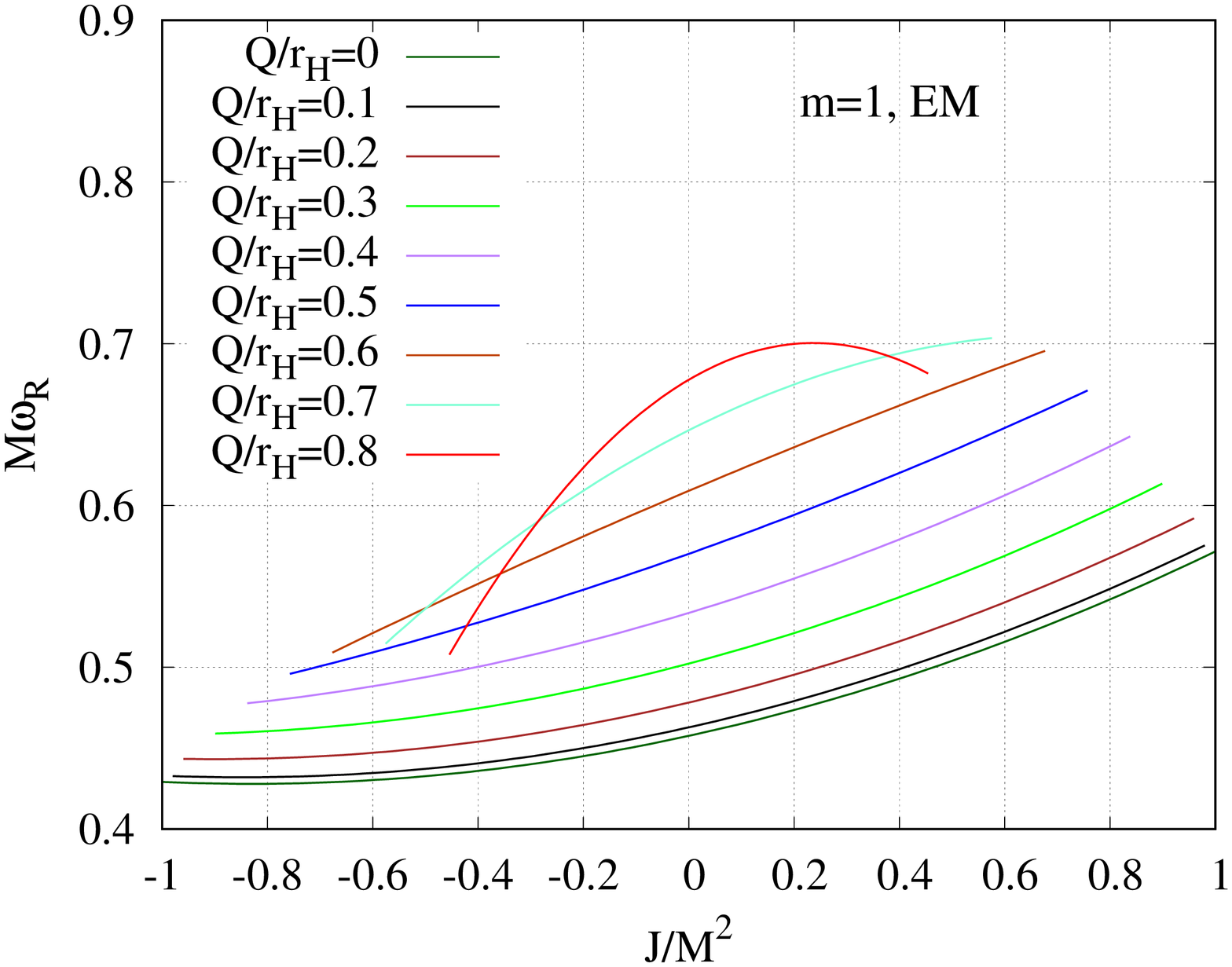}
		}
		\subfloat[]{
		\includegraphics[width=0.37\textwidth,angle=-90]{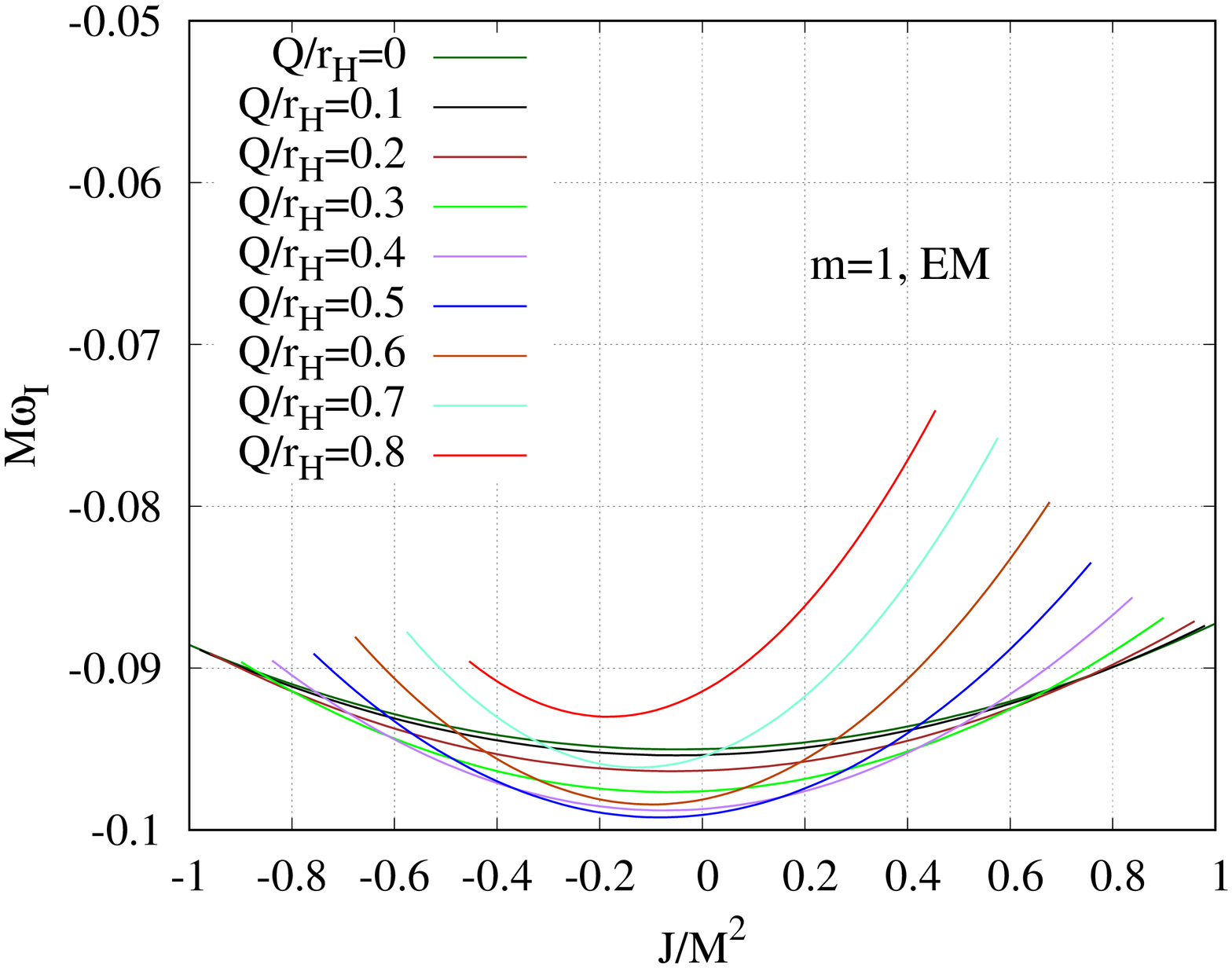}
		} \\
		\subfloat[]{
		\includegraphics[width=0.37\textwidth,angle=-90]{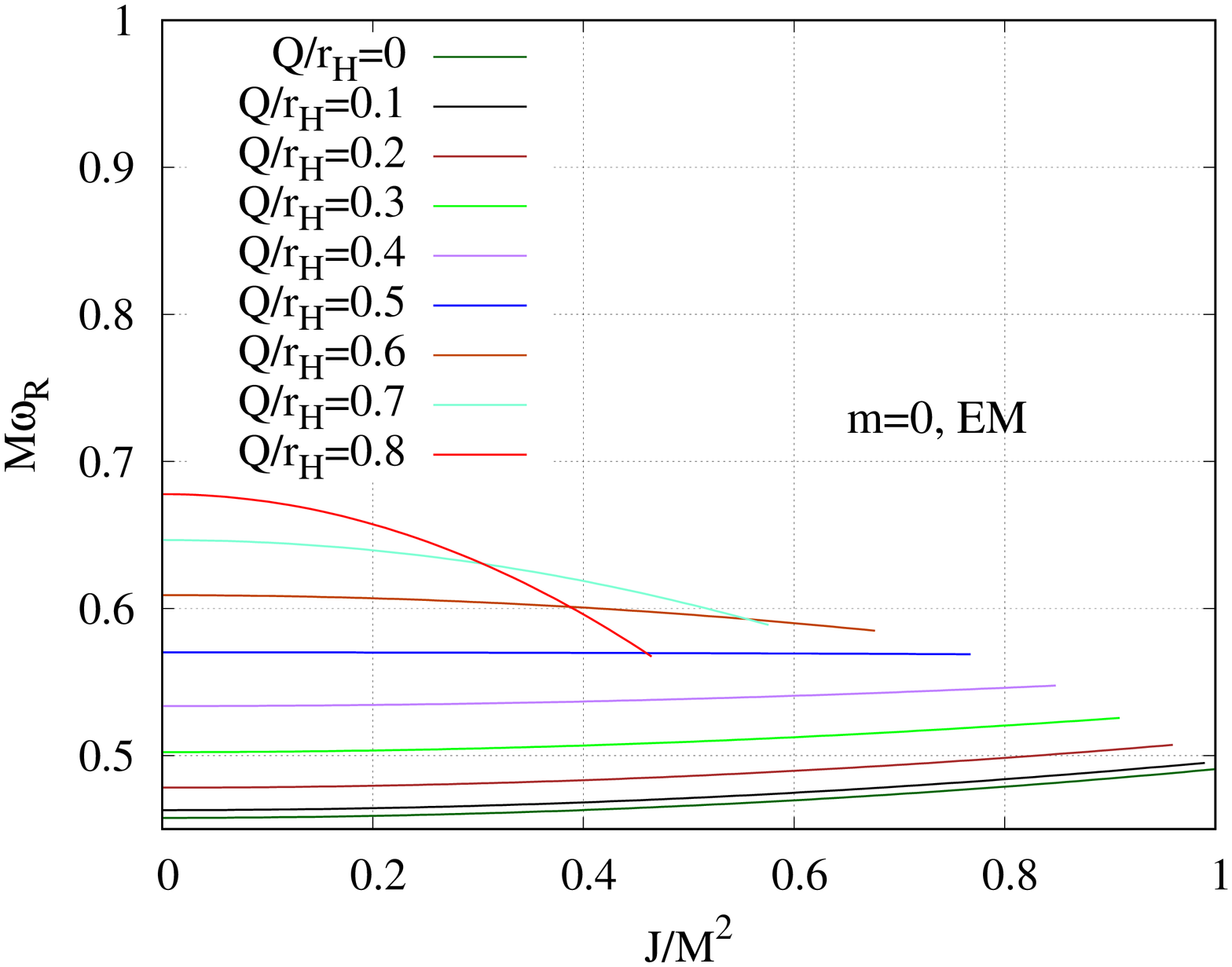}
		}
		\subfloat[]{
		\includegraphics[width=0.37\textwidth,angle=-90]{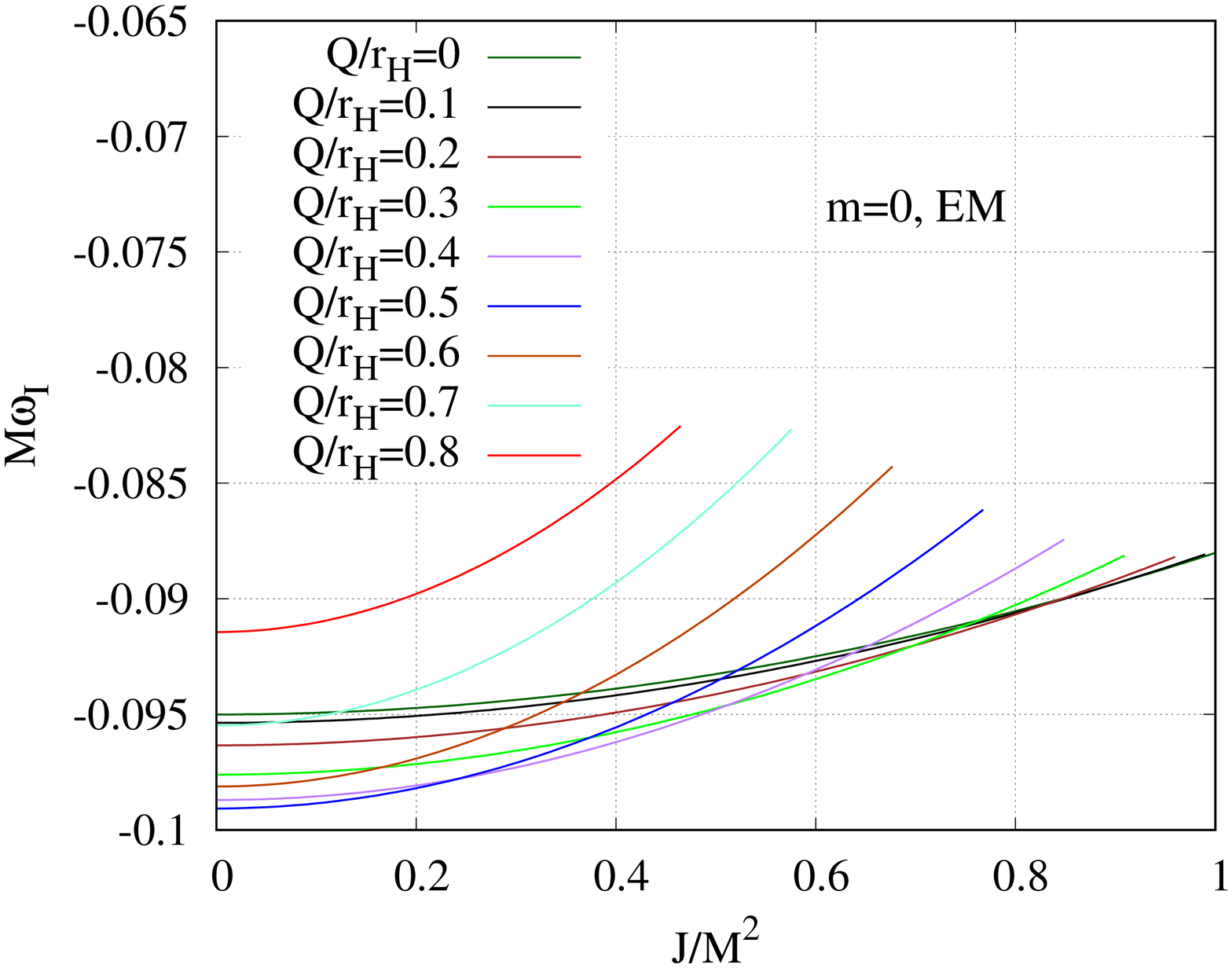}
		}
		\caption{Electromagnetic $\mathrm{l} = 2, \mathrm{m}=2,1,0$ quasinormal modes: Real frequency $\omega_R$ (a),(c),(e) and imaginary frequency $\omega_I$ (b),(d),(f) scaled with mass $M$ versus angular momentum $J$ scaled with inverse mass-squared. Different values of charge $Q$ divided by the Schwarzschild radius $r_H$ are in different colors, with the exact Kerr solution in dotted grey. }
		\label{Fig_l2EM_pol}
	\end{figure}	

\begin{figure}[]
		\centering
		\subfloat[]{
		\includegraphics[width=0.37\textwidth,angle=-90]{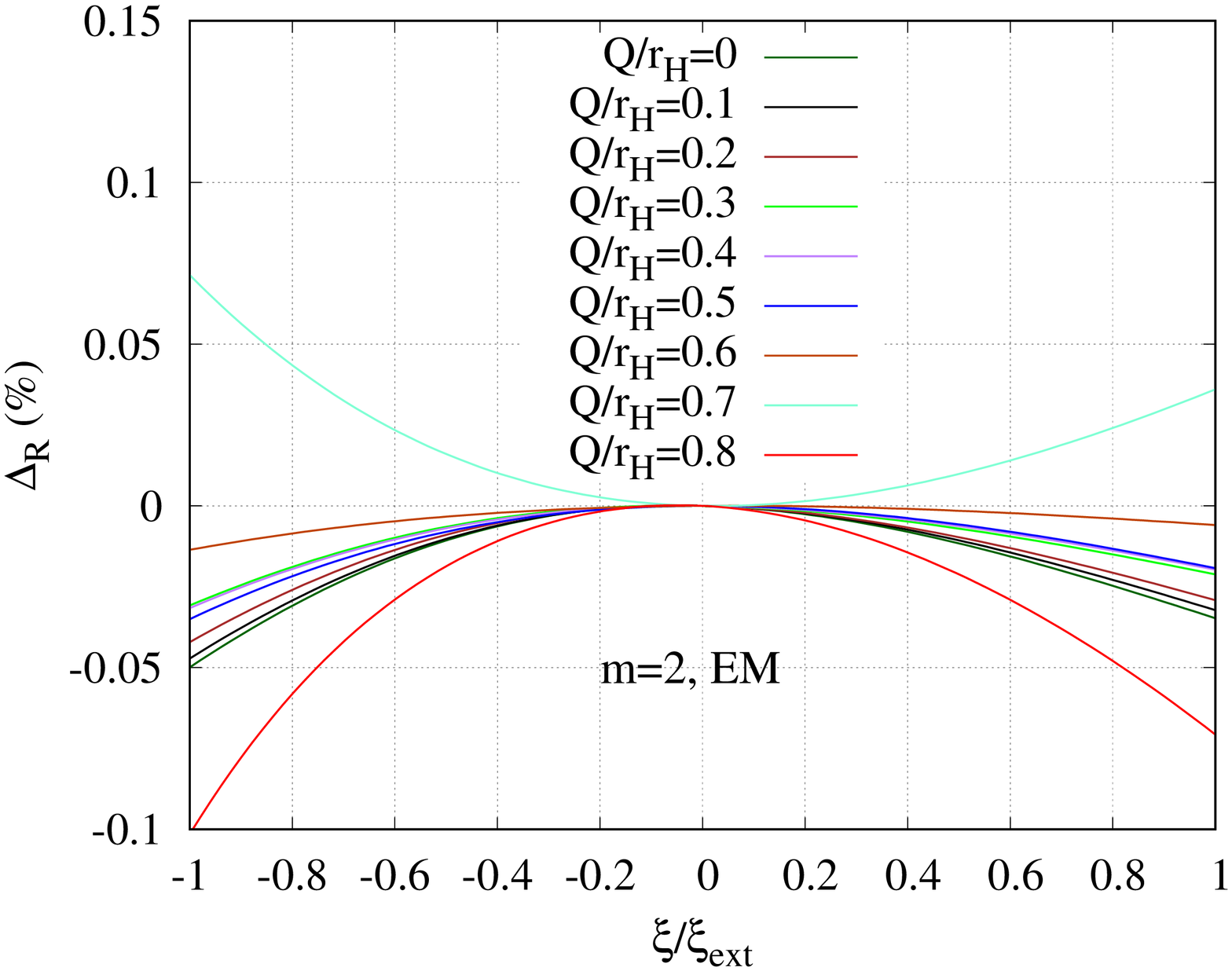}
		}
		\subfloat[]{
		\includegraphics[width=0.37\textwidth,angle=-90]{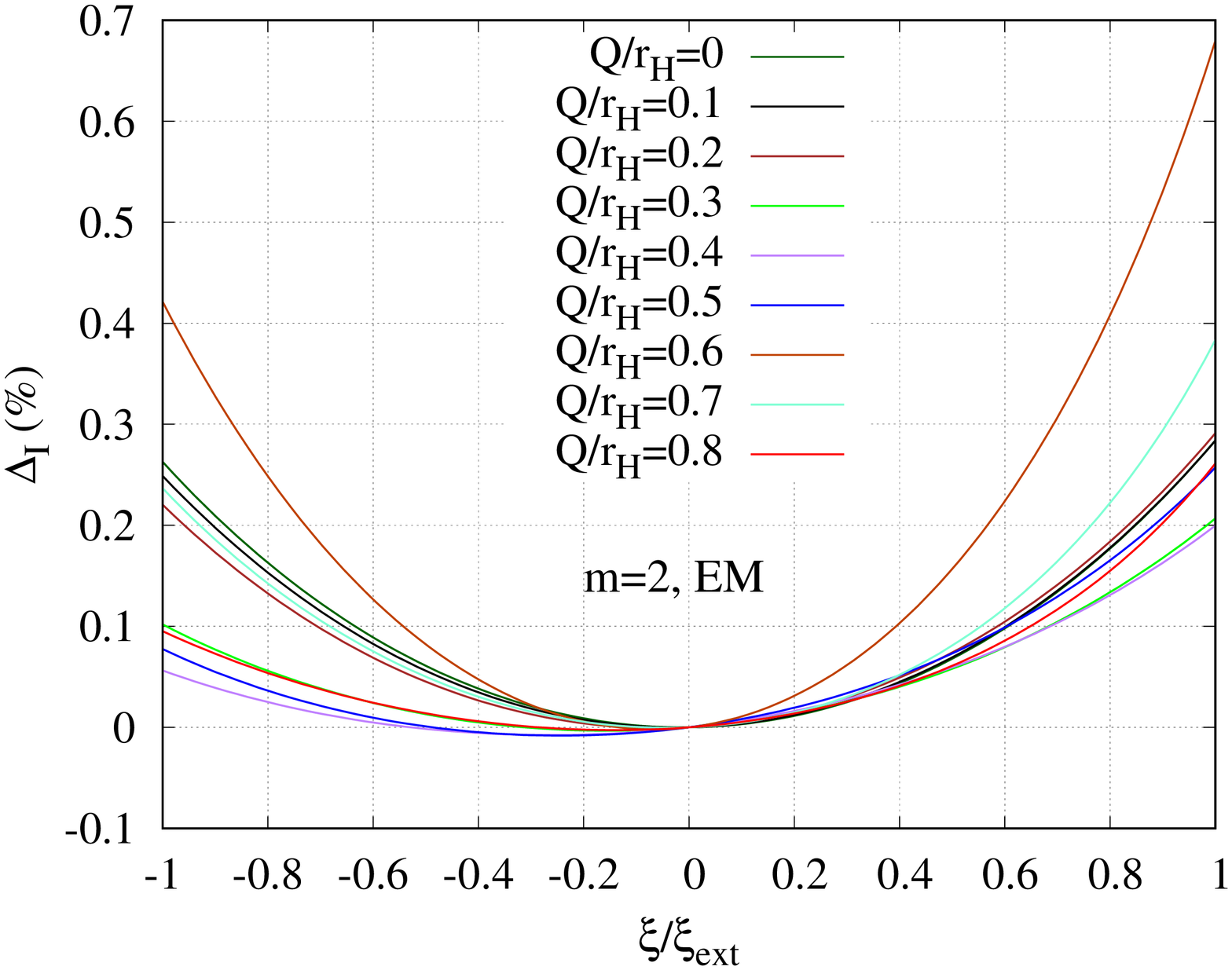}
		}\\
		\subfloat[]{
		\includegraphics[width=0.37\textwidth,angle=-90]{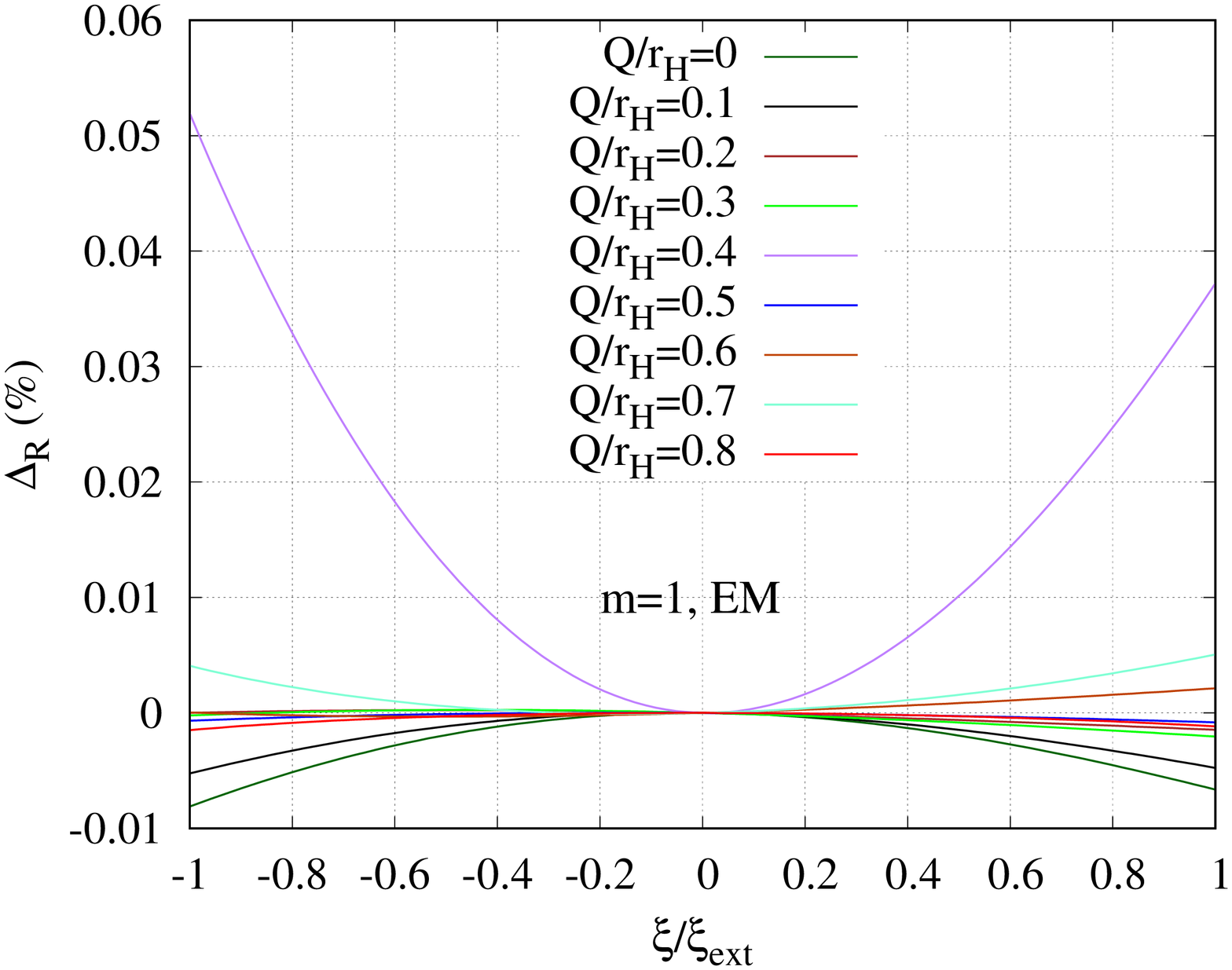}
		}
		\subfloat[]{
		\includegraphics[width=0.37\textwidth,angle=-90]{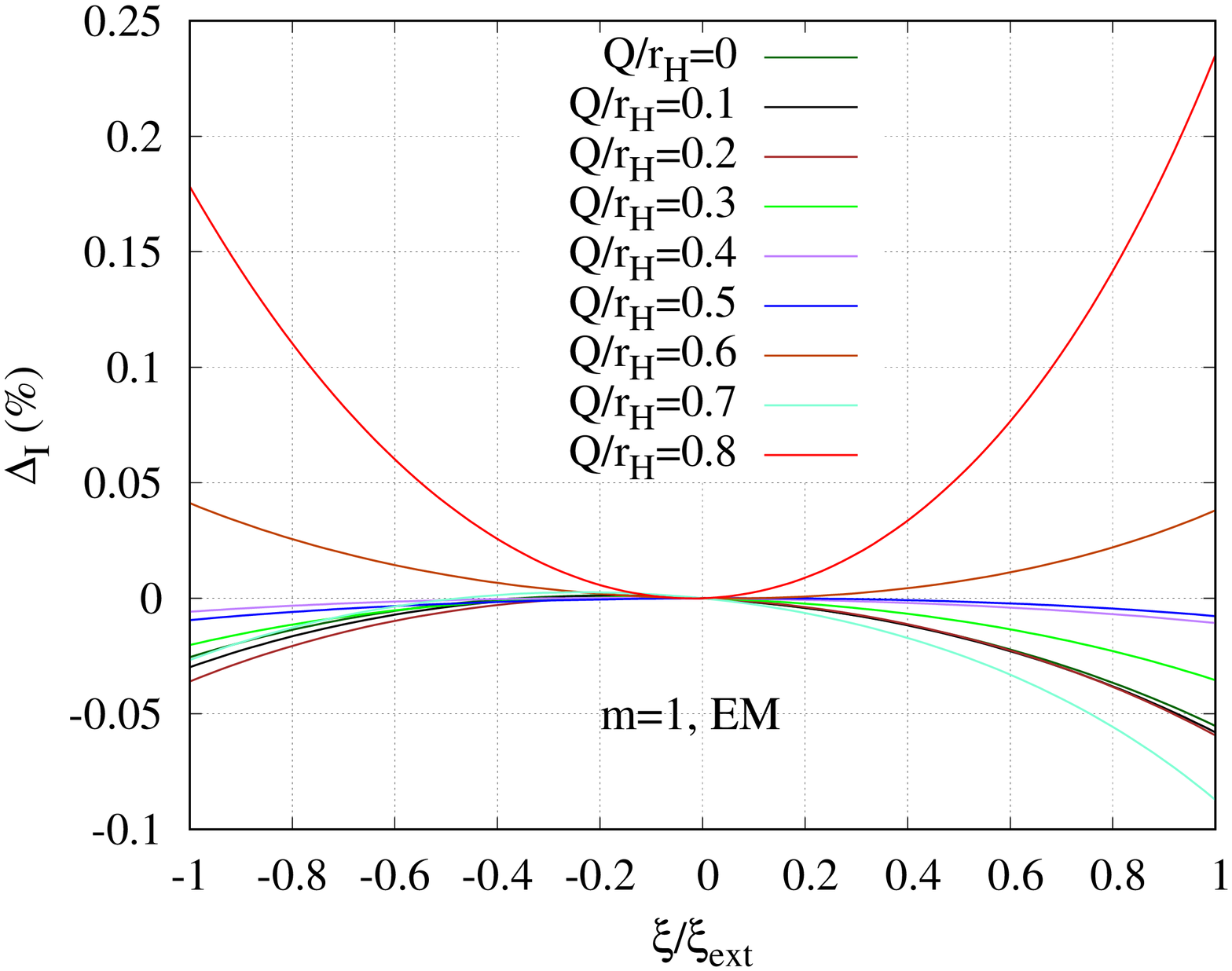}
		} \\
		\subfloat[]{
		\includegraphics[width=0.37\textwidth,angle=-90]{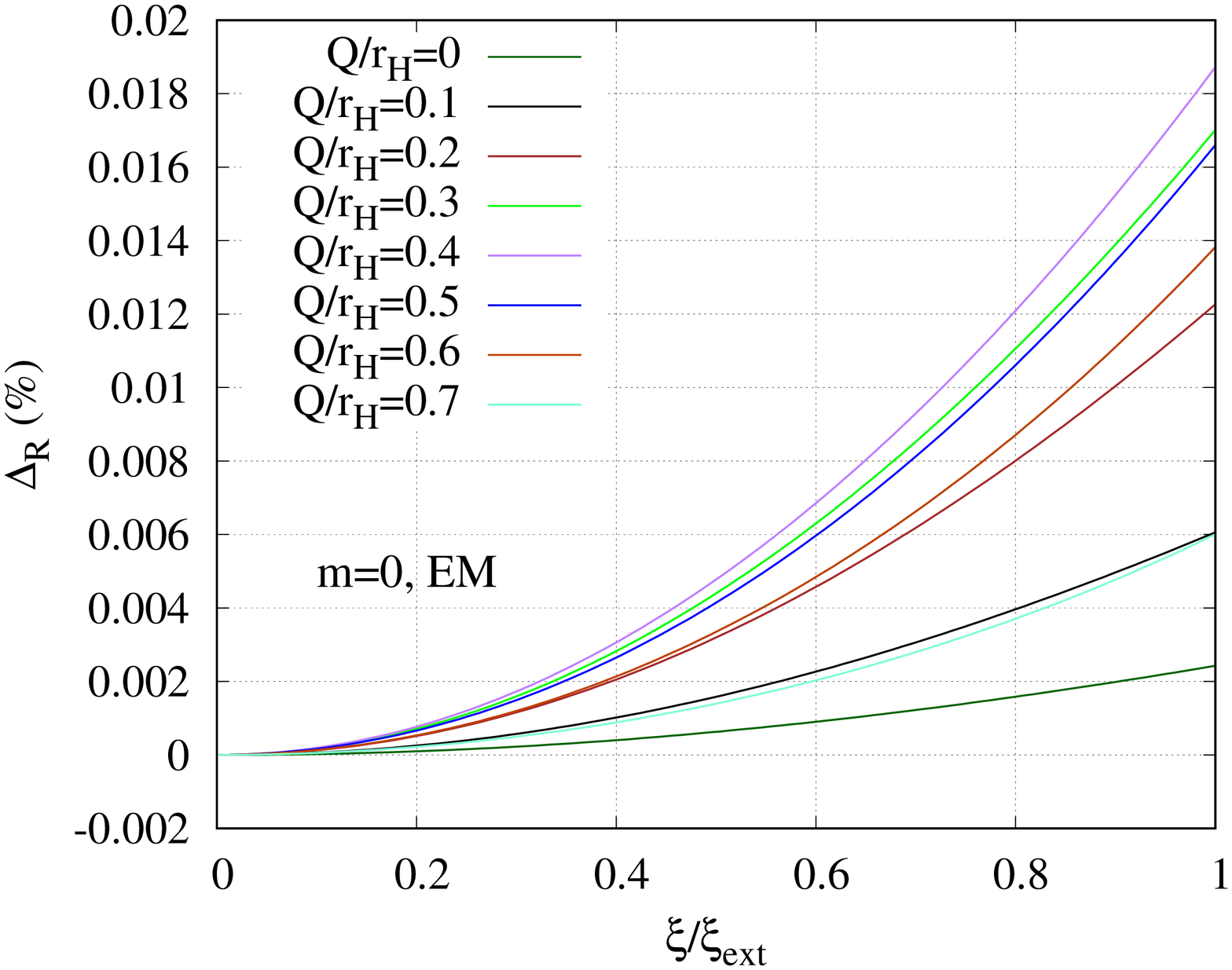}
		}
		\subfloat[]{
		\includegraphics[width=0.37\textwidth,angle=-90]{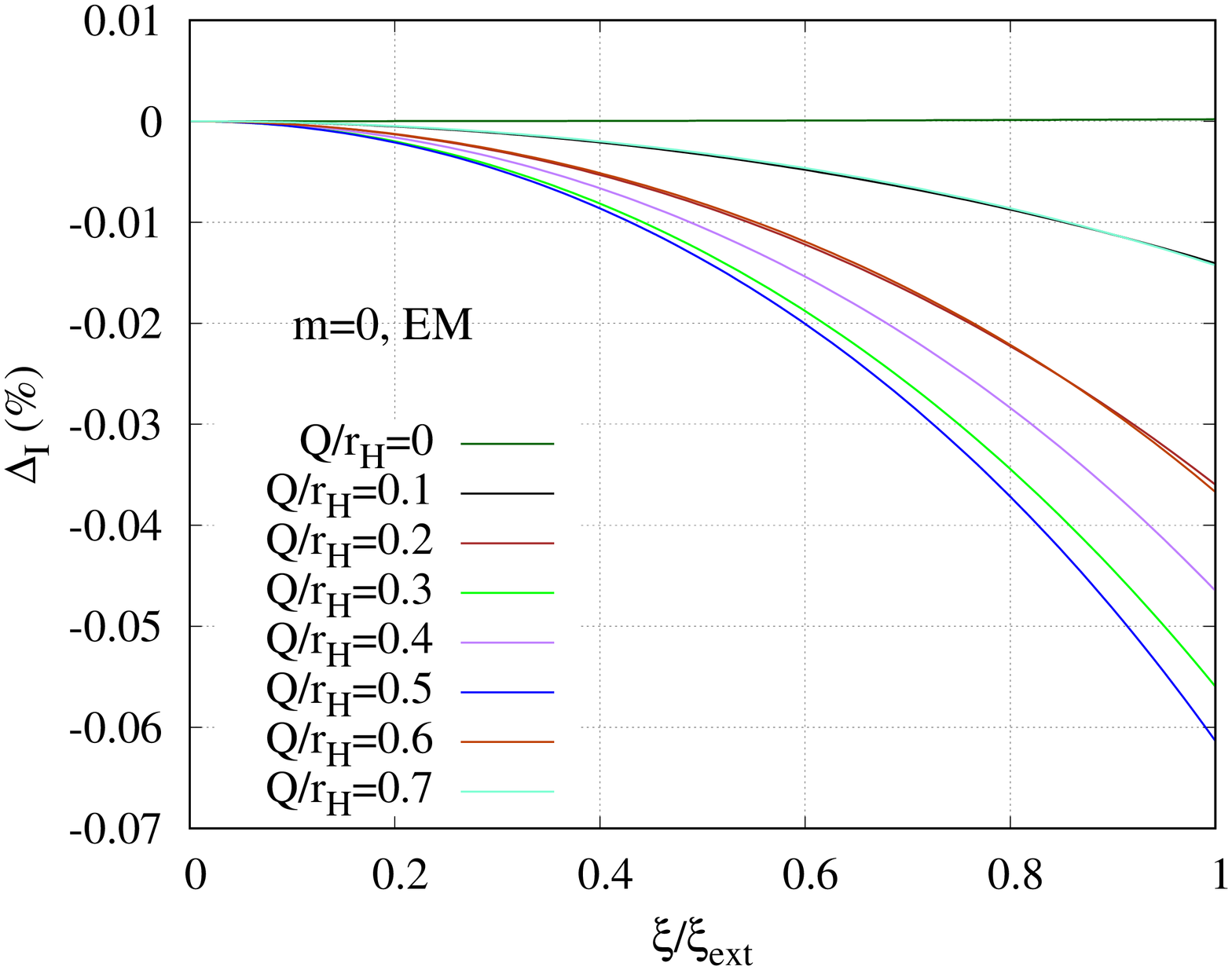}
		}
		\caption{Isospectrality for electromagnetic $\mathrm{l}= 2, \mathrm{m}=2,1,0$ quasinormal modes: 
  Percentage of difference in the real frequency $\omega_R$ (a),(c),(e) and imaginary frequency $\omega_I$ (b),(d),(f) versus the dimensionless angular momentum $\xi$ scaled with its extremal value. Different values of charge $Q$ divided by the Schwarzschild radius $r_H$ are in different colors. }
		\label{Fig_l2EM_ISO}
	\end{figure}

We present in the following the results for electromagnetic quasinormal modes for $\mathrm{m}=2,1,0$.
The corresponding coefficients for the quadratic description of the modes are given in Tables \ref{m2_em}, \ref{m1_em} and \ref{m0_em} in Appendix \ref{append_coeff}.

Following the same style as in the previous section, we show in Figure \ref{Fig_l2EM_pol} the scaled values of the quasinormal modes as a function of the scaled angular momentum, on the left for the real part and on the right for the imaginary part. In contrast to the gravitational modes, the quadratic coefficient of the real part does not increase monotonically with the charge. This means that for large enough charge the shape of the curves in Figures \ref{Fig_l2EM_pol} a), c) and e) change from convex to concave. As for the imaginary part, Figures \ref{Fig_l2EM_pol} b), d) and f)  show 
a similar general behavior as the gravitational modes. It is again the case that the linear coefficient is proportional to $\mathrm{m}$, and increases monotonically with the charge in both the real and imaginary parts.

We show the isospectrality of the electromagnetic modes in Figure \ref{Fig_l2EM_ISO}, respectively for $\mathrm{m}=2,1,0$. 
The isospectrality of the electromagnetic modes across all the considered $\mathrm{m}$ values fares better than the isospectrality of the gravitational modes, with the largest deviation of only 0.7\% coming from the imaginary frequencies of $\mathrm{m}=2$. In general, the deviation increases as we increase the scaled angular momenta.

\subsection{Scalar modes}

\begin{figure}[]
		\centering
		\subfloat[]{
		\includegraphics[width=0.37\textwidth,angle=-90]{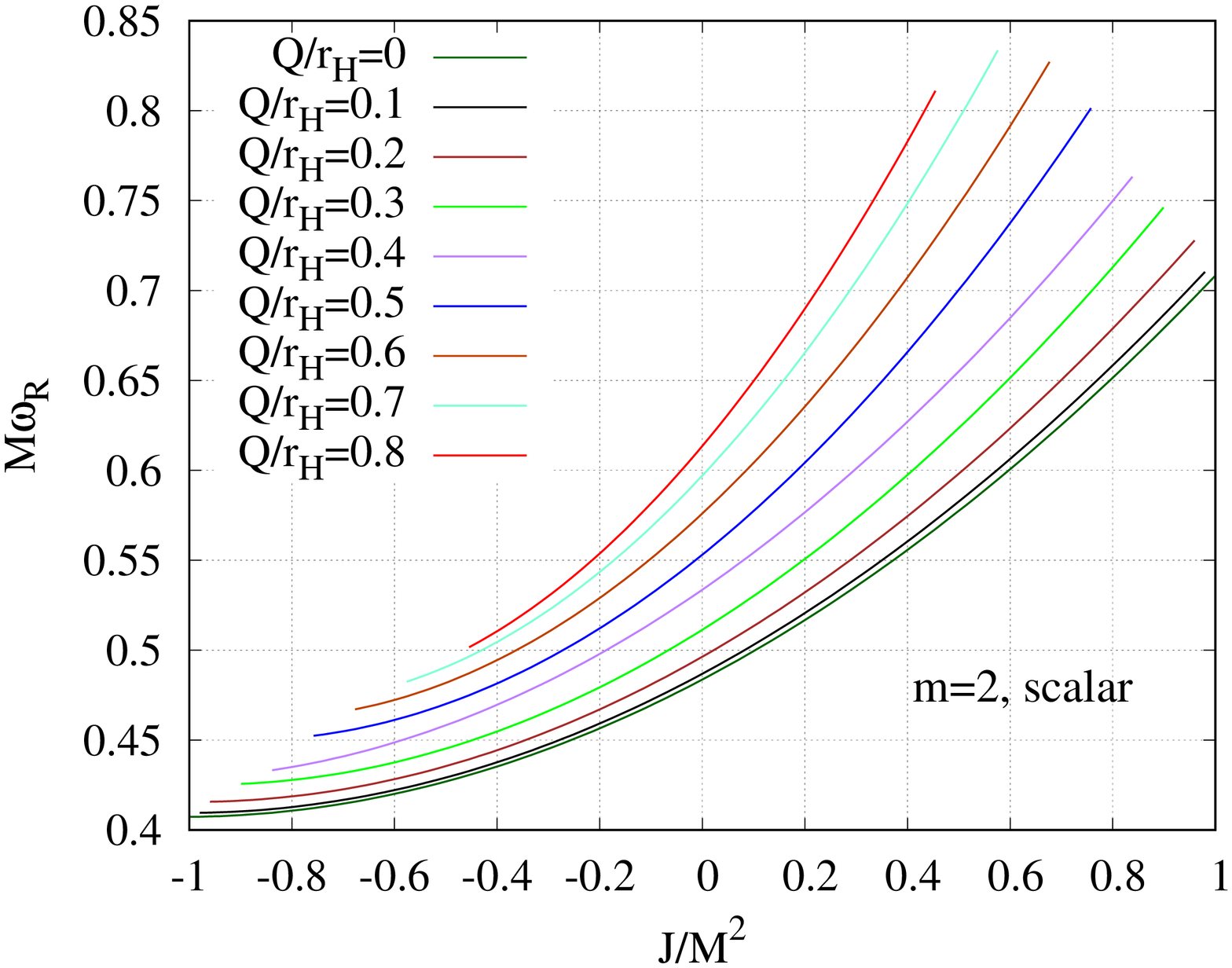}
		}
		\subfloat[]{
		\includegraphics[width=0.37\textwidth,angle=-90]{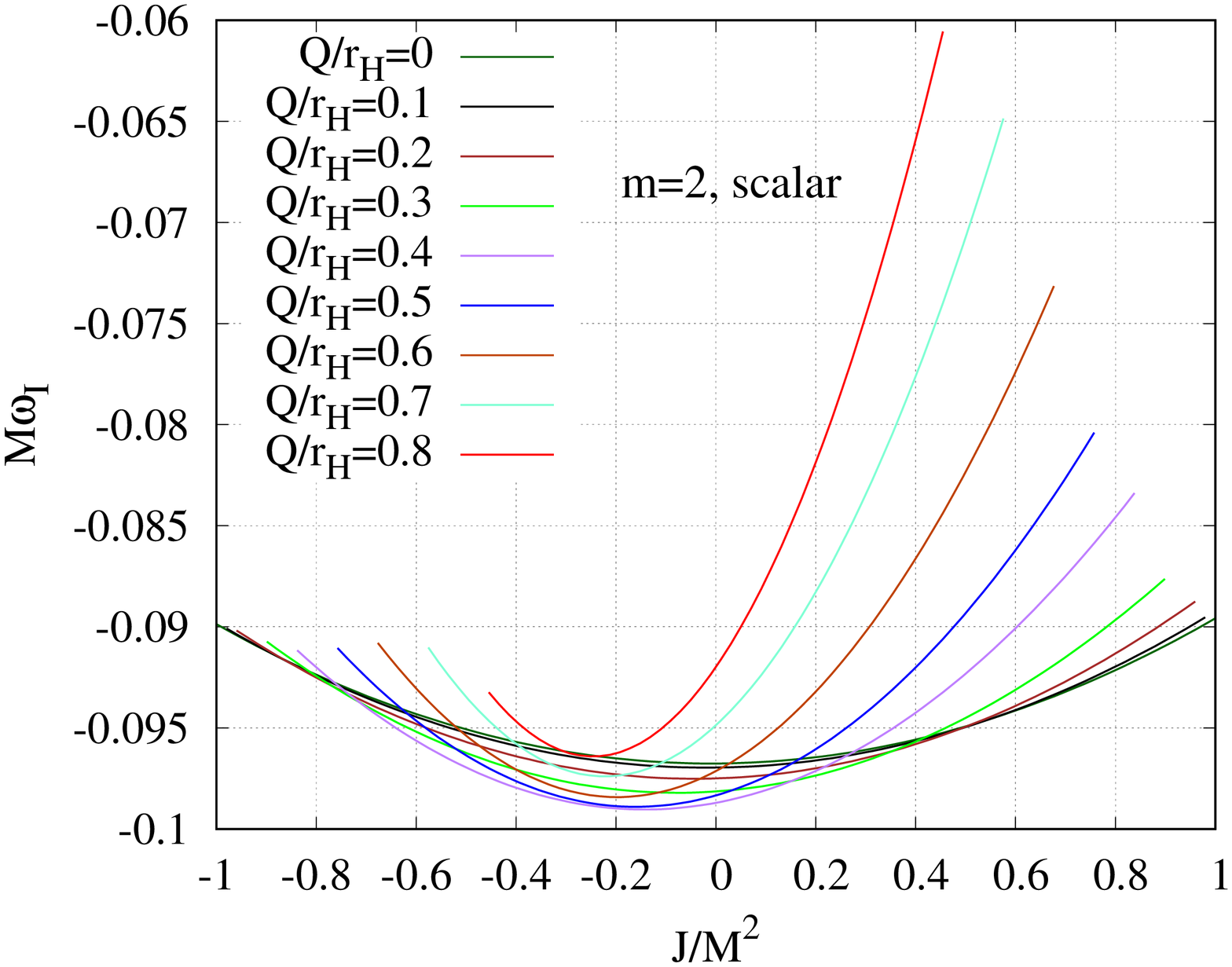}
		} \\
		\subfloat[]{
		\includegraphics[width=0.37\textwidth,angle=-90]{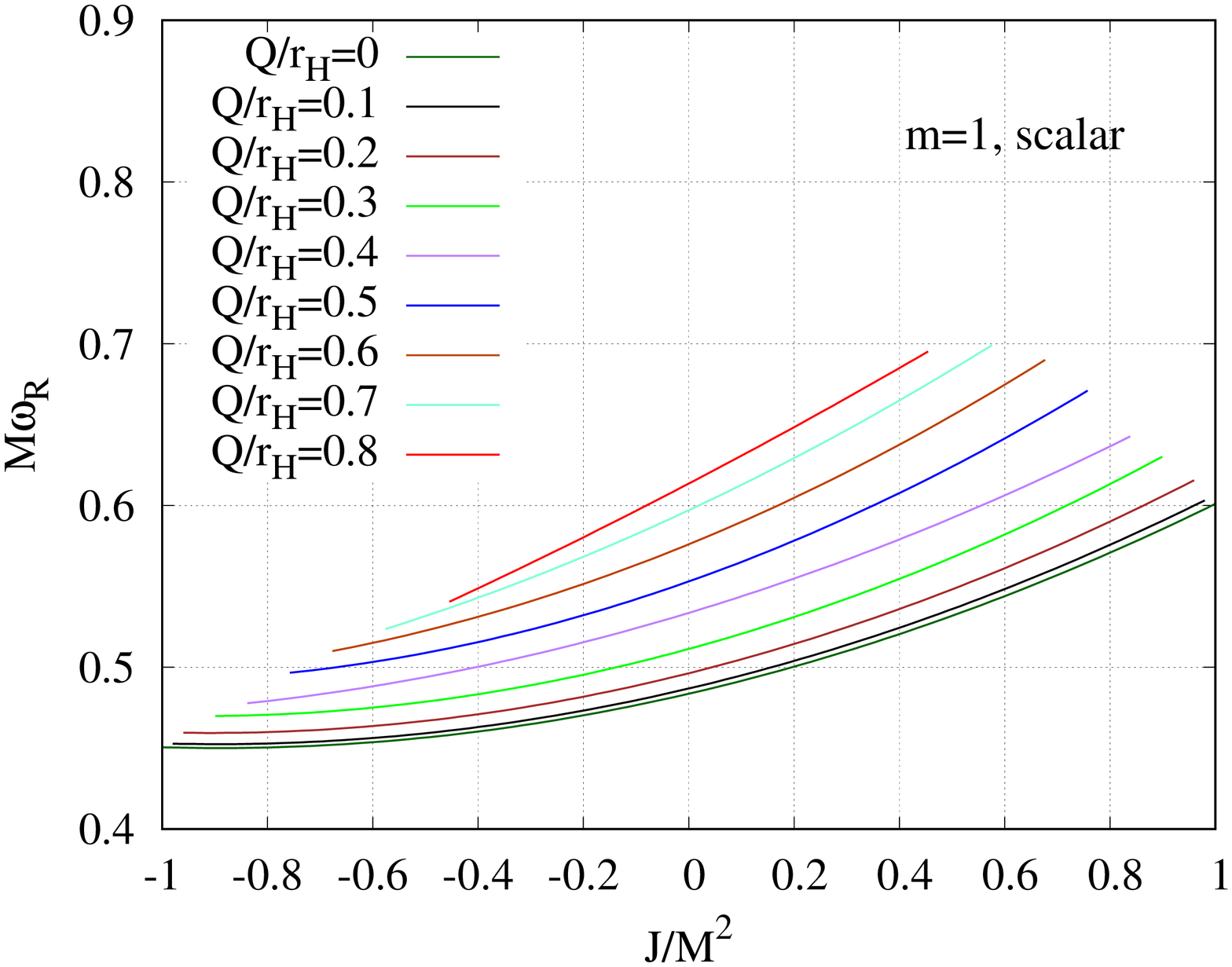}
		}
		\subfloat[]{
		\includegraphics[width=0.37\textwidth,angle=-90]{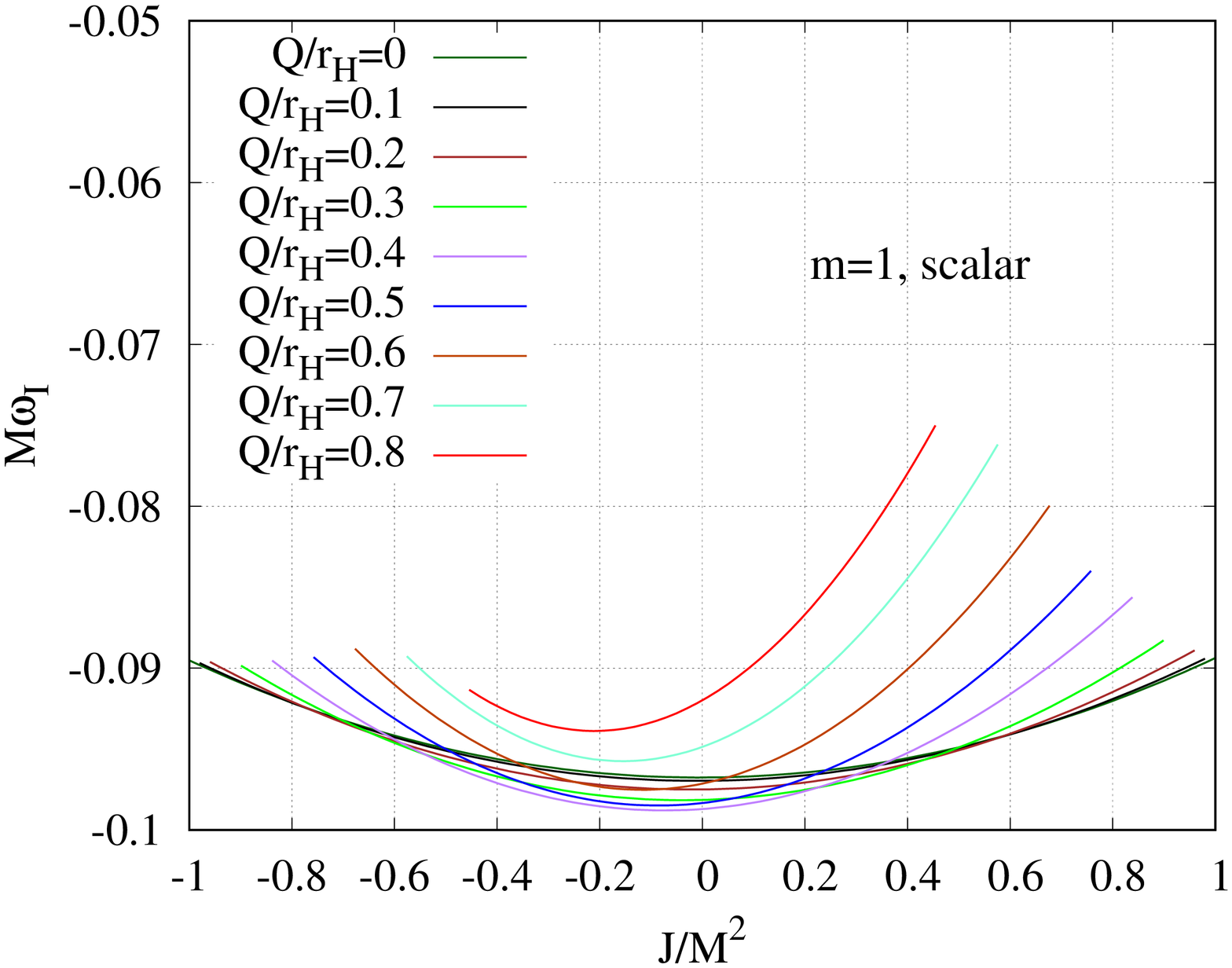}
		} \\
		\subfloat[]{
		\includegraphics[width=0.37\textwidth,angle=-90]{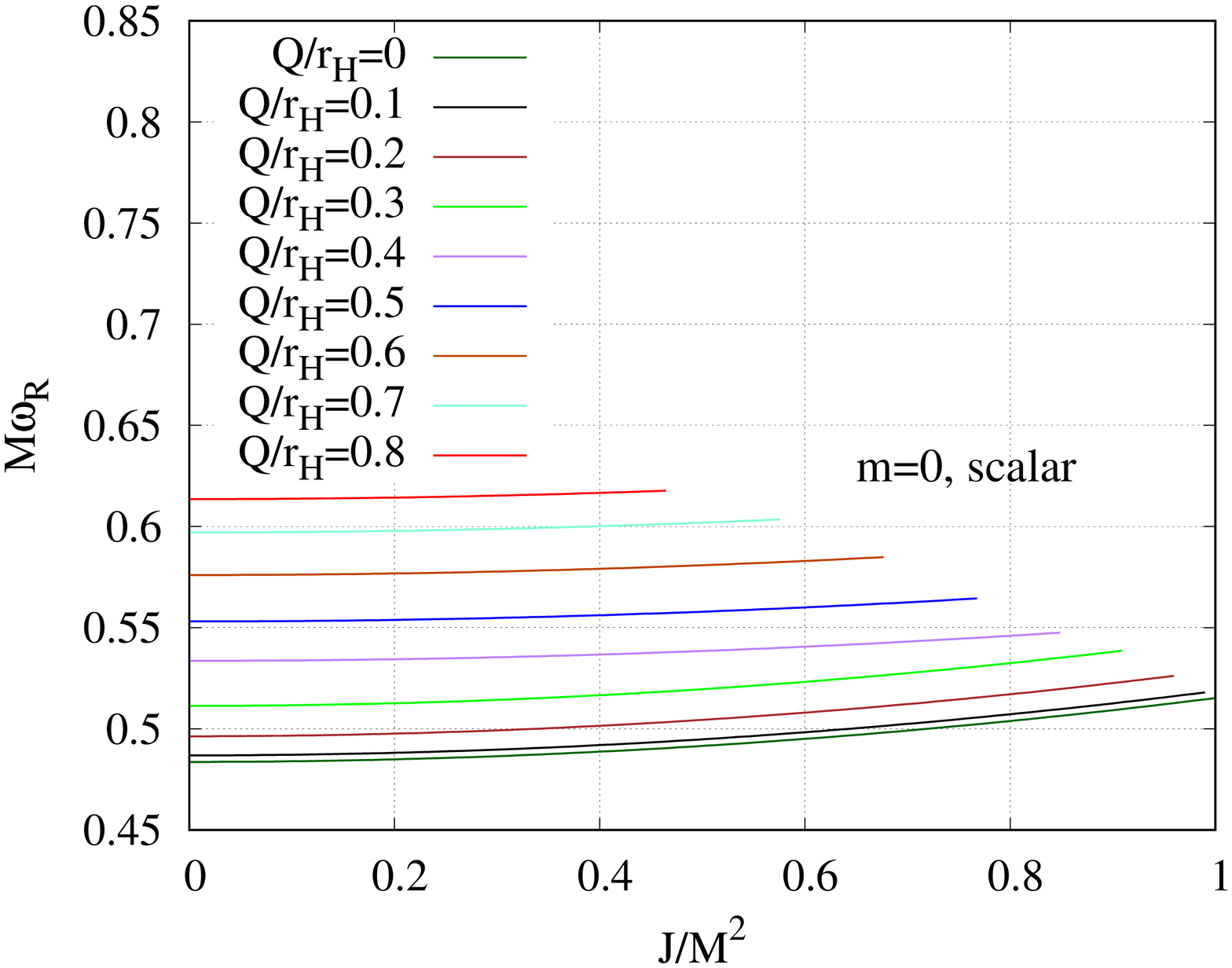}
		}
		\subfloat[]{
		\includegraphics[width=0.37\textwidth,angle=-90]{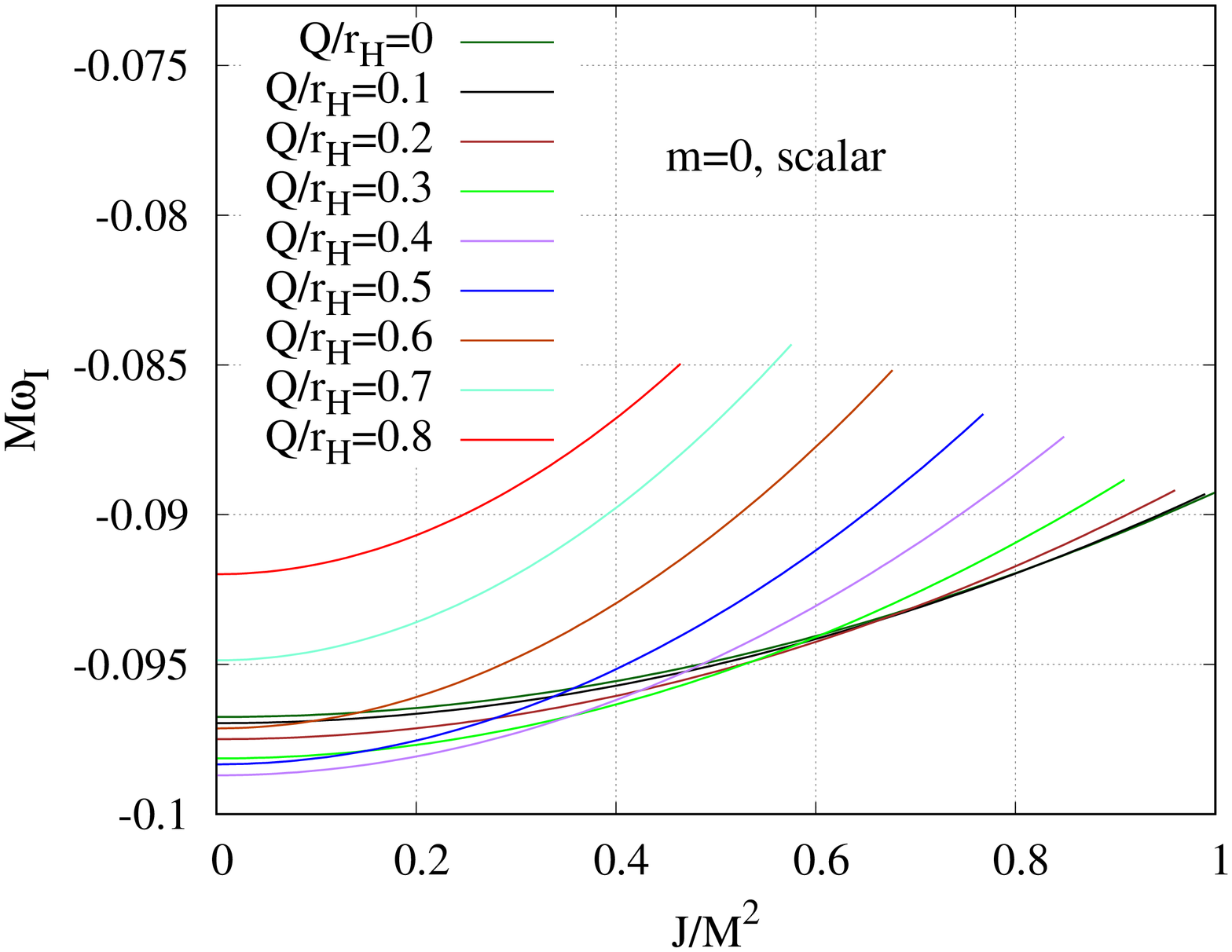}
		}
		\caption{Scalar $\mathrm{l} = 2, \mathrm{m}=2,1,0$ quasinormal modes: Real frequency $\omega_R$ (a),(c),(e) and imaginary frequency $\omega_I$ (b),(d),(f) scaled with mass $M$ versus angular momentum $J$ scaled with inverse mass-squared. Different values of charge $Q$ divided by the Schwarzschild radius $r_H$ are in different colors, with the exact Kerr solution in dotted grey. }
	\label{Fig_l2scalar_pol}
	\end{figure}

This section details the results for scalar quasinormal modes for $\mathrm{m}=2,1,0$.
The corresponding coefficients for the quadratic fit of the modes are given in Tables \ref{m2_sc}, \ref{m1_sc} and \ref{m0_sc} in Appendix \ref{append_coeff}.

The scalar modes in Figure \ref{Fig_l2scalar_pol}
for the real frequency part (a)
and for the imaginary frequency part (b) for $\mathrm{m}=2$
exhibit a similar dependence on the scaled angular momentum $J/M^{2}$ as the electromagnetic family of modes $\mathrm{m}=2$,
except for $Q/r_H=0.8$. Here the real frequencies {for $Q/r_H=0.8$} 
behave like the {real frequencies} for the other charges.

For the scalar modes of $\mathrm{m}=1$, shown in Figure \ref{Fig_l2scalar_pol}(c), the scaled real frequency for $Q/r_H = 0.8$ exhibits a rather linear behavior with respect to $J/M^2$, in contrast to the quadratic behavior of the smaller scaled charges. 
In Figure \ref{Fig_l2scalar_pol}(d) a slight increase in the minimum of the scaled imaginary frequency is noticeable for the lower charges.

Figure \ref{Fig_l2scalar_pol}(e) and (f) show the results for the scalar modes of $m=0$.
The scaled real frequency for the scalar modes presents a similar pattern as the electromagnetic modes of $\mathrm{m}=0$.
Each dimensionless charge $Q/r_H$ takes a {distinct} scaled frequency value at each $J/M^2$.
{The scaled imaginary frequency (Figure \ref{Fig_l2scalar_pol}(f)) on the other hand is comparable with the scaled imaginary frequency of the $\mathrm{m}=1,2$ scalar mode, although the rise of the curve as the angular momentum increases is slower the smaller the value of $\mathrm{m}$ is.}

In general, in all the three types of modes,
for large values of the charge $Q$, the associated QNMs
tend to be longer-lived, especially as we increase the angular momentum for the $\mathrm{l}=\mathrm{m}$ co-rotating case. Such an effect was pointed out in \cite{Glampedakis:2001js} for the QNMs of Kerr, arguing that the $\mathrm{l}=\mathrm{m}$ co-rotating waves are subjected to maximal inertial dragging effects, hence the longer damping times. On the other hand, let us also note that in
\cite{Hod:2008se}, it was shown that the damping time is inversely related to the temperature of the near-extremal
black hole, suggesting that generically one would expect the modes to be longer-lived as we increase the background charges towards extremality.

\section{Conclusions}

In this paper we have calculated the spectrum of quasinormal modes of slowly rotating Kerr-Newman black holes for $\mathrm{l} = 2$ perturbations, 
extending the previous works in slow-rotation limit to second order.
We have used a perturbative double expansion method. The background Kerr-Newman metric has been approximated in rotation up to second order, employing a Schwarzschild-like coordinate. 
{Based on} this metric, we have introduced first order non-radial perturbations. By 
systematically
simplifying the field equations, we have obtained a system of linear homogeneous equations that describe polar-led and axial-led perturbations {for} a slowly rotating charged black hole. 
After obtaining the approximate wave-like solutions to this system of equations near the horizon and in the far-field region, we have calculated numerically the
gravitational, electromagnetic and {(minimally coupled)} scalar quasinormal modes of the Kerr-Newman black holes. 

We have reproduced
the previous results, confirming that our double-expansion method correctly estimates with very good precision the {(slow rotation)} spectrum of quasinormal modes. 
In particular, for the most interesting family of modes, the gravitational modes, we have seen that the frequency only deviates up to $10\%$ for configurations close to $80\%$ of the extremal angular momentum. 
Meanwhile, the damping time is more sensitive to the approximation, but nonetheless can be accurately calculated within $10\%$ of deviation for configurations close to $60\%$ of the extremal angular momentum.
We have also checked the isospectrality of the modes for the Kerr-Newman spacetime. {Here} we have found that the spectra of axial- and polar-led perturbations are approximately equal with high accuracy for slowly rotating configurations.

Calculations of quasinormal modes {of rotating black holes} in alternative gravity theories as, for instance, in Einstein-Maxwell-dilaton theories \cite{Gibbons:1987ps,Garfinkle:1990qj,Kleihaus:2003sh} are left for future work. 
{In the case of Einstein-Maxwell-scalar theories the present approach will only need the additional inclusion of the respective appropriate (non-minimal) coupling functions in order to obtain the corresponding spectra.}
{Likewise, the presented formalism will allow us to obtain the quasinormal mode spectra of rotating black holes in Einstein-scalar-Gauss-Bonnet theories with various types of coupling functions that lead to interesting sets of rotating black holes that might be of dilatonic type \cite{Kanti:1995vq,Kleihaus:2011tg} (see \cite{Pierini:2021jxd,Pierini:2022eim}) or of spontaneously scalarized type \cite{Doneva:2017bvd,Silva:2017uqg,Antoniou:2017acq,Cunha:2019dwb,Collodel:2019kkx}.
When, however, slowly rotating solutions do not exist as, for instance, for the spin-induced rotating Einstein-scalar-Gauss-Bonnet black holes \cite{Dima:2020yac,Herdeiro:2020wei,Berti:2020kgk},  other methods will be needed that are more appropriate for the rapidly rotating case (see e.g. \cite{Dias:2015wqa,Dias:2015nua,Carullo:2021oxn,Dias:2021yju, Dias:2022oqm} and \cite{Li:2022pcy}).
}

\section{Acknowledgements}

We would like to gratefully acknowledge support by the DFG Research Training Group 1620 \textit{Models of Gravity}, 
DFG project Ku612/18-1, 
FCT project PTDC/FIS-AST/3041/2020, COST Actions CA15117 and CA16104, and MICINN project PID2021-125617NB-I00 ``QuasiMode".
JLBS gratefully acknowledges support from Santander-UCM project PR44/21‐29910.
We are grateful to Jutta Kunz, Luis Manuel Gonz\'alez-Romero and Francisco Navarro-L\'erida for discussions and comments on the manuscript.
We thank \'Oscar J. C. Dias, Mahdi Godazgar and Jorge E. Santos for providing data.
Part of the computations were performed on the HPC Cluster CARL funded by the DFG under INST 184/157-1 FUGG.

\newpage

\appendix

\section{Coefficients of the quadratic fit}
\label{append_coeff}

The quadratic expression for the quasinormal modes is given by
\begin{equation} 
M\omega_{R,I} = M\omega^{(0)}_{R,I} + M\delta\omega^{(1)}_{R,I} \, \left(\frac{J}{M^2}\right) + M \delta\omega^{(2)}_{R,I} \, \left(\frac{J}{M^2}\right)^2 \,.
\label{quad_fit}
\end{equation}
The coefficients for the modes studied in this paper are given in the following tables.

\begin{table}[h!]
\begin{center}
\begin{tabular}{ || c | c c c | c c c || }
 \hline
$Q/r_H$ & $M\omega^{(0)}_R$ & $M\delta\omega^{(1)}_R$ & $M\delta\omega^{(2)}_R$ &
$M\omega^{(0)}_I$ & $M\delta\omega^{(1)}_I$ & $M\delta\omega^{(2)}_I$ 
\\ 
  \hline
0 & 0.37367 &  0.12576 & 0.07153 
& -0.08896 & 0.00202 & 0.00511
\\  
0.1 & 0.37473 & 0.12844 & 0.07417  
& -0.08907 & 0.00207 & 0.00537
\\  
0.2 & 0.37804 &  0.13650 & 0.08235  
& -0.08937 & 0.00235 & 0.00623
\\  
0.3 & 0.38379 & 0.15000 & 0.09700 
& -0.08973 & 0.00317 & 0.00805
\\  
0.4  & 0.39175 & 0.16905 & 0.11992  
& -0.08991 & 0.00508 & 0.01162
\\  
0.5 & 0.40121  & 0.19339 & 0.15375  
& -0.08965 & 0.00903 & 0.01867
\\  
0.6  & 0.41104 & 0.22166 & 0.20073  
& -0.08871 & 0.01637 & 0.03317
\\  
0.7  & 0.41987 & 0.25023 & 0.25769 
& -0.08709 & 0.02830 & 0.06286
\\  
0.8  & 0.42646 & 0.27292 & 0.30466  
& -0.08526 & 0.04336 & 0.11203
\\  
 \hline
\end{tabular}
\end{center}
\caption{$\mathrm{m}=2$ gravitational modes}
\label{m2_grav}
\end{table}

\begin{table}[h!]
\begin{center}
\begin{tabular}{ || c | c c c | c c c || }
 \hline
$Q/r_H$ & $M\omega^{(0)}_R$ & $M\delta\omega^{(1)}_R$ & $M\delta\omega^{(2)}_R$ &
$M\omega^{(0)}_I$ & $M\delta\omega^{(1)}_I$ & $M\delta\omega^{(2)}_I$ 
\\ 
  \hline
0 & 0.37367 & 0.06289 & 0.04485
& -0.08896 & 0.00100 & 0.00610
\\  
0.1 & 0.37473 & 0.06423 & 0.04638 
& -0.08907 & 0.00102 & 0.00639
\\  
0.2 & 0.37804 & 0.06825 & 0.05109 
&  -0.08937 & 0.00115 & 0.00735
\\  
0.3 & 0.38379 & 0.07499 & 0.05948 
& -0.08972 & 0.00156 & 0.00918
\\  
0.4  & 0.39175 & 0.08449 & 0.07282
& -0.08991 & 0.00252 & 0.01228
\\  
0.5 & 0.40121 & 0.09663 & 0.09390
& -0.08965 & 0.00451 & 0.01725
\\  
0.6  & 0.41104 & 0.11075 & 0.12946
&  -0.08871 & 0.00819 & 0.02498
\\  
0.7  & 0.41987 & 0.12506 & 0.19994
&  -0.08709 & 0.01416 & 0.03650
\\  
0.8  & 0.42646 & 0.13647 & 0.39656
& -0.08526 & 0.02163 & 0.05099
\\  
 \hline
\end{tabular}
\end{center}
\caption{$\mathrm{m}=1$ gravitational modes}
\label{m1_grav}
\end{table}

\begin{table}[h!]
\begin{center}
\begin{tabular}{ || c | c c c | c c c || }
 \hline
$Q/r_H$ & $M\omega^{(0)}_R$ & $M\delta\omega^{(1)}_R$ & $M\delta\omega^{(2)}_R$ &
$M\omega^{(0)}_I$ & $M\delta\omega^{(1)}_I$ & $M\delta\omega^{(2)}_I$ 
\\ 
  \hline
0 & 0.37368 & 0.00000 & 0.03592
& -0.08896 & 0.00000 & 0.00638
\\  
0.1 & 0.37473 & 0.00000 & 0.03708 
& -0.08907 & 0.00000 & 0.00668
\\  
0.2 & 0.37804 & 0.00000 & 0.04062
&  -0.08937 & 0.00000 & 0.00765
\\  
0.3 & 0.38379 & 0.00000 & 0.04689 
& -0.08973 & 0.00000 & 0.00948
\\  
0.4  & 0.39175 & 0.00000 & 0.05701
& -0.08991 & 0.00000 & 0.01242
\\  
0.5 & 0.40121 & 0.00000 & 0.07390
& -0.08965 & 0.00000 & 0.01666
\\  
0.6  & 0.41104 & 0.00000 & 0.10582
&  -0.08871 & 0.00000 & 0.02198
\\  
0.7  & 0.41986 & 0.00000 & 0.18087
&  -0.08710 & 0.00000 & 0.02699
\\  
0.8  & 0.42645 & 0.00001 & 0.42502
& -0.08525 & 0.00001 & 0.02606
\\  
 \hline
\end{tabular}
\end{center}
\caption{$\mathrm{m}=0$ gravitational modes}
\label{m0_grav}
\end{table}

\begin{table}[h!]
\begin{center}
\begin{tabular}{ || c | c c c | c c c || }
 \hline
$Q/r_H$ & $M\omega^{(0)}_R$ & $M\delta\omega^{(1)}_R$ & $M\delta\omega^{(2)}_R$ &
$M\omega^{(0)}_I$ & $M\delta\omega^{(1)}_I$ & $M\delta\omega^{(2)}_I$ 
\\ 
  \hline
0 & 0.45760 & 0.14251 & 0.07164
& -0.09501 & 0.00131 & 0.00720
\\  
0.1 & 0.46286 & 0.14557 & 0.07293 
& -0.09537 & 0.00149 & 0.00765
\\  
0.2 & 0.47816 & 0.15506 & 0.07687
& -0.09634 & 0.00205 & 0.00908
\\  
0.3 & 0.50228 & 0.17178 & 0.08347  
& -0.09761 & 0.00302 & 0.01183
\\  
0.4  & 0.53362 & 0.19681 & 0.09194
& -0.09871 & 0.00464 & 0.01626
\\  
0.5 & 0.57013 & 0.23125 & 0.09979
& -0.09907 & 0.00740 & 0.02312
\\  
0.6  & 0.60903 & 0.27560 & 0.09990
& -0.09812 & 0.01229 & 0.03299
\\  
0.7  & 0.64652 & 0.32822 & 0.07028
& -0.09548 & 0.02081 & 0.04846
\\  
0.8  & 0.67774 & 0.38230 & -0.07677
& -0.09144 & 0.03413 & 0.06612
\\  
 \hline
\end{tabular}
\end{center}
 \caption{$\mathrm{m}=2$ electromagnetic modes}
 \label{m2_em}
\end{table}

\begin{table}[h!]
\begin{center}
\begin{tabular}{ || c | c c c | c c c || }
 \hline
$Q/r_H$ & $M\omega^{(0)}_R$ & $M\delta\omega^{(1)}_R$ & $M\delta\omega^{(2)}_R$ &
$M\omega^{(0)}_I$ & $M\delta\omega^{(1)}_I$ & $M\delta\omega^{(2)}_I$ 
\\ 
  \hline
0 & 0.45760 & 0.07124 & 0.04279
& -0.09501 & 0.00066 & 0.00710
\\  
0.1 & 0.46286 & 0.07277 & 0.04289
& -0.09537 & 0.00076 & 0.00754
\\  
0.2 & 0.47816 & 0.07752 & 0.04297
& -0.09634 & 0.00103 & 0.00896
\\  
0.3 & 0.50228 & 0.08588 & 0.04205
& -0.09761 & 0.00152 & 0.01157
\\  
0.4 & 0.53362 & 0.09840 & 0.03788
& -0.09871 & 0.00233 & 0.01583
\\  
0.5 & 0.57013 & 0.11562 & 0.02336
& -0.09907 & 0.00371 & 0.02227
\\  
0.6  & 0.60903 & 0.13780 & -0.01459
& -0.09812 & 0.00616 & 0.03104
\\  
0.7  & 0.64652 & 0.16411 & -0.11308
& -0.09548 & 0.01042 & 0.04136
\\  
0.8  & 0.67774 & 0.19115 & -0.40182
& -0.09144 & 0.01706 & 0.04652
\\  
 \hline
\end{tabular}
\end{center}
 \caption{$\mathrm{m}=1$ electromagnetic modes}
 \label{m1_em}
\end{table}

\begin{table}[h!]
\begin{center}
\begin{tabular}{ || c | c c c | c c c || }
 \hline
$Q/r_H$ & $M\omega^{(0)}_R$ & $M\delta\omega^{(1)}_R$ & $M\delta\omega^{(2)}_R$ &
$M\omega^{(0)}_I$ & $M\delta\omega^{(1)}_I$ & $M\delta\omega^{(2)}_I$ 
\\ 
  \hline
0 & 0.45760 & 0.00000 & 0.03311
& -0.09501 & 0.00000 & 0.00699
\\  
0.1 & 0.46286 & 0.00000 & 0.03281
& -0.09537 & 0.00000 & 0.00744
\\  
0.2 & 0.47816 & 0.00000 & 0.03161
& -0.09634 & 0.00000 & 0.00885
\\  
0.3 & 0.50228 & 0.00000 & 0.02821
& -0.09761 & 0.00000 & 0.01146
\\  
0.4  & 0.53362 & 0.00000 & 0.01938
& 0.09871 & 0.00000 & 0.01566
\\  
0.5 & 0.57013 & 0.00000 & -0.00219
& -0.09907 & 0.00000 & 0.02193
\\  
0.6  & 0.60903 & 0.00000 & -0.05283
& -0.09812 & 0.00000 & 0.03023
\\  
0.7  & 0.64652 & 0.00000 & -0.17417
& -0.09548 & 0.00000 & 0.03863
\\  
0.8  & 0.67774 & 0.00000 & -0.51129
& -0.09144 & 0.00000 & 0.04129
\\  
 \hline
\end{tabular}
\end{center}
 \caption{$\mathrm{m}=0$ electromagnetic modes}
 \label{m0_em}
\end{table}

\begin{table}[h!]
\begin{center}
\begin{tabular}{ || c | c c c | c c c || }
 \hline
$Q/r_H$ & $M\omega^{(0)}_R$ & $M\delta\omega^{(1)}_R$ & $M\delta\omega^{(2)}_R$ &
$M\omega^{(0)}_I$ & $M\delta\omega^{(1)}_I$ & $M\delta\omega^{(2)}_I$ 
\\ 
  \hline
0 & 0.48364 & 0.15050 & 0.07415
& -0.09676 & 0.00014 & 0.00704
\\  
0.1 & 0.48687 & 0.15351 & 0.07608
& -0.09696 & 0.00028 & 0.00745
\\  
0.2 & 0.49631 & 0.16266 & 0.08197
& -0.09750 & 0.00075 & 0.00872
\\  
0.3 & 0.51133 & 0.17829 & 0.09228 
& -0.09814 & 0.00173 & 0.01109
\\  
0.4  & 0.53362 & 0.19681 & 0.09194
& -0.09871 & 0.00464 & 0.01626
\\  
0.5 & 0.55305 & 0.23032 & 0.12864
& -0.09834 & 0.00703 & 0.02198
\\  
0.6  & 0.57601 & 0.26601 & 0.15525
& -0.09714 & 0.01304 & 0.03311
\\  
0.7  & 0.59707 & 0.30485 & 0.18383
& -0.09487 & 0.02273 & 0.05106
\\  
0.8  & 0.61354 & 0.34035 & 0.20704
& -0.09199 & 0.03595 & 0.07305
\\  
 \hline
\end{tabular}
\end{center}
 \caption{$\mathrm{m}=2$ scalar modes}
 \label{m2_sc}
\end{table}

\begin{table}[h!]
\begin{center}
\begin{tabular}{ || c | c c c | c c c || }
 \hline
$Q/r_H$ & $M\omega^{(0)}_R$ & $M\delta\omega^{(1)}_R$ & $M\delta\omega^{(2)}_R$ &
$M\omega^{(0)}_I$ & $M\delta\omega^{(1)}_I$ & $M\delta\omega^{(2)}_I$ 
\\ 
  \hline
0 & 0.48364 & 0.07525 & 0.04214
& -0.09676 & 0.00007 & 0.00731
\\  
0.1 & 0.48687 & 0.07675 & 0.04284
& -0.09696 & 0.00014 & 0.00772
\\  
0.2 & 0.49631 & 0.08133 & 0.04482
& -0.09750 & 0.00037 & 0.00895
\\  
0.3 & 0.51133 & 0.08914 & 0.04783
& -0.09814 & 0.00086 & 0.01123
\\  
0.4  & 0.53362 & 0.09840 & 0.03788
& -0.09871 & 0.00233 & 0.01583
\\  
0.5 & 0.55305 & 0.11516 & 0.05359
& -0.09834 & 0.00352 & 0.02035
\\  
0.6  & 0.57601 & 0.13301 & 0.05239
& -0.09714 & 0.00652 & 0.02785
\\  
0.7  &  0.59707 & 0.15243 & 0.04312
& -0.09487 & 0.01136 & 0.03663
\\  
0.8  & 0.61354 & 0.17017 & 0.02102
& -0.09199 & 0.01797 & 0.04268
\\  
 \hline
\end{tabular}
\end{center}
 \caption{$\mathrm{m}=1$ scalar modes}
 \label{m1_sc}
\end{table}

\begin{table}[h!]
\begin{center}
\begin{tabular}{ || c | c c c | c c c || }
 \hline
$Q/r_H$ & $M\omega^{(0)}_R$ & $M\delta\omega^{(1)}_R$ & $M\delta\omega^{(2)}_R$ &
$M\omega^{(0)}_I$ & $M\delta\omega^{(1)}_I$ & $M\delta\omega^{(2)}_I$ 
\\ 
  \hline
0 & 0.48364 & 0.00000 & 0.03157
& -0.09676 & 0.00000 & 0.00751
\\  
0.1 & 0.48687 & 0.00000 & 0.03176
& -0.09696 & 0.00000 & 0.00781
\\  
0.2 &  0.49631 & 0.00000 & 0.03244
& -0.09750 & 0.00000 & 0.00902
\\  
0.3 & 0.51133 & 0.00000 & 0.03302
& -0.09814 & 0.00000 & 0.01126
\\  
0.4  & 0.53362 & 0.00000 & 0.01934
& -0.09871 & 0.00000 & 0.01571
\\  
0.5 & 0.55305 & 0.00000 & 0.01934
& -0.09834 & 0.00000 & 0.01984
\\  
0.6  &0.57601 & 0.00000 & 0.01934
& -0.09714 & 0.00000 & 0.02611
\\  
0.7  &  0.59707 & 0.00000 & 0.01934
& -0.09487 & 0.00000 & 0.03184
\\  
0.8  & 0.61354 & 0.00000 & 0.01934
& -0.09199 & 0.00000 & 0.03255
\\  
 \hline
\end{tabular}
\end{center}
 \caption{$\mathrm{m}=0$ scalar modes}
 \label{m0_sc}
\end{table}

\newpage

\bibliographystyle{unsrt}

\end{document}